\documentclass[10pt,twocolumn,letterpaper]{article}

\usepackage{cvpr}              

\usepackage{graphicx}
\usepackage{amsmath}
\usepackage{amssymb}
\usepackage{booktabs}
\newcommand{\raw}{M}
\newcommand{\rawl}{M}
\newcommand{\rawr}{S}
\newcommand{\rgb}{O}
\newcommand{\sellname}{StereoISP}
\newcommand{\height}{p}
\newcommand{\width}{r}

\usepackage[pagebackref,breaklinks,colorlinks]{hyperref}

\usepackage[capitalize]{cleveref}
\crefname{section}{Sec.}{Secs.}
\Crefname{section}{Section}{Sections}
\Crefname{table}{Table}{Tables}
\crefname{table}{Tab.}{Tabs.}

\def\cvprPaperID{10044} 
\def\confName{CVPR}
\def\confYear{2023}

\begin{document}

\title{StereoISP: Rethinking Image Signal Processing for Dual Camera Systems}
\author{
Ahmad Bin Rabiah\\
Purdue University\\
{\tt\small abinrabi@purdue.edu}
\and
Qi Guo\\
Purdue University\\
{\tt\small guo675@purdue.edu}
}
\maketitle

\begin{abstract}

Conventional image signal processing (ISP) frameworks are designed to reconstruct an RGB image from a single raw measurement. As multi-camera systems become increasingly popular these days, it is worth exploring improvements in ISP frameworks by incorporating raw measurements from multiple cameras. This manuscript is an intermediate progress report of a new ISP framework that is under development, \sellname. It employs raw measurements from a stereo camera pair to generate a demosaicked, denoised RGB image by utilizing disparity estimated between the two views. We investigate \sellname~by testing the performance on raw image pairs synthesized from stereo datasets. Our preliminary results show an improvement in the PSNR of the reconstructed RGB image by at least 2dB on KITTI 2015 and drivingStereo datasets using ground truth sparse disparity maps. 

\end{abstract}

\section{Introduction}
\label{sec:intro}
\noindent Multi-camera systems have been unprecedentedly useful on various artificial platforms such as mobile phones, autonomous vehicles, and drones. For example, the dual camera system on the latest iPhone 14 can synthesize large aperture photos and achieve continuous, hybrid-optical-and-digital zooming. Given their growing popularity, it is worth asking how image signal processing (ISP) framework can be improved by leveraging raw measurements from multiple cameras.

\begin{figure}[!t]
\begin{subfigure}{.48\textwidth}
     \centering
    \includegraphics[width=\linewidth]{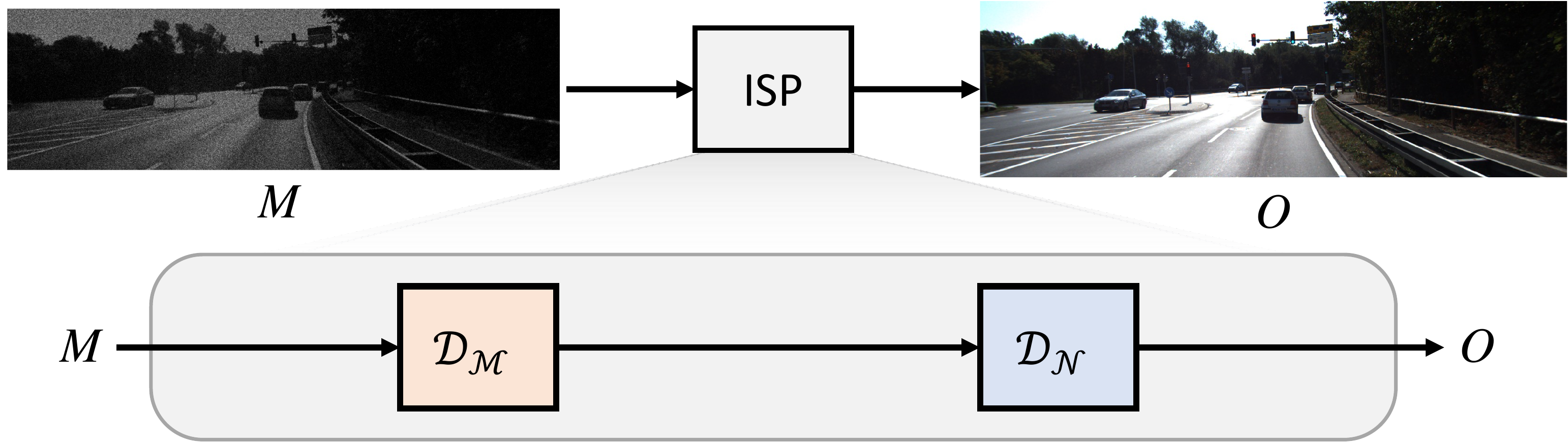}
    \caption{Conventional ISP}
    \label{fig:conventional ISP}  
\end{subfigure}
\medskip

\begin{subfigure}{.48\textwidth}
     \centering
    \includegraphics[width=\linewidth]{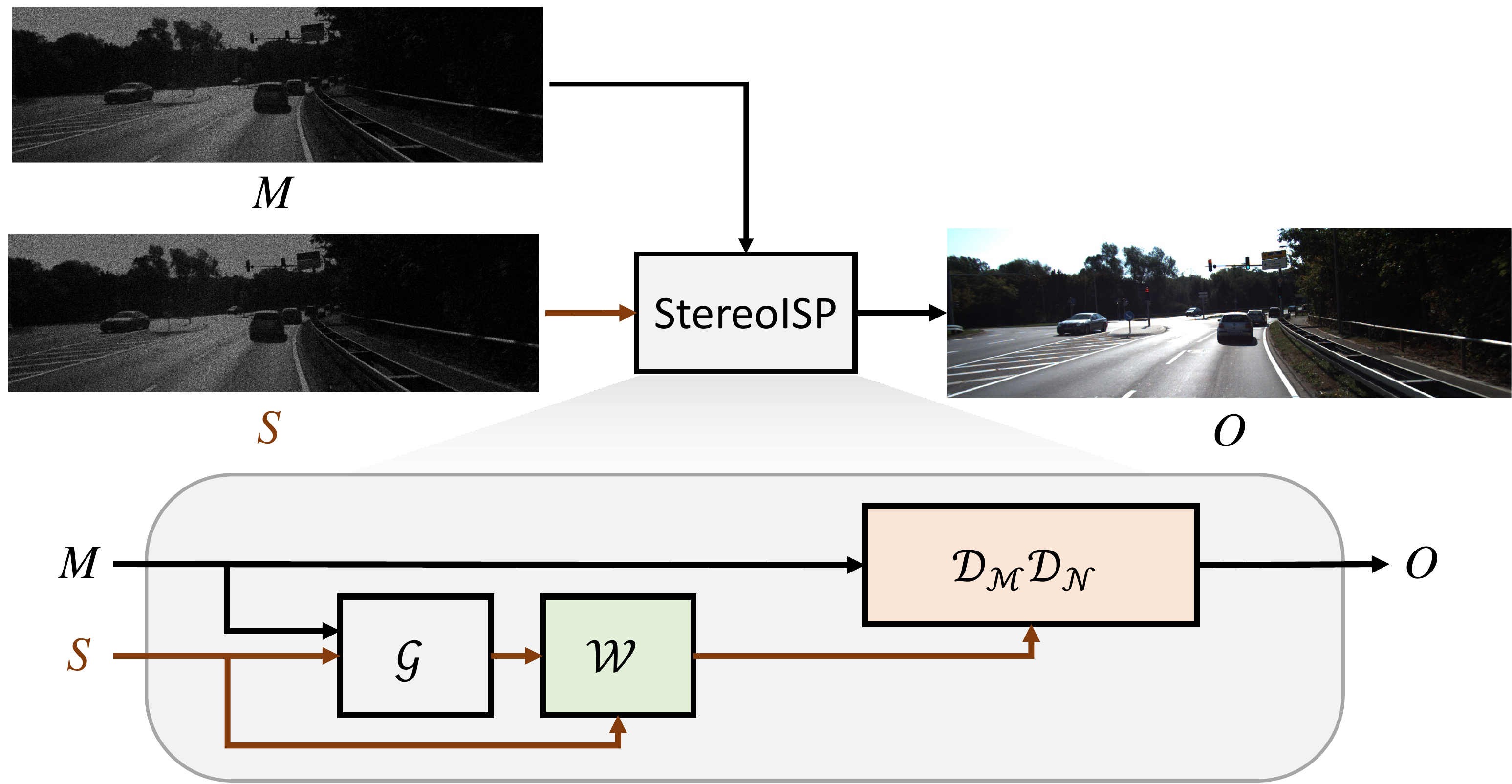}
    \caption{\sellname}
    \label{fig:stereoISP}  
\end{subfigure}

\caption{\sellname~vs. conventional ISP. (a) Conventional image signal processing (ISP) recovers an RGB image from a single raw measurement by sequentially performing demosaicking $\mathcal{D_M}$ and denoising $\mathcal{D_N}$. (b) \sellname~is specialized for dual camera systems. It utilizes raw measurements $M$ and $S$ captured by the cameras to output an RGB image with improved quality. The diagram shows the ultimate framework. It feeds the raw image pair $\rawl$ and $\rawr$ into a stereo matching module $\mathcal{G}$ to estimate the disparity between raw images. The warping module $\mathcal{W}$ warps $S$ according to the estimated disparity to align with $M$. Finally, the framework sends the raw measurement $\rawl$ and the raw measurement warped from $\rawr$ into a joint demosaicking and denoising module $\mathcal{D_MD_N}$ to generate the RGB image $\rgb$. At the current stage, we haven't successfully developed the stereo matching module $\mathcal{G}$, so our experiment uses ground truth disparity maps between $M$ and $S$.}

\label{fig: frameworks}

\end{figure}

In digital photography, the ISP framework transforms a color-mosaicked, noisy raw measurement $\raw$ to an RGB image $\rgb$. The conventional ISP framework sequentially performs demosaicking and denoising on the raw image $\raw$ (\cref{fig:conventional ISP}). Since the birth of digital photography, there have been numerous methods on improving this ISP framework~\cite{liu2014fast, gharbi2016deep, godard2018deep, rong2020burst, liang2020decoupled, qian2022rethinking, li2022efficient}.
Some most recent ones include: switching the sequence of demosaicking and denoising~\cite{qian2022rethinking}, performing demosaicking and denoising simultaneously~\cite{gharbi2016deep, xing2021end}, and using burst captures to improve the quality of the final output~\cite{li2022efficient, liang2020decoupled}. But to the best of our knowledge, there have been very limited studies on designing ISP frameworks specialized for multi-camera systems, despite their recent popularity. 






We propose \sellname. It is an ISP framework that takes in raw measurements $\rawl, \rawr$ from a dual camera system and outputs an RGB image $\rgb$
(\cref{fig:stereoISP}). \sellname~utilizes a deep stereo-matching model to find the disparity map between the raw image pairs. It warps one of the raw measurement according to the estimated disparity to align with the other, and then feeds both raw measurements into a deep neural network for joint denoising and demosaicking. 

There are two key components that empowers \sellname. First, the stereo-matching architecture in \sellname~uses raw stereo pairs as inputs. We demonstrate experimentally that the accuracy of the output disparity map is close to that of using RGB image pairs as inputs. Second, \sellname~explicitly performs alignment of raw image pairs via warping before denoising and demosaicking. This is different from recent work in burst photography~\cite{li2022efficient, mildenhall2018burst} that directly use unwarped images for restorations. We show via ablation study that the output image quality of~\sellname~drops to the level of  single image ISP if the warping operation is removed, which suggests its importance. This is probably because the disparity between stereo image pairs are systematically larger than the disparity among burst images. It is more challenging to implicitly learn the pixel correspondences between images while performing denoising and demosaicking when the disparity is large. 

At the current stage, we test the performance of \sellname~on raw image pairs synthesized from stereo datasets, KITTI 2015~\cite{menze2015object} and DrivingStereo~\cite{yang2019drivingstereo}. Our current experiment shows that \sellname~boosts the PSNR of the output RGB image by 2.32dB and 2.43dB on KITTI 2015 and drivingStereo datasets using the ground truth sparse disparity map. As discussed above, we also provide comprehensive ablation studies to validate stereo matching on raw images, the importance of the warping operation for the framework.

The contribution of this papers include:
\begin{itemize}
    \item A novel ISP framework specialized for dual camera systems
    \item A comprehensive experimental study to understand the proposed ISP framework
\end{itemize}

The code and the pre-trained models of~\sellname~will be published online for the community to use and research.

\begin{figure*}[h]
    \centering
    \includegraphics[width=\linewidth]{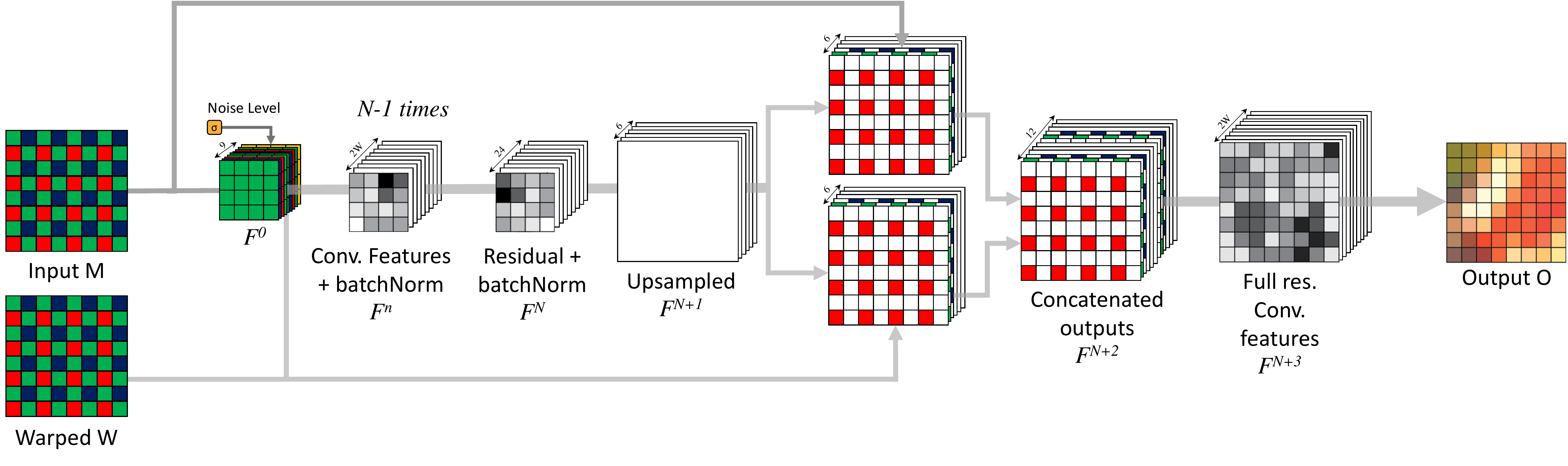}
    \caption{Stereo DemosaicNet inspired by original DemosaicNet \cite{gharbi2016deep} that provides joint demosaicking and denoising for a single camera. Without the loss of generality of StereoISP, we extended on DemosaicNet to explicitly align the warped image. Note that the model architecture follows \cref{eq: framework} and assumed that output of  $\mathcal{G}$ has been founded and warped $W$ obtained from $\mathcal{W}$.}
    \label{fig: stereo demosaicnet}
\end{figure*}

\section{Related Work}
\label{sec: related work}

\textbf{Deep stereo matching.} Designing a good structure for stereo matching is challenging, as it consists of multiple stages that need to be integrated: (1) feature networks extracting general context from 2D input images. (2) cost volume computation which stores the distance between 2D features. (3) 3D matching networks. (4) a powerful disparity regression module to estimate the continuous disparity map \cite{kendall2017end}. MC-CNN \cite{zbontar2016stereo} is the first deep learning-based stereo matching method. It replaces the first stage of stereo matching (the matching cost computation) by learning a similarity measure on small image patches exploiting convolutional neural networks. The first End-to-End network, namely DispNet, has been developed in \cite{mayer2016large}. Authors in \cite{chang2018pyramid} proposed PSMNet, a stereo matching model that exploits spatial pyramid pooling module \cite{he2015spatial} for effective incorporation of the global context of the stereo pair. Recently, LEAStereo \cite{cheng2020hierarchical} is proposed which is an end-to-end hierarchical neural architecture search stereo matching network. the main contribution is the volumetric stereo-matching pipeline which allows the network to automatically select the optimal structures for feature extracting and matching networks. 

\textbf{Single image denoising.} Single image denoising methods can be categorized into two categories: model-based and learning-based methods. BM3D \cite{dabov2007image} falls into a model-based category, which is regarded as a denoising benchmark. In 2017, DnCNN \cite{zhang2017beyond} was proposed by exploiting deep convolutional nets and adopting residual learning for blind zero-mean noise denoising. Later on, different learning-based methods were proposed in different domains, not only in supervised learning manner \cite{lehtinen2018noise2noise,guo2019toward,krull2019noise2void}.


\textbf{Joint denoising and demosaicking.} Demosaicking becomes complicated in the presence of noise. Practically, the output of image sensors will affect the quality of reconstructed images. Estimating edge orientation in noisy data is sensitive which may lead to annoying artifacts in the restored images. The first deep joint demosaicking and denoising work was done in \cite{gharbi2016deep} in which authors proposed DemosaicNet, a deep CNN-based demosaicking and denoising algorithm. DemosaicNet is trained on millions of carefully selected image patches and set a milestone on joint deep learning-based demosaicking and denoising approaches. In 2019, an extensive study on the previous work was done in \cite{ehret2019joint}. Authors in \cite{qian2022rethinking} proposed a trinity network (TENet) to jointly solve this composite problem with the super-resolution output. The authors analyzed the mixture problem of demosaicking, denoising, and super-resolution and proposed a new pipeline by moving demosaicking process after the denoising stage. Most recently, authors in \cite{xing2021end} proposed a deep learning-based end-to-end method for joint image demosaicking, denoising, and super-resolution based on a specially designed deep convolutional neural network (CNN). All previous works assumed that a noise level is known in advance or properly estimated.

\section{Framework}
\label{sec:framework}

\subsection{\sellname}

We address our problem in the case of a two-camera ISP. Assume that we have raw image pairs: primary image $\rawl$ and secondary image $\rawr$ such that $\rawl, \rawr \in \mathbb{R}^{\height \times \width}$ where $\height$ and $\height$ are the height and width of images, respectively. $M$ and $S$ can be summarized by a Poisson model such that $M$ and $S \sim$ Poisson($\lambda$), where $\lambda$ is the average total number of photons seen by the sensor \cite{chan2022does}. Our framework proposed that the output denoised image $O$ becomes a function of the primary image and the warped secondary image with the guidance of the disparity map $D$. \cref{eq: framework} demonstrates the proposed framework:
\begin{equation}
    \label{eq: framework}
    O = \mathcal{D_MD_N}(M,\mathcal{W}(S,D)))
\end{equation}

where $\mathcal{W}$ is a warping function results projection of secondary image $S$ onto the primary image $M$ guided by the disparity map $D$. $\mathcal{D_M}$ and $\mathcal{D_N}$ are demosaicking and denoising processes, respectively. So that the primary image and warped image become concatenated to be passed into demosaicking $\mathcal{D_MD_N}$. The output denoised image $O \in \mathbb{R}^{3\times p \times r}$ is an RGB image.

For an end-to-end solution, the disparity map $D$ can be considered as a stereo matching mapping function $g$ exploits the primary and secondary images to perform disparity map $D$, i.e., $D = \mathcal{G}(M, S) \in \mathbb{R}^{p \times r}$. The output $O$ can be re-written as a function of primary and secondary images:

\begin{equation}
    \label{eq: framework M S}
    O = \mathcal{D_MD_N}(M,\mathcal{W}(S,\mathcal{G}(M,S))))
\end{equation}

For generalization purposes, \cref{fig:stereoISP} illustrates the end-to-end novel ISP formulated in \cref{eq: framework M S}. For studying the feasibility of the idea and investigating its practicality, we will continue our discussion based on \cref{eq: framework} to study the upper bound of the framework.

\subsection{Stereo DemosaicNet}
\label{subsec: Stereo DemosaicNet}

\begin{figure*}[!htb]
 \begin{subfigure}{0.33\textwidth}
     \includegraphics[width=\textwidth]{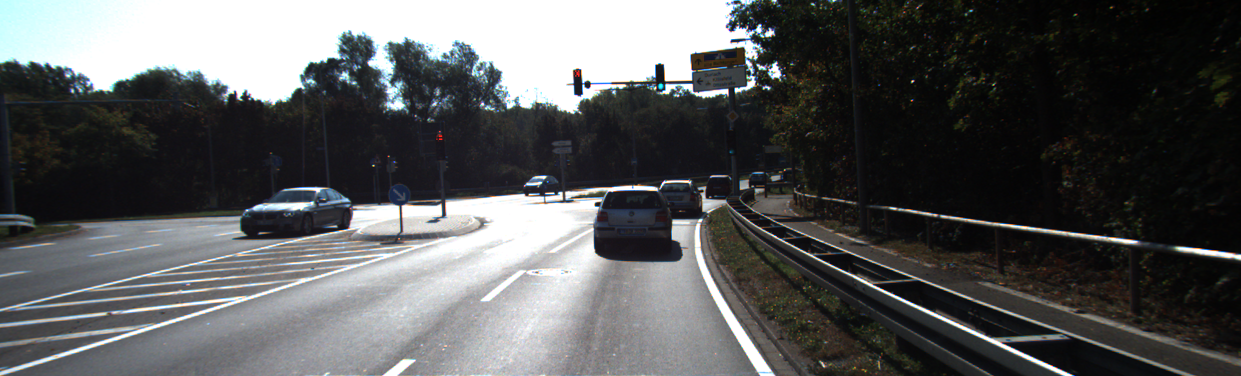}
     \caption{left image (ground truth)}
     
 \end{subfigure}
 \hfill
 \begin{subfigure}{0.33\textwidth}
     \includegraphics[width=\textwidth]{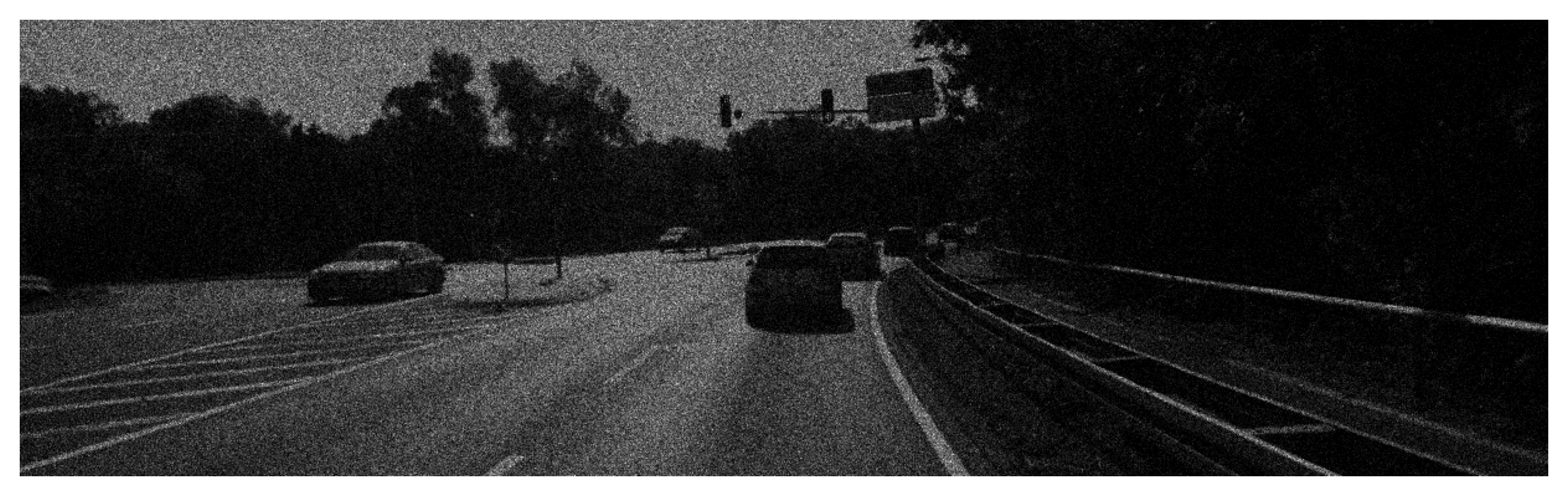}
     \caption{Left raw image}
     
 \end{subfigure}
 \hfill
 \begin{subfigure}{0.33\textwidth}
     \includegraphics[width=\textwidth]{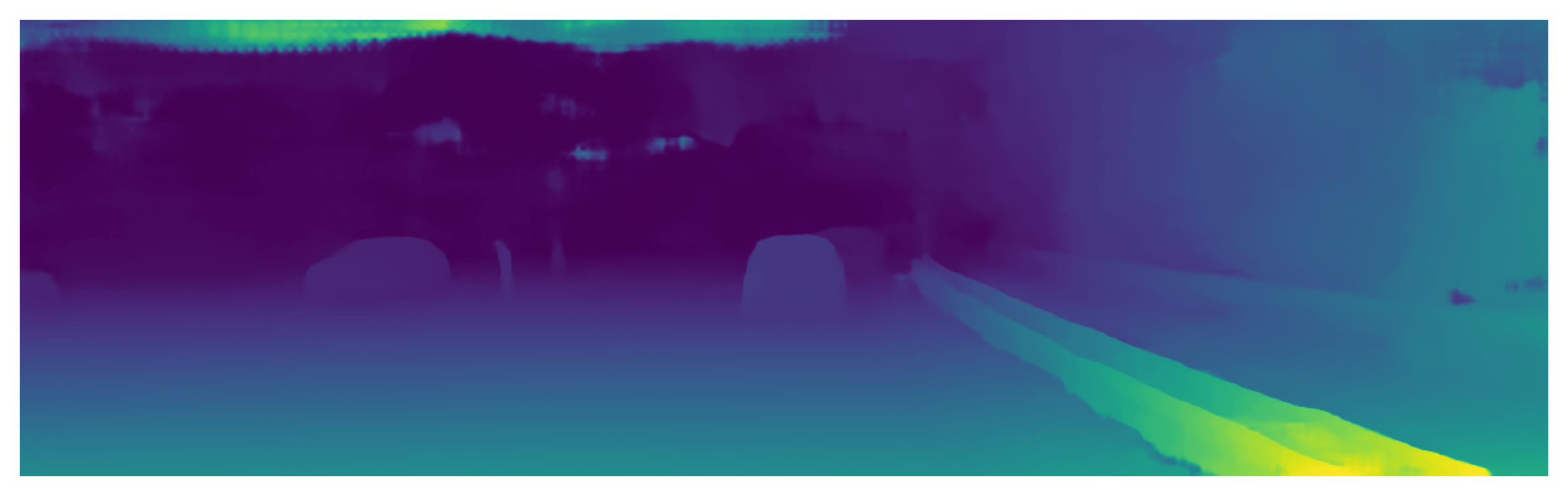}
     \caption{Estimated disparity map from raw images}
     
 \end{subfigure}
  
 \medskip
 
 \begin{subfigure}{0.33\textwidth}
     \includegraphics[width=\textwidth]{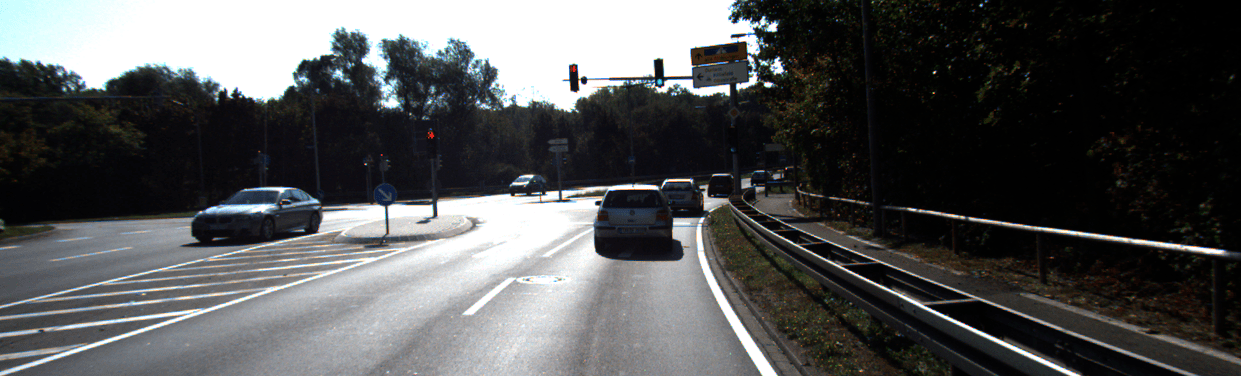}
     \caption{Right image (ground truth)}
     
 \end{subfigure}
 \hfill
 \begin{subfigure}{0.33\textwidth}
     \includegraphics[width=\textwidth]{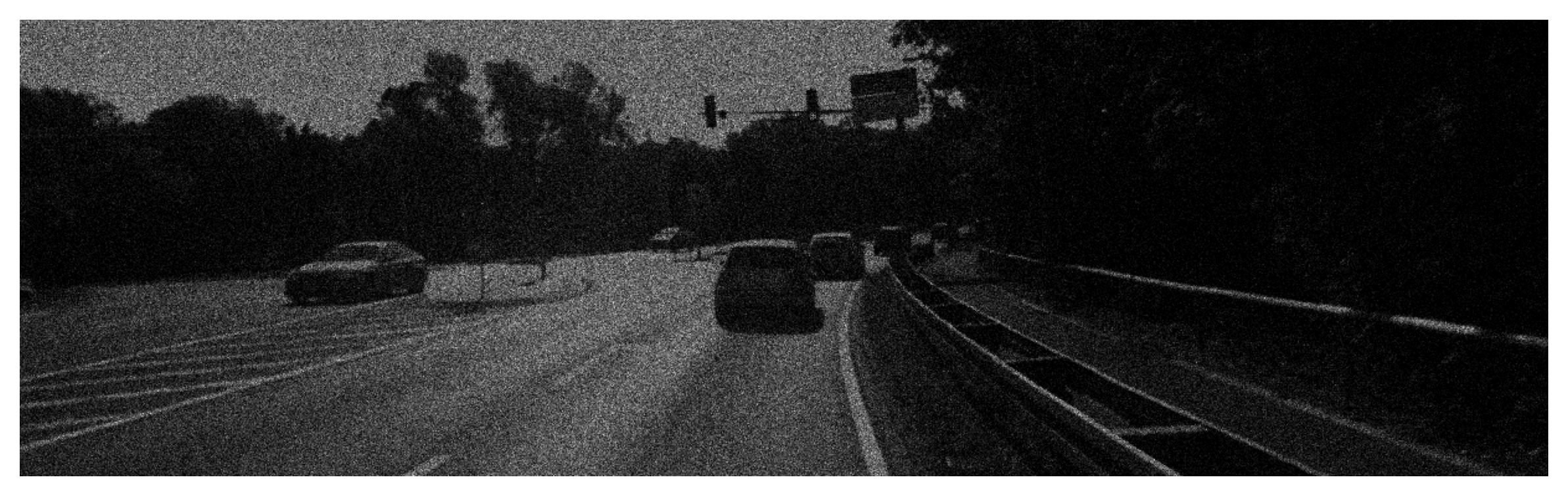}
     \caption{Right raw image}
     
 \end{subfigure}
 \hfill
 \begin{subfigure}{0.33\textwidth}
     \includegraphics[width=\textwidth]{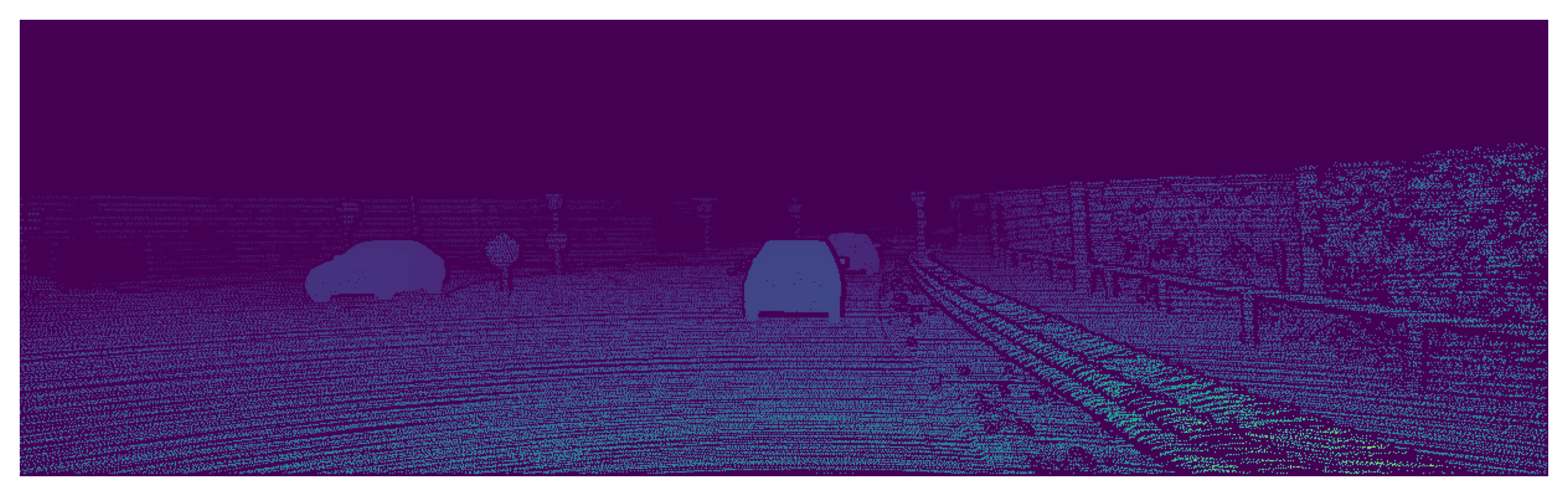}
     \caption{Ground truth disparity map}
     
 \end{subfigure}

 \caption{Disparity estimation using modified PSMNet \cite{chang2018pyramid} with KITTI 2015 \cite{menze2015object} by using raw stereo images.}
 \label{fig: stereo result 1}

\end{figure*}

\begin{figure*}
 \begin{subfigure}{0.33\textwidth}
     \includegraphics[width=\textwidth]{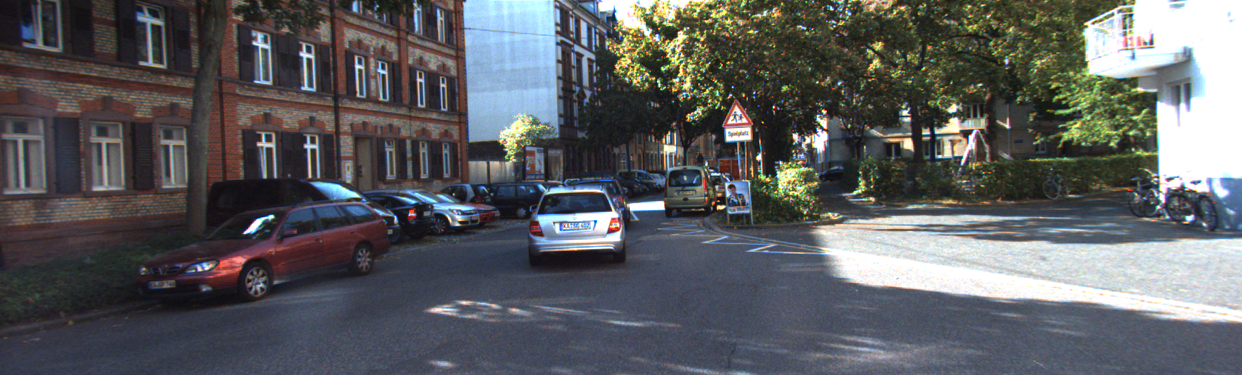}
     \caption{left image (ground truth)}
     
 \end{subfigure}
 \hfill
 \begin{subfigure}{0.33\textwidth}
     \includegraphics[width=\textwidth]{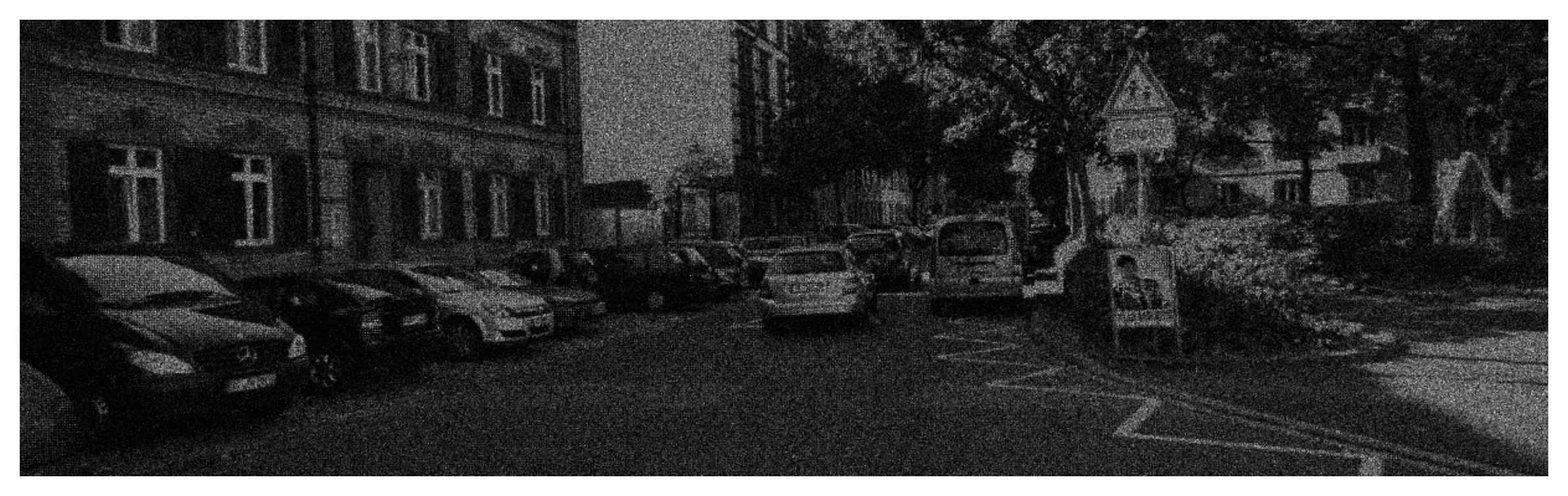}
     \caption{Left raw image}
     
 \end{subfigure}
 \hfill
 \begin{subfigure}{0.33\textwidth}
     \includegraphics[width=\textwidth]{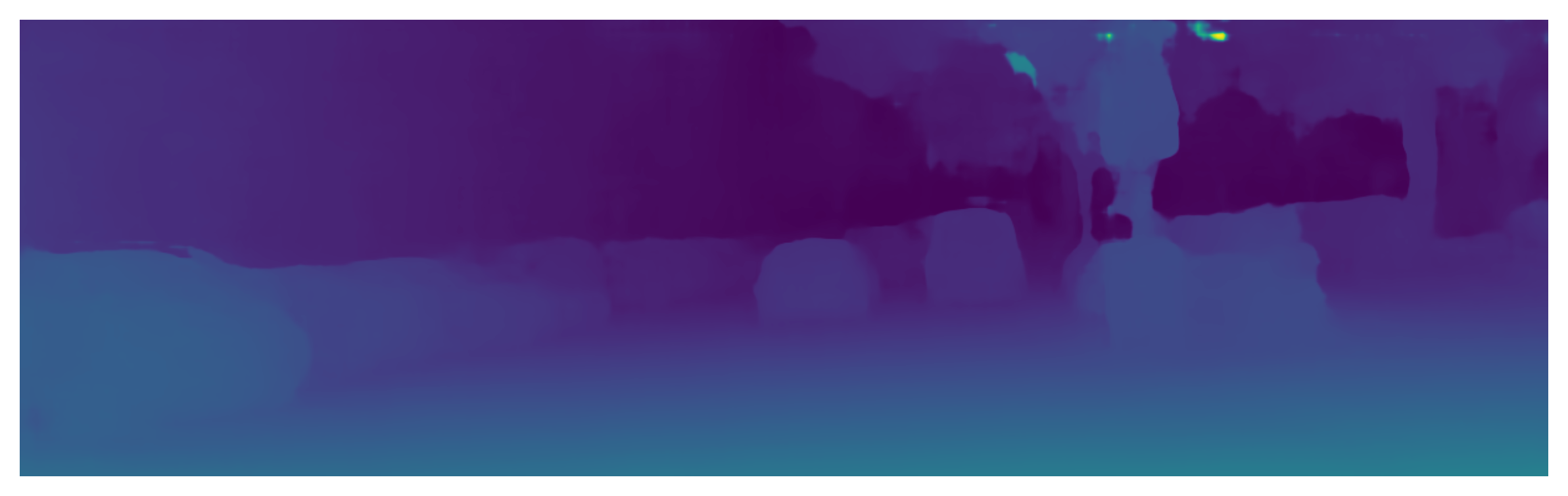}
     \caption{Estimated disparity map from raw images}
     
 \end{subfigure}
  
 \medskip
 
 \begin{subfigure}{0.33\textwidth}
     \includegraphics[width=\textwidth]{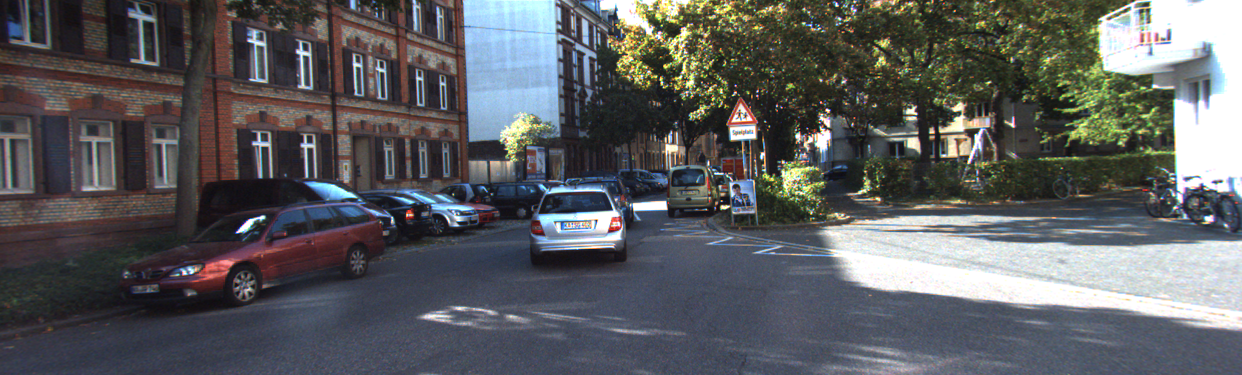}
     \caption{Right image (ground truth)}
     
 \end{subfigure}
 \hfill
 \begin{subfigure}{0.33\textwidth}
     \includegraphics[width=\textwidth]{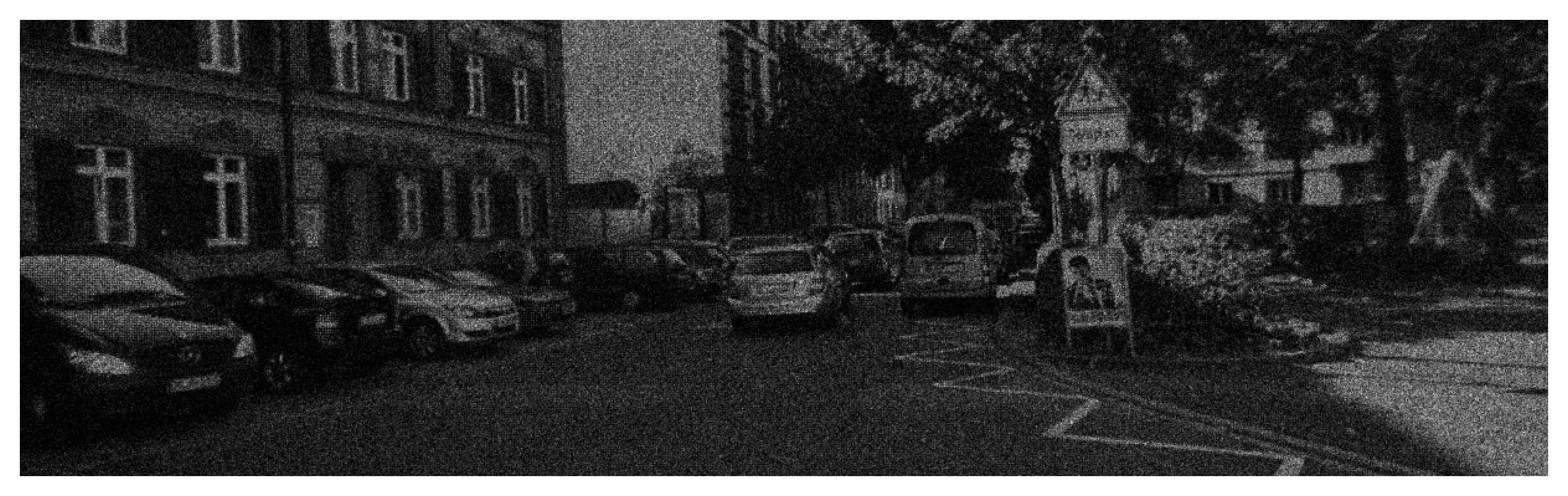}
     \caption{Right raw image}
     
 \end{subfigure}
 \hfill
 \begin{subfigure}{0.33\textwidth}
     \includegraphics[width=\textwidth]{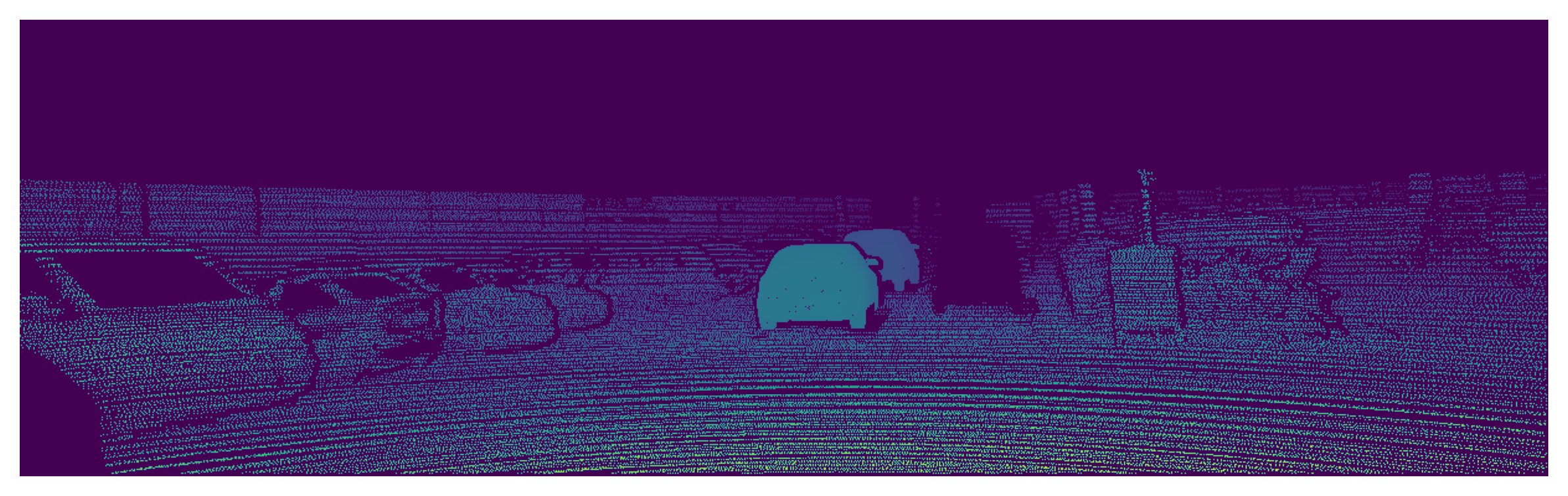}
     \caption{Ground truth disparity map}
     
 \end{subfigure}

 \caption{Disparity estimation using modified PSMNet \cite{chang2018pyramid} with KITTI 2015 \cite{menze2015object} by using raw stereo images.}
 \label{fig: stereo result 2}

\end{figure*}

Inspired by \cite{gharbi2016deep}, we extend on DemosaicNet that is originally designed to obtain denoised RGB images using a single image. In Stereo DemosaicNet, we use a standard feed-forward architecture to implement our stereo operator that has been illustrated in \cref{fig: stereo demosaicnet}. Our network is composed of $N+3$ convolutional layers. We denote W as the number of outputs per input. In the case of the dual camera model, each convolutional layer has 2W outputs using the kernel size $K\times K$, followed by batch normalization layer to prevent the problem of internal covariate shift \cite{ioffe2015batch}. 

We first preprocess a warped image by applying backwarping on the secondary image guided by the disparity map. With the guidance of disparity map $D$, we warp the secondary image to the primary image by creating a rectilinear grid with size $p\times r$. We set the interpolator to interpolate the locations of warped images linearly. We denote the output of the warping process as $W$. We rearrange the samples of the Bayer input mosaic to obtain multi-frame multi-channel images and concatenate them together. This process makes the spatial pattern translation invariant with a period of 1 pixel reducing the computational cost in the following steps. 

After we prepared the data to be passed into the denoising operator (in this case it is stereo DemosaicNet), we pack the primary image and the warped image together, which results in an 8-channel layer $F^0$. We assume the noise level of input images is known or has been computed by a method that estimates the noise level. Since a combination of Poisson and Gaussian noises in linear space accurately models camera noise, we use the Poisson Gaussian noise model in this pipeline \cite{foi2008practical}. We added a 9-th channel to the stack of $F^0$. 

The bulk of the feature extraction process is performed at this lower resolution by the next $N$ blocks. Each block consists of a filter bank of spatial footprints $K\times K$ followed by a batch normalization layer. The output of batchNorm passes through a point-wise ReLU non-linearity $f(\cdot) =$ max(0,$\cdot$). The last layer in low-resolution feature extraction, $F^N$, has 24 channels that correspond to the color samples of a $2\times2$ neighborhood. 

We upsample the output of $F^N$ to a full resolution in $F^{N+1}$. It should be noted that the first 12 channels correspond to the primary image $M$, whereas the last 12 channels correspond to warped input $W$. We split the upsamples output and concatenate masked copies of $M$ and $W$ to their corresponding upsampled features. We stack the split layers where we place the masked input and its corresponding features extracted in the front of $F^{N+2}$ which results in a 12 channels layer. For full-resolution reconstruction, we feed $F^{N+2}$ into a full-resolution convolutional features layer followed by a batch normalization operator and a point-wise ReLU non-linearity function. The final output $O$ on the network is an affine combination of the last feature maps that contain features maps extracted on the lower resolution and forward masked inputs $F^{N+3}$. We opted for a stereo DemosaicNet with W = 15 and D = 64 per input, which is similar to the original work \cite{gharbi2016deep}.
\section{Experiments}
\label{sec: experiments}

\subsection{Training procedure}
\label{subsec: procedure}
In this experiment, assume that the disparity map between stereo pairs is calculated by one of the methods mentioned in \cref{sec: related work}, so we use the ground truth disparity maps. We use a dataset $\mathcal{D} = \{(\mathrm{M}_i, \mathrm{S}_i, \mathrm{D}_i, \mathrm{I}_i)\}_i$ where $\mathrm{M}_i$ is the mosaicked and noisy version of an $i$-th example of RGB image $I_i$. We fix  $\sigma$ for the training procedure to be 10, where $\sigma$ is the average total number of photons seen by the sensor as mentioned in \cref{sec:framework}, We optimize the weights and biases by minimizing $L_2$ norm on the training set:
\begin{equation}
    \mathcal{L}\left(\left\{w^{(n)}, b^{(n)}\right\}_n\right)=\frac{1}{(p \times r)|\mathcal{D}|} \sum_i\left\|\mathrm{O}_i-\mathrm{I}_i\right\|^2
\end{equation}

\noindent such that $w^{(n)}$ and $b^{(n)}$ are weights and biases of $n$-th layer, respectively, and $p \times r$ is the patch size for the training samples. We assess our measurement by utilizing peak signal-to-noise ratio (PSNR) to measure the quality of output image $O_i$ with respect to the ground truth $\mathrm{I}_i$. Note that the maximum value for RGB is 255 per channel as each pixel consists of 8 bits.

\begin{equation}
    \mathrm{PSNR} = 20\log_{10}\left(\frac{\mathrm{MSE}_{O_i}}{\sqrt{\mathrm{MAX}_{I_i}}}\right)
    \label{eq: psnr}
\end{equation}

We exploited KITTI 2015 \cite{menze2015object} that contains 200 pairs, where we split the dataset into 160 training examples and 40 testing examples. to improve image quality, we also considered a large-scale dataset for stereo matching drivingStereo \cite{yang2019drivingstereo} that contains 174,437 training pairs and 7,751 testing pairs. To the best of our knowledge, there is no raw stereo image dataset in the literature. Therefore, we create the mosaicked and noisy images $M$ and $S$ by retaining only one color channel per pixel according to the Bayer pattern from the sRGB images \cite{gharbi2016deep}.

The training has been done in two stages: in the first stage, we pass $\mathrm{M}_i$ twice with different noise structures in both input ports (i.e., $M$ and $W$). The reason is that we want to let the model perform denoising tasks and demosaicing with the ideal disparity between primary and secondary images. That means the projection of the secondary image is identical to the primary image except for the effect of noise. After the model is converged, we replace the input of the $W$ port with a warping block with the sake of letting the model inherits the information of the objects in the secondary image and being added to the primary image which results increasing in the PSNR.

In all experiments, we use a patch size $p = 256$ by $r = 512$ pixels and we initialized all biases $b^{(n)}$ to 0. We fixed the noise level $\sigma = 10$ such that the noise follows Poisson distribution which depicts shot noise. The optimizer is carried out by ADAM  \cite{kingma2014adam}. We use a batch size to be 8 and an initial learning rate of $10^{-4}$. The training is performed with PyTorch on a single NVIDIA V100 GPU with 16 GB memory.

\subsection{Stereo with raw images}

Since stereo-matching for raw images has not been yet well studied, we aim to investigate current stereo-matching architecture to assess the feasibility to do disparity estimation tasks in the raw domain. We use PSMNet \cite{chang2018pyramid} on KITTI 2015 dataset. We present our results for disparity map estimation on raw stereo images and exploiting PSMNet \cite{chang2018pyramid}. Although results needed to be improved, it seems it is promising to develop the feature extraction modules in stereo matching architecture to handle demosaicing/denoising artifacts s shown in figures \ref{fig: stereo result 1} and \ref{fig: stereo result 2}. As our goal in this paper is to prove our hypothesis that we want to investigate the feasibility to inherit more information from the secondary image, our discussion in the following section will focus on using the \textit{ground truth disparity map.}

\subsection{DemosaicNet vs Stereo DemosaicNet}

\begin{table}
  \centering
  
  \begin{tabular}{@{}lc@{}c@{}}
    \toprule
    \toprule
    & KITTI 2015 & drivingStereo\\
    \toprule
    Method & \multicolumn{2}{c}{PSNR (dB)} \\
    \midrule
    DemosaicNet \cite{gharbi2016deep}  & 24.47 & 29.59\\
    Ours (\cref{subsec: Stereo DemosaicNet})  & \textbf{26.79 (+9.48\%)}& \textbf{32.02 (+8.2\%)} \\
    \bottomrule
    \bottomrule
  \end{tabular}
  \caption{Evaluation on KITTI 2015 \cite{gharbi2016deep} and drivingStereo \cite{yang2019drivingstereo} our proposed model architecture based on proposed \sellname~ framework. }
  \label{tab: results}
\end{table}

\begin{figure}
 \begin{subfigure}{0.09\textwidth}
     \includegraphics[width=\textwidth]{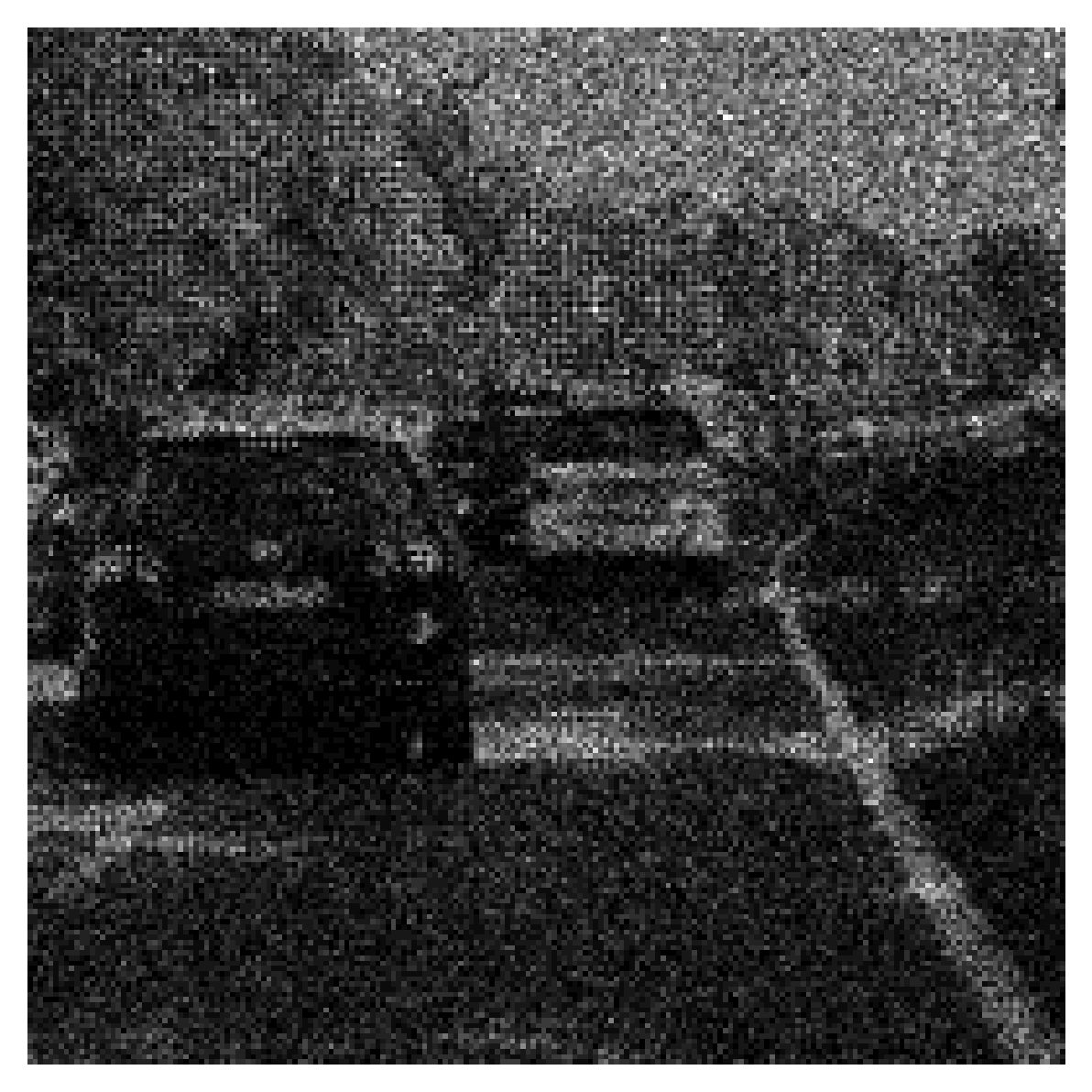}
     
 \end{subfigure}
 \hfill
 \begin{subfigure}{0.09\textwidth}
     \includegraphics[width=\textwidth]{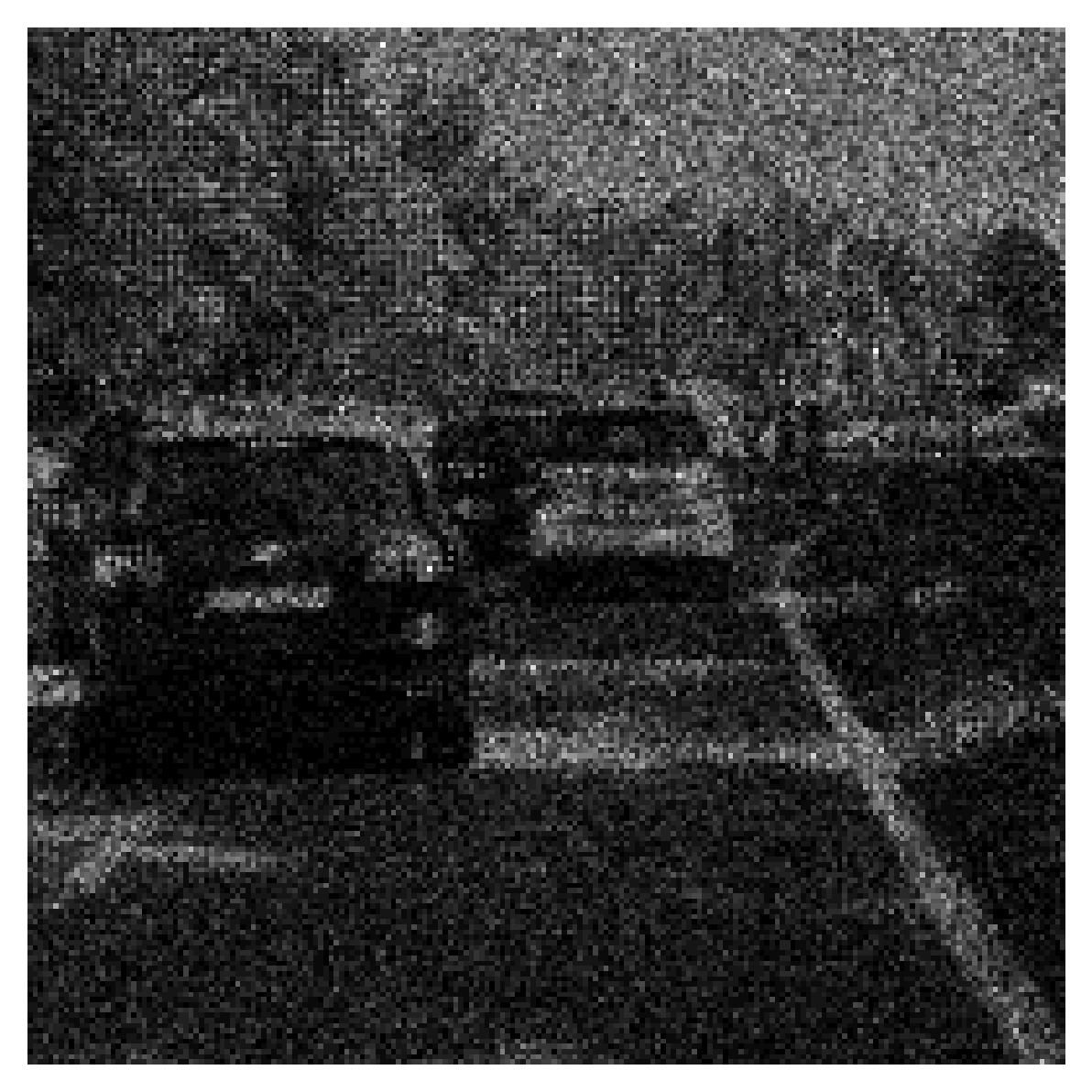}
     
 \end{subfigure}
 \hfill
 \begin{subfigure}{0.09\textwidth}
     \includegraphics[width=\textwidth]{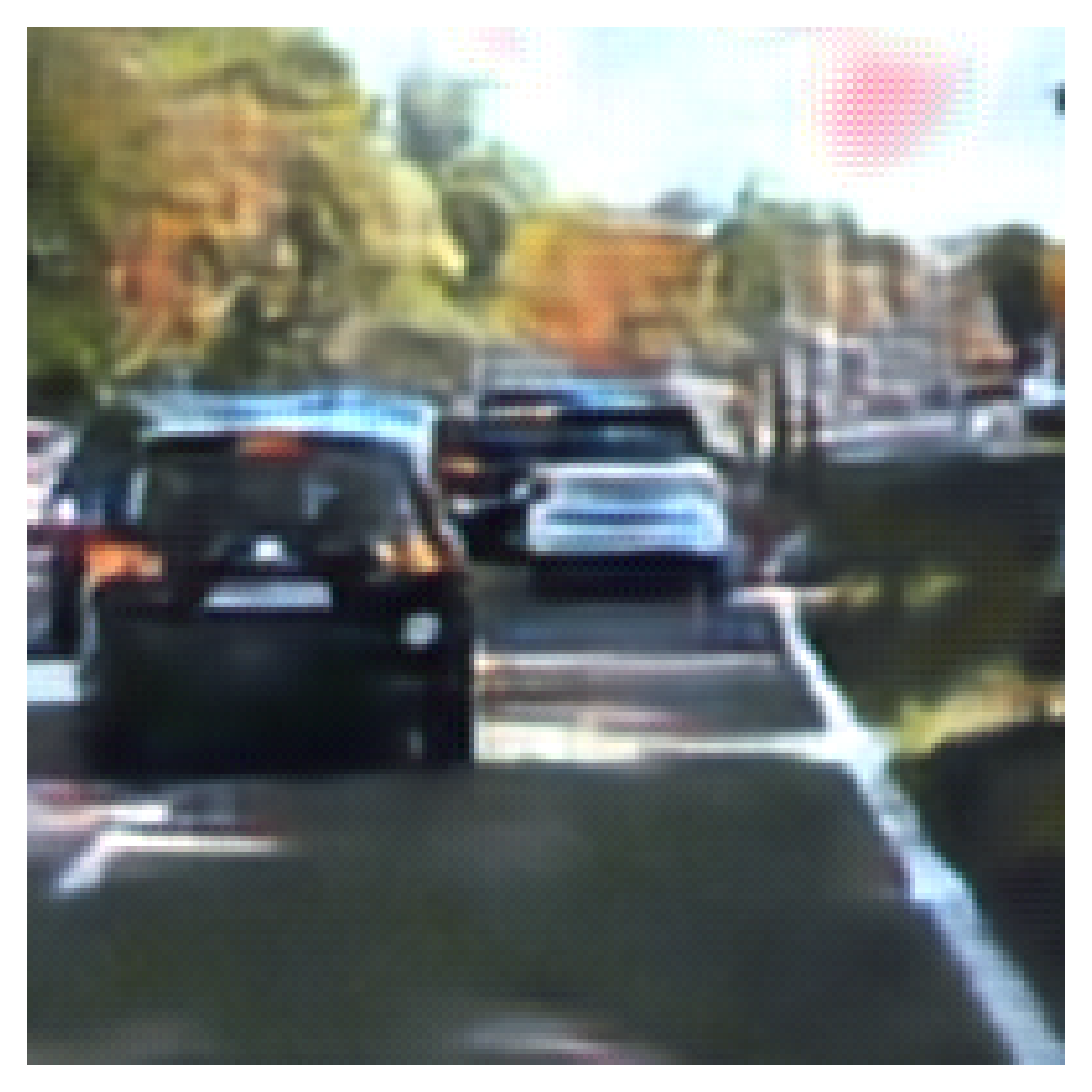}
     
 \end{subfigure}
 \hfill
 \begin{subfigure}{0.09\textwidth}
     \includegraphics[width=\textwidth]{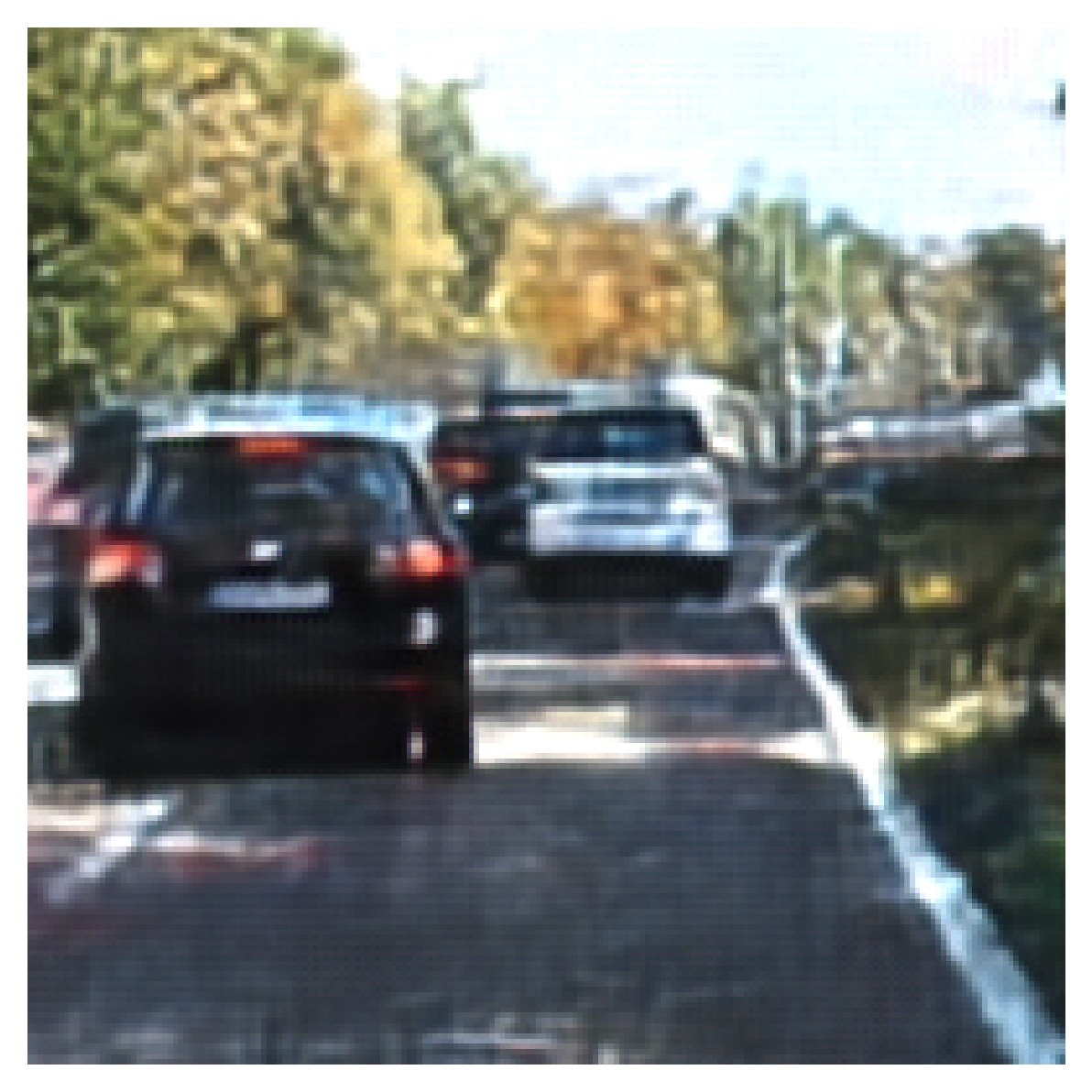}
     
 \end{subfigure}  
 \hfill
 \begin{subfigure}{0.09\textwidth}
     \includegraphics[width=\textwidth]{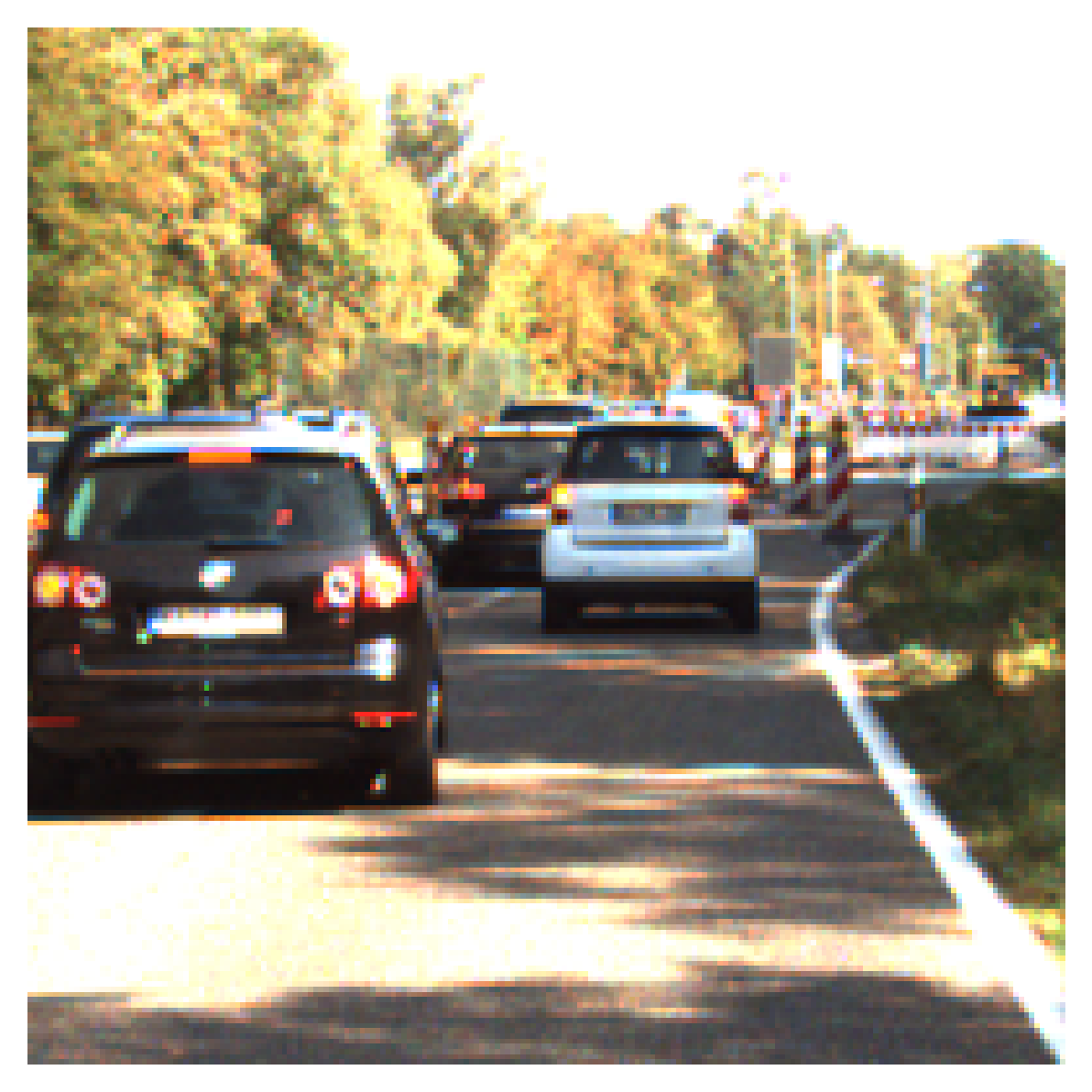}
     
 \end{subfigure}  
 \smallskip
 \begin{subfigure}{0.09\textwidth}
     \includegraphics[width=\textwidth]{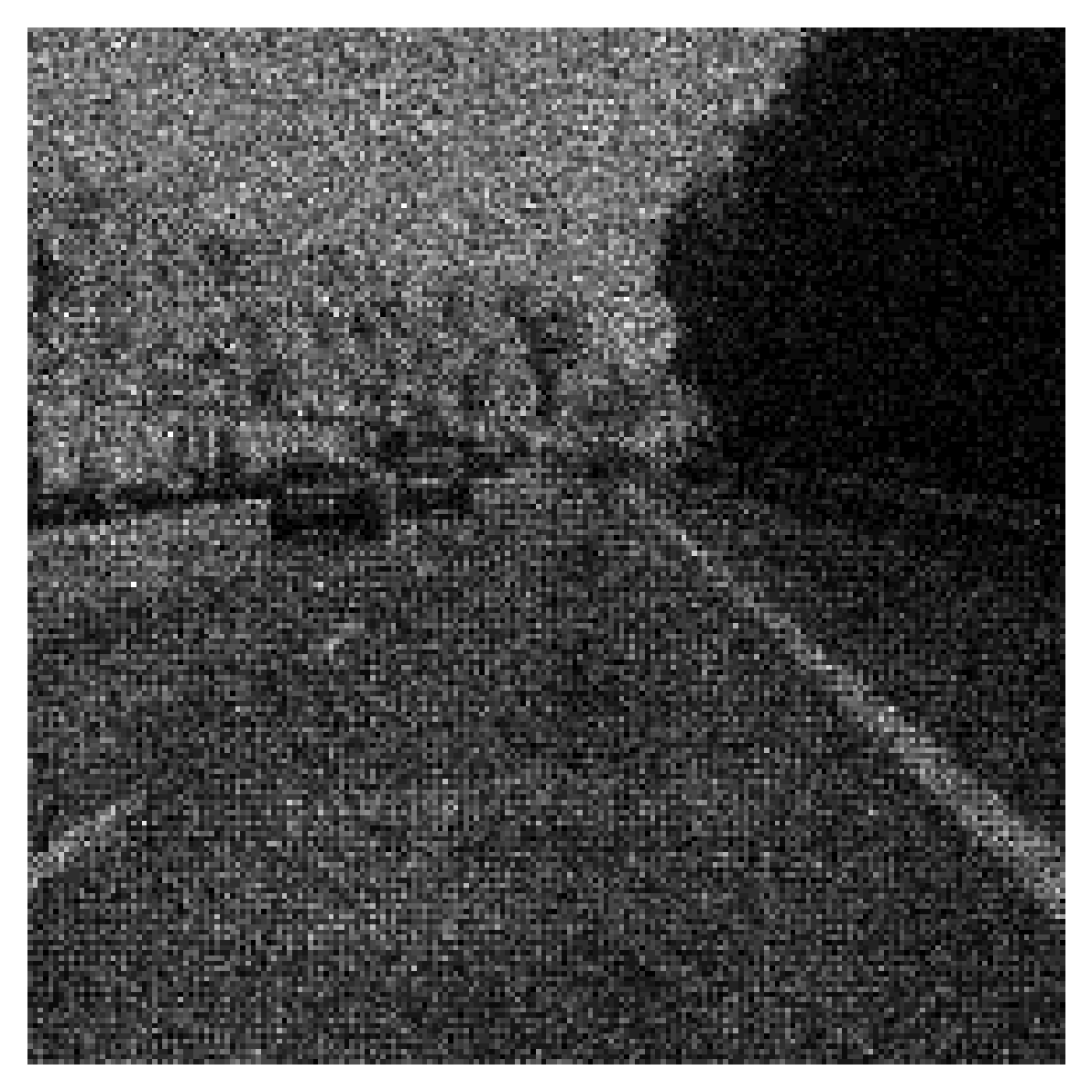}
     
 \end{subfigure}
 \hfill
 \begin{subfigure}{0.09\textwidth}
     \includegraphics[width=\textwidth]{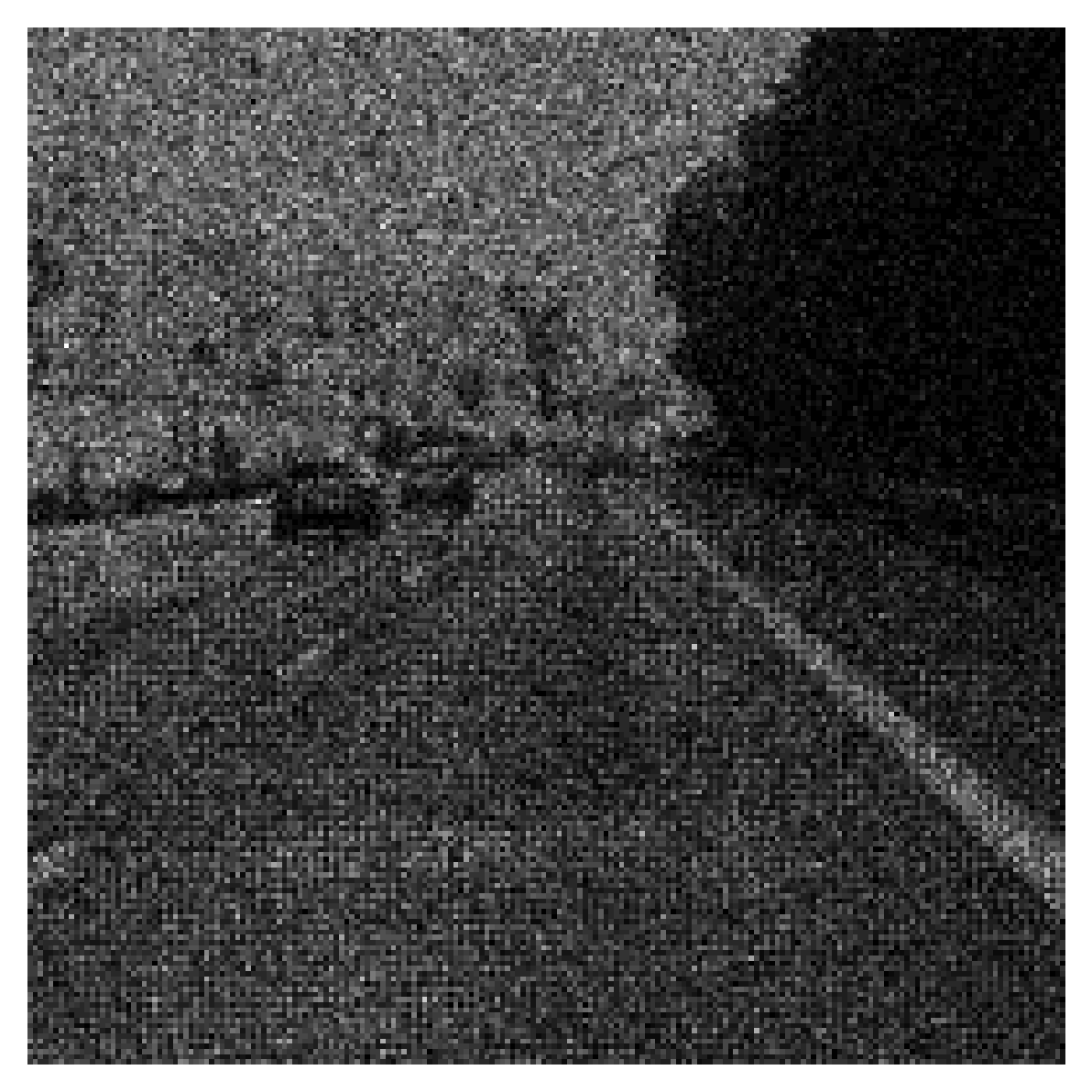}
     
 \end{subfigure}
 \hfill
 \begin{subfigure}{0.09\textwidth}
     \includegraphics[width=\textwidth]{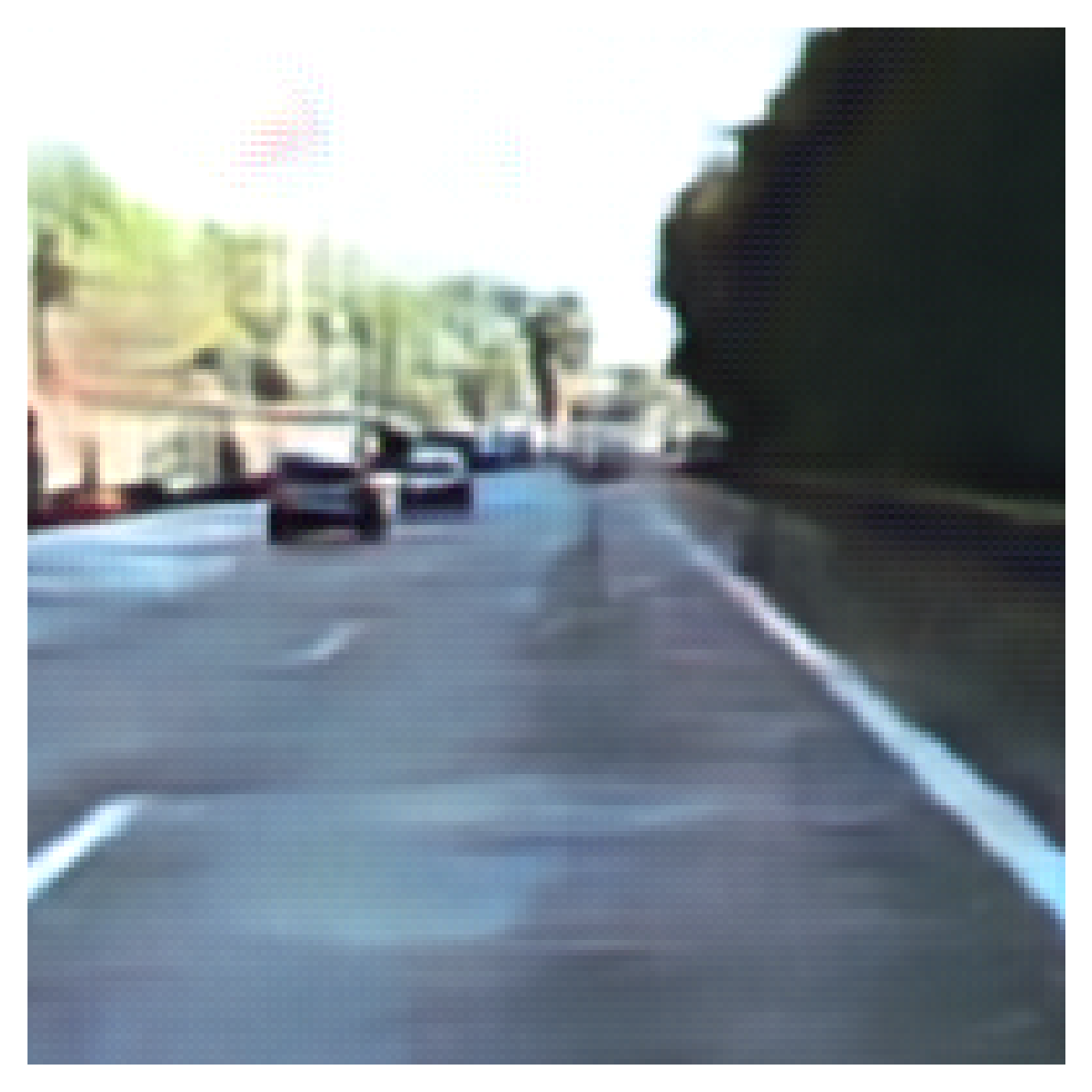}
     
 \end{subfigure}
 \hfill
 \begin{subfigure}{0.09\textwidth}
     \includegraphics[width=\textwidth]{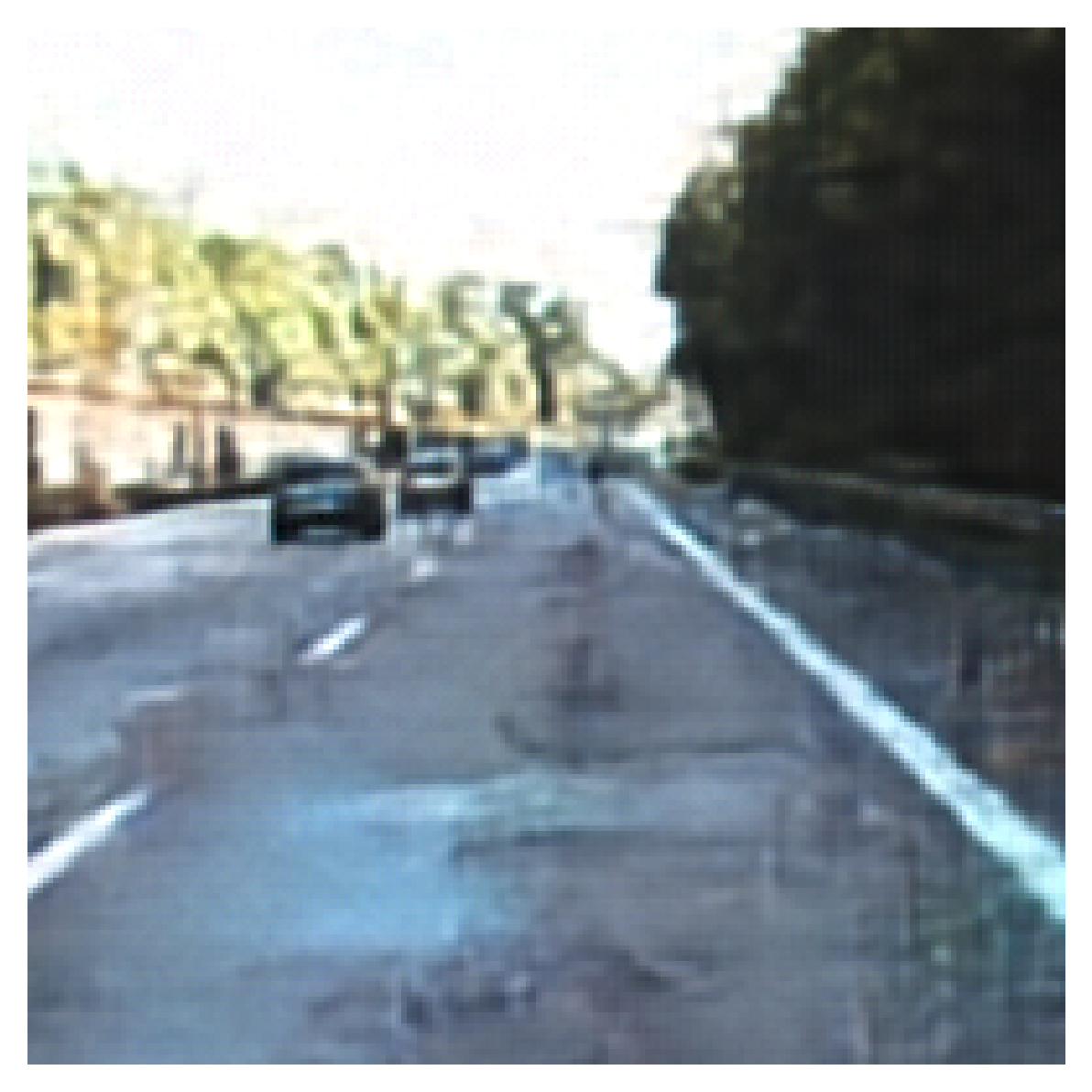}
     
 \end{subfigure}  
 \hfill
 \begin{subfigure}{0.09\textwidth}
     \includegraphics[width=\textwidth]{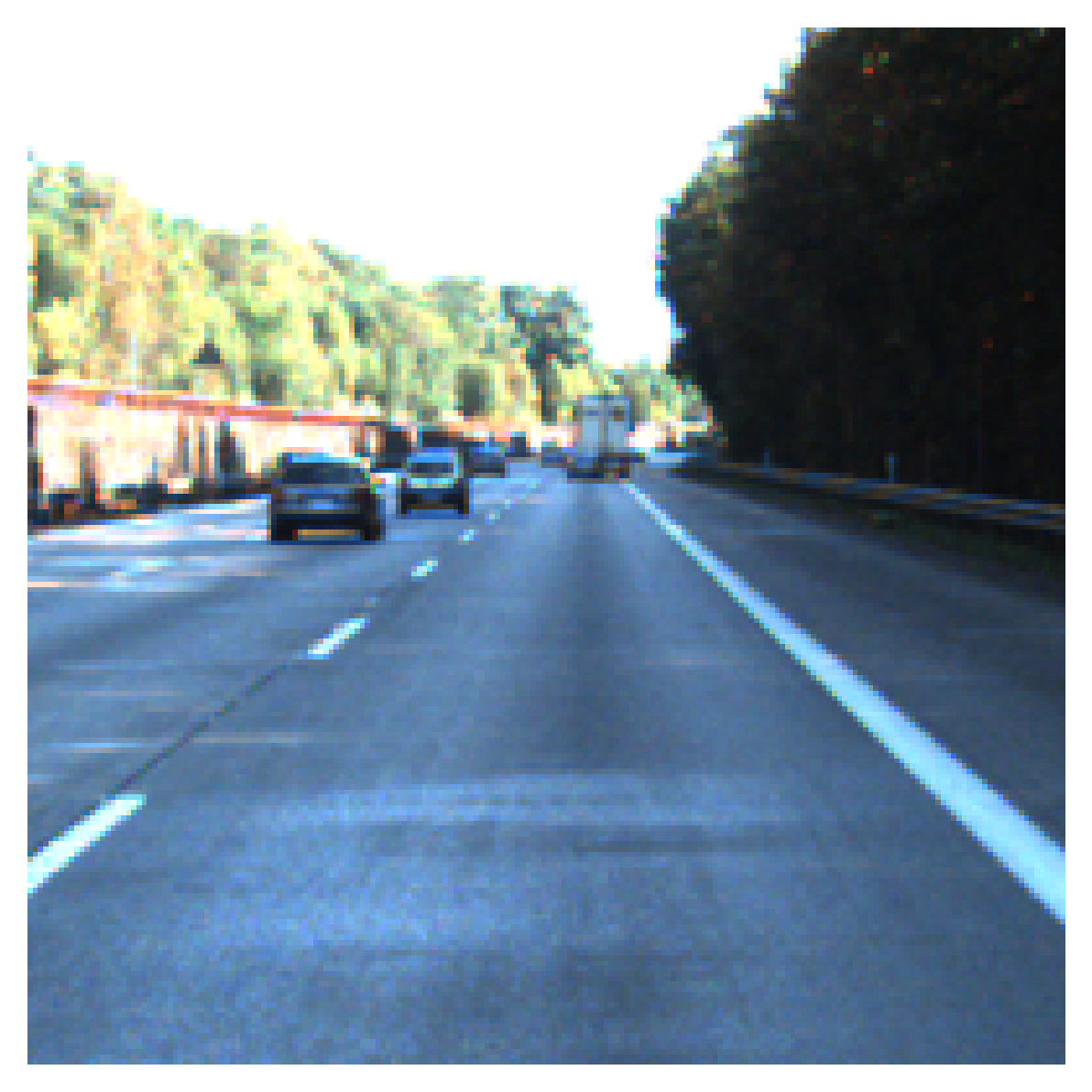}
     
 \end{subfigure} 
 \medskip
 \begin{subfigure}{0.09\textwidth}
     \includegraphics[width=\textwidth]{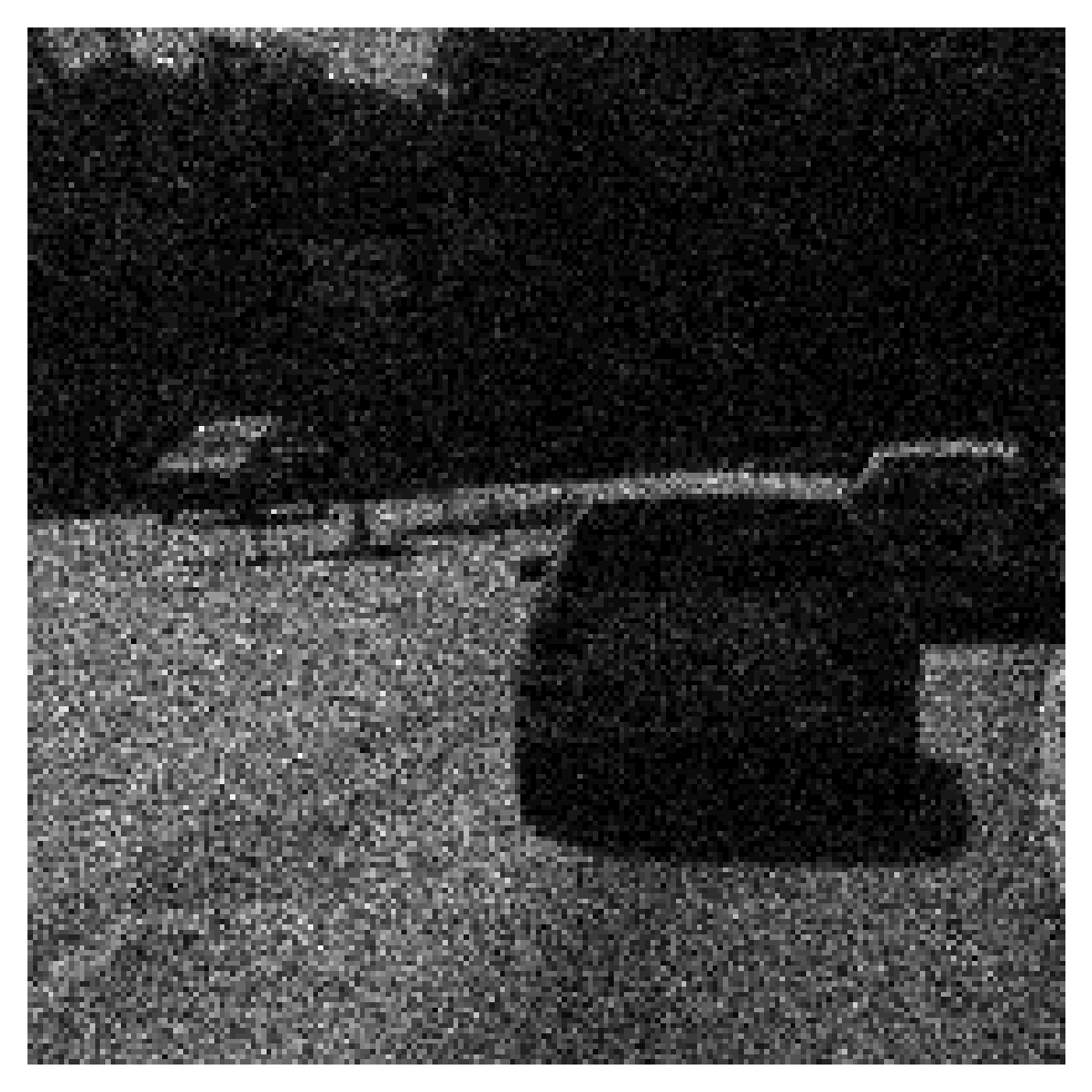}
     
 \end{subfigure}
 \hfill
 \begin{subfigure}{0.09\textwidth}
     \includegraphics[width=\textwidth]{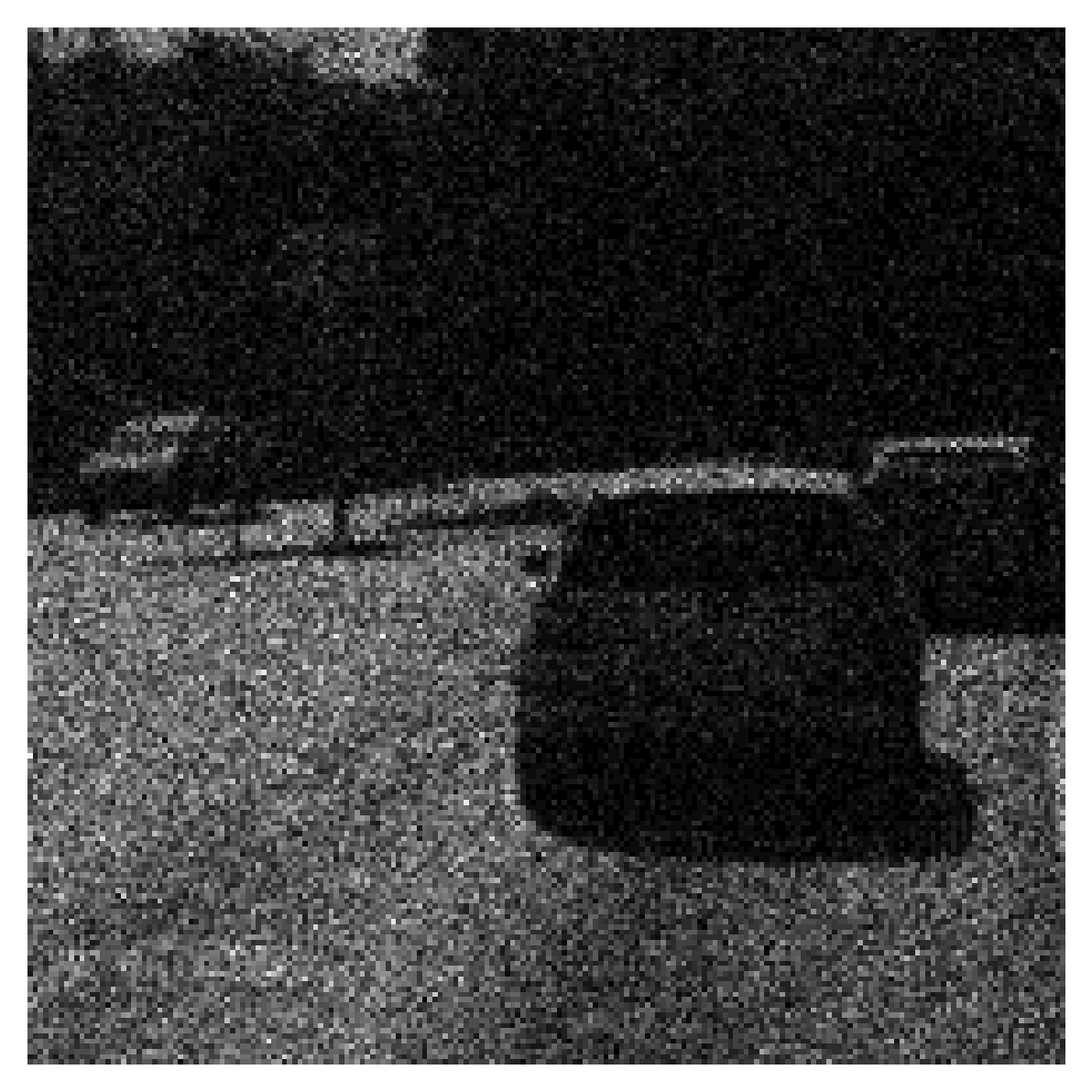}
     
 \end{subfigure}
 \hfill
 \begin{subfigure}{0.09\textwidth}
     \includegraphics[width=\textwidth]{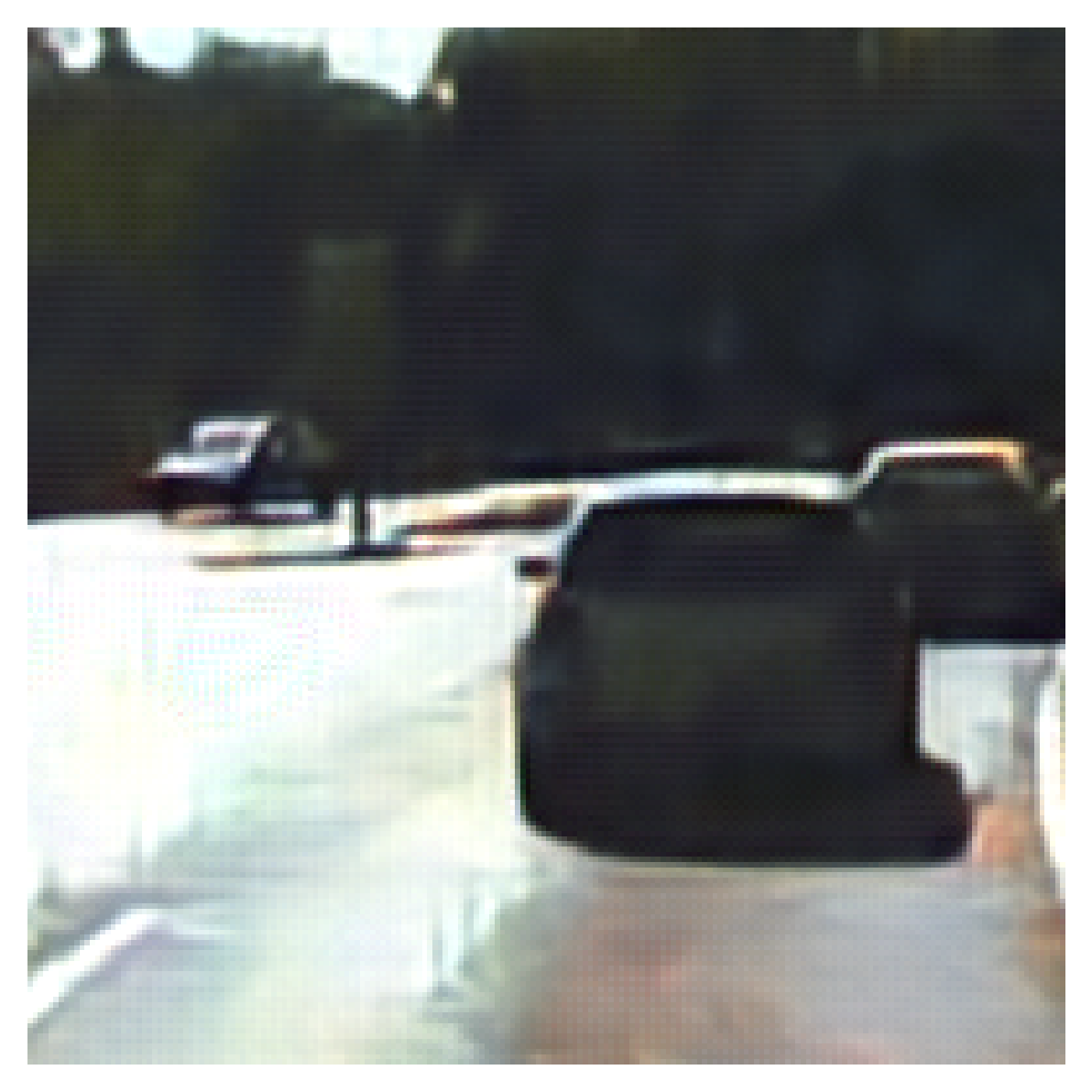}
     
 \end{subfigure}
 \hfill
 \begin{subfigure}{0.09\textwidth}
     \includegraphics[width=\textwidth]{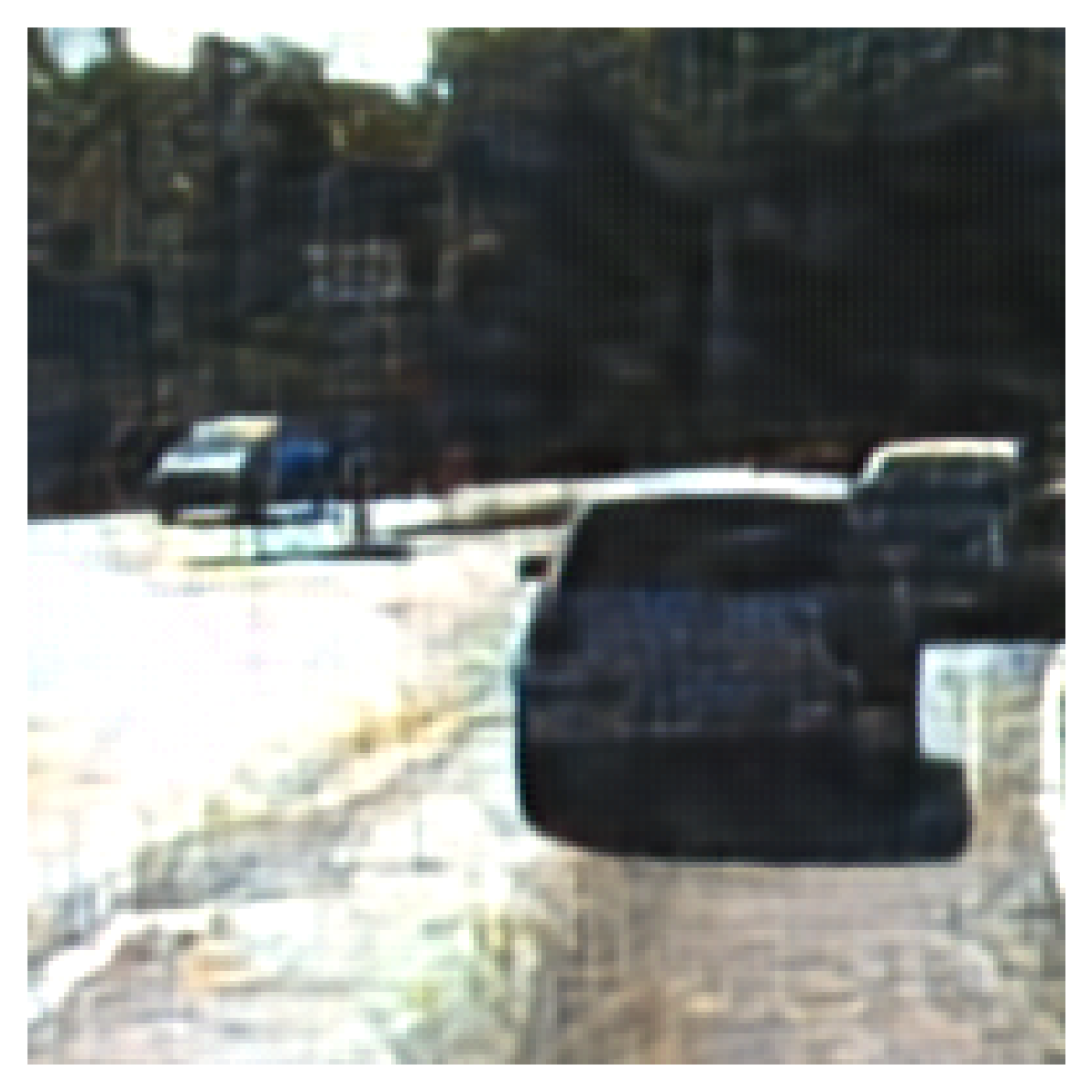}
     
 \end{subfigure}  
 \hfill
 \begin{subfigure}{0.09\textwidth}
     \includegraphics[width=\textwidth]{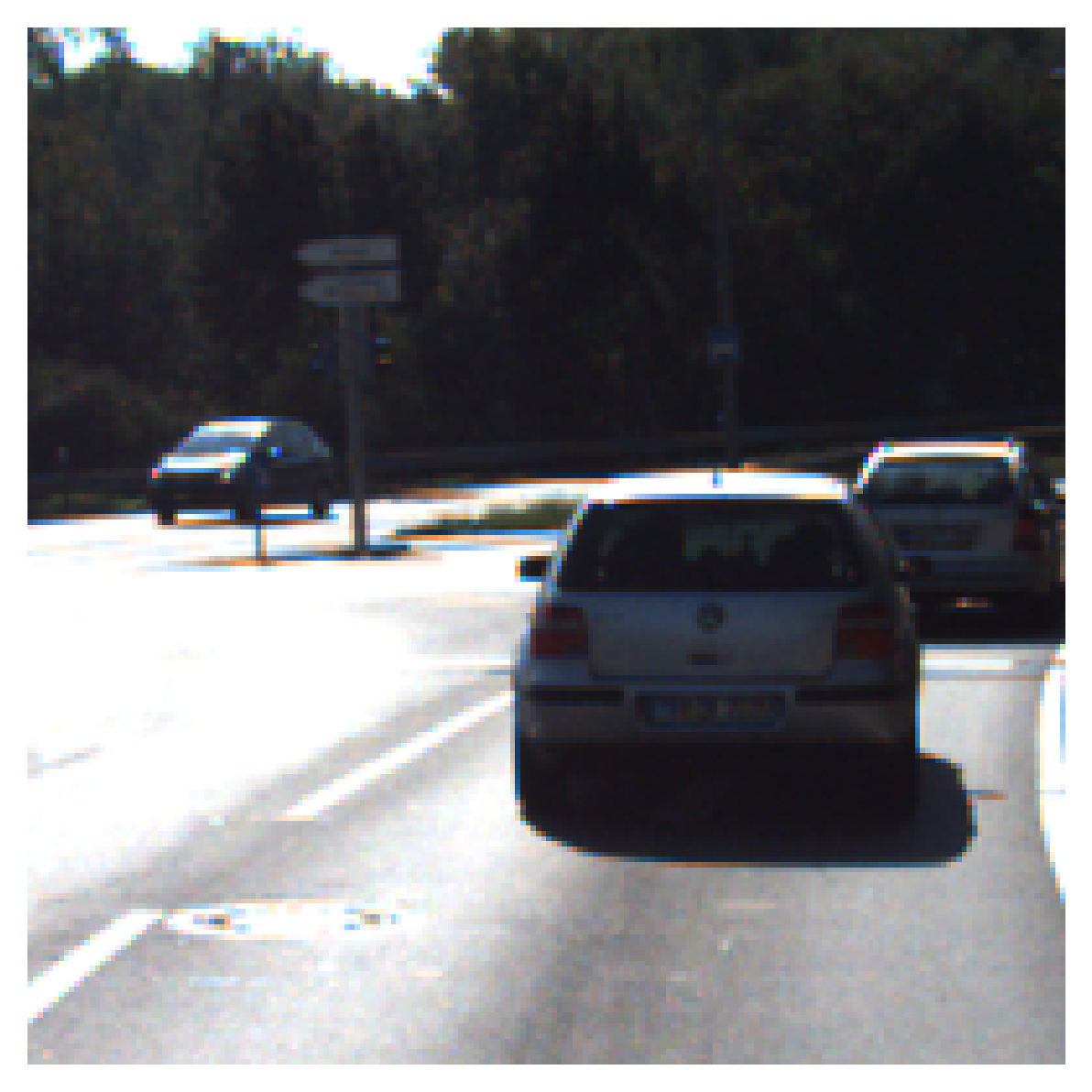}
     
 \end{subfigure}

\vspace{-10pt}
 \caption{Evaluation on KITTI 2015 \cite{menze2015object} dataset with dense disparity map. From left to right: left mosaicked noisy image $M$, right mosaicked noisy image $S$, DemosaicNet \cite{gharbi2016deep}, ours proposed in \cref{subsec: Stereo DemosaicNet}, ground truth $I$.}
 \label{fig: denoising results kitti}

\end{figure}

\begin{figure}
 \begin{subfigure}{0.09\textwidth}
     \includegraphics[width=\textwidth]{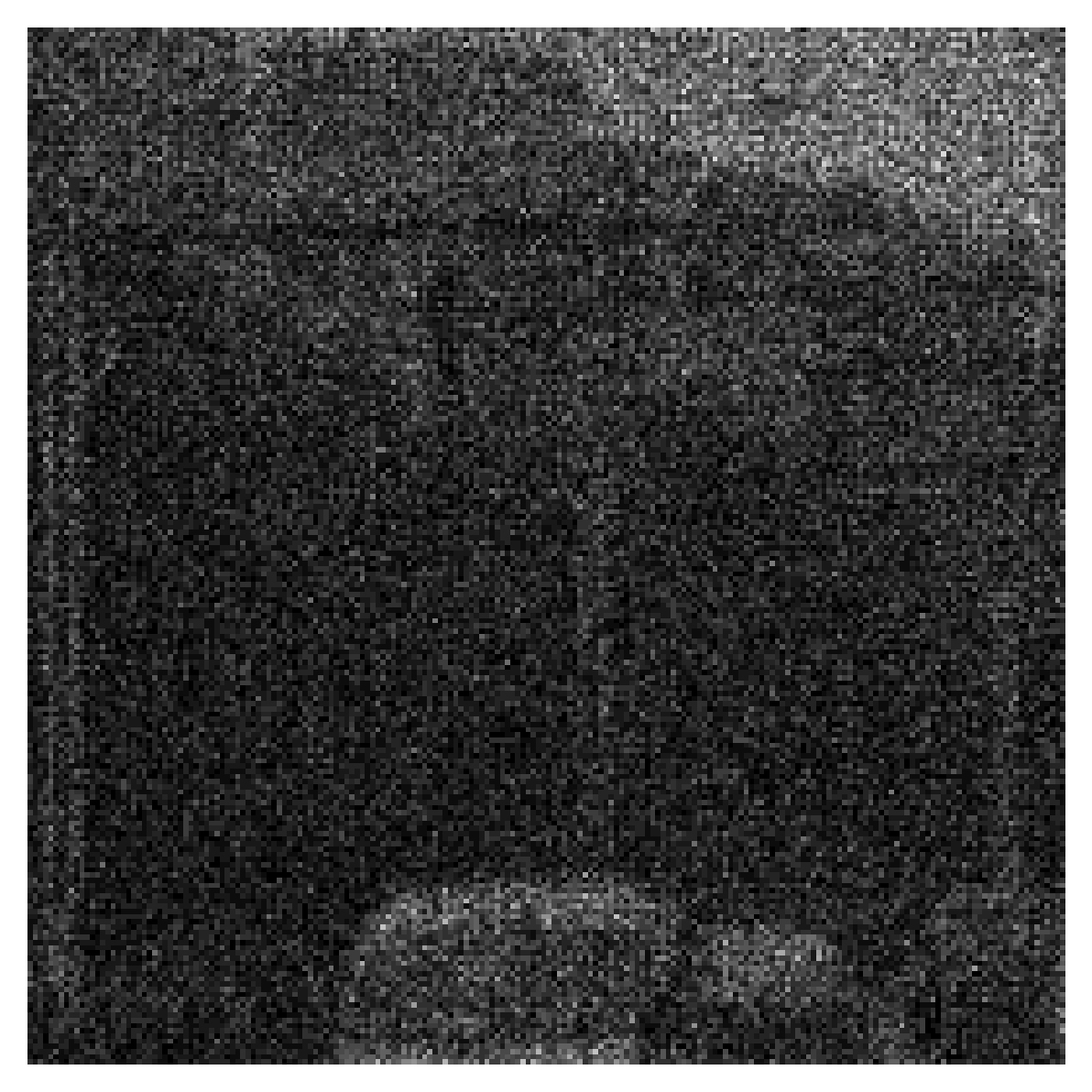}
     
 \end{subfigure}
 \hfill
 \begin{subfigure}{0.09\textwidth}
     \includegraphics[width=\textwidth]{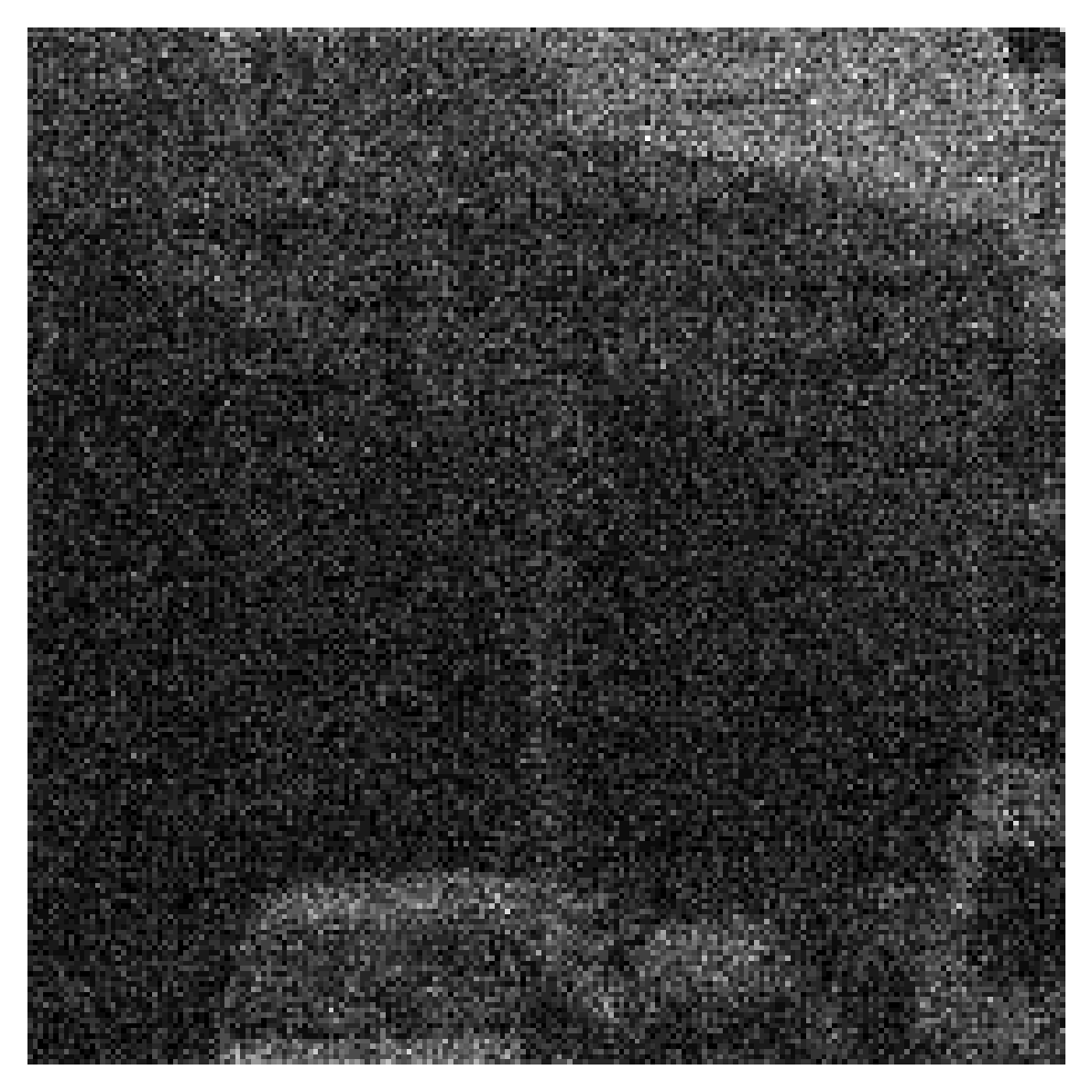}
     
 \end{subfigure}
 \hfill
 \begin{subfigure}{0.09\textwidth}
     \includegraphics[width=\textwidth]{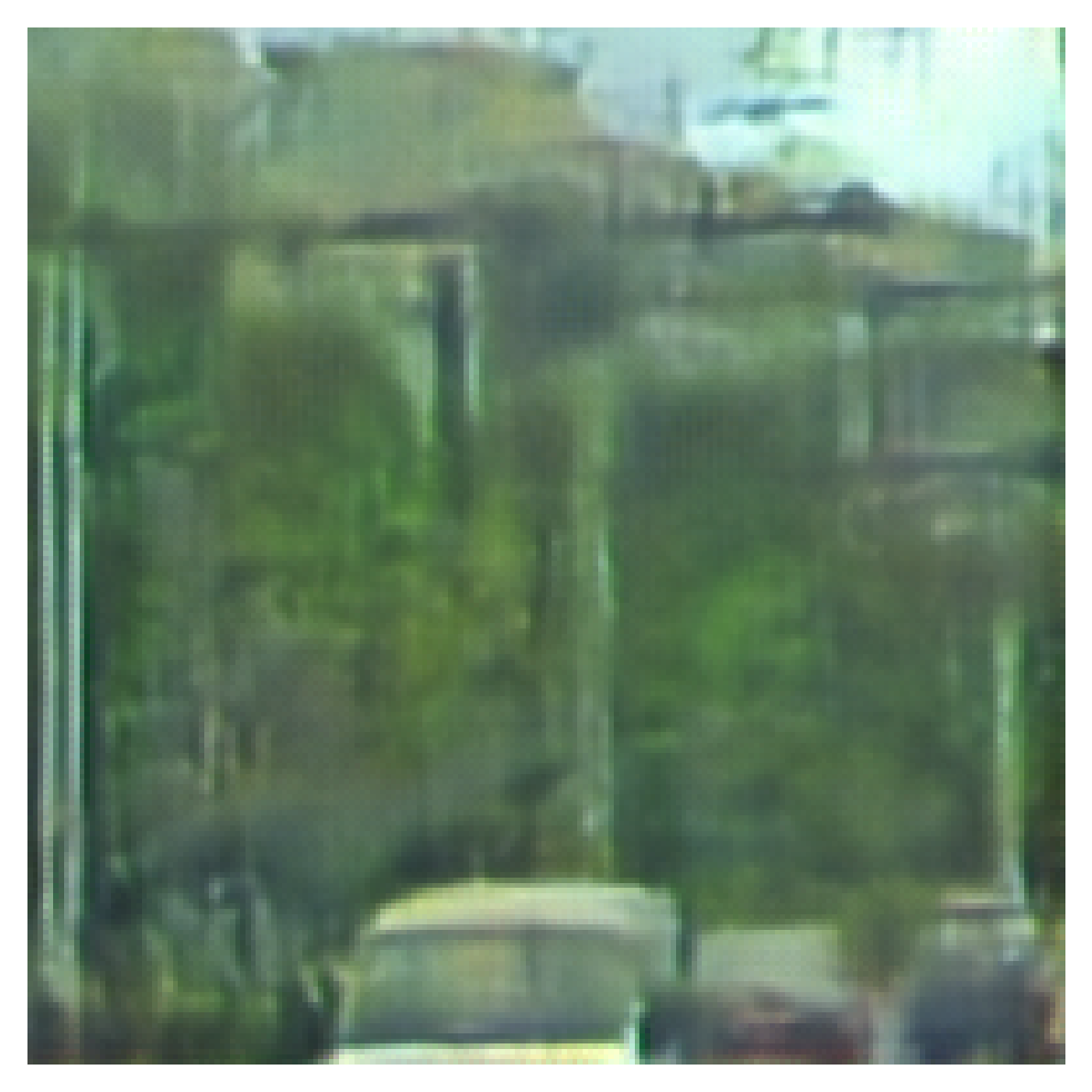}
     
 \end{subfigure}
 \hfill
 \begin{subfigure}{0.09\textwidth}
     \includegraphics[width=\textwidth]{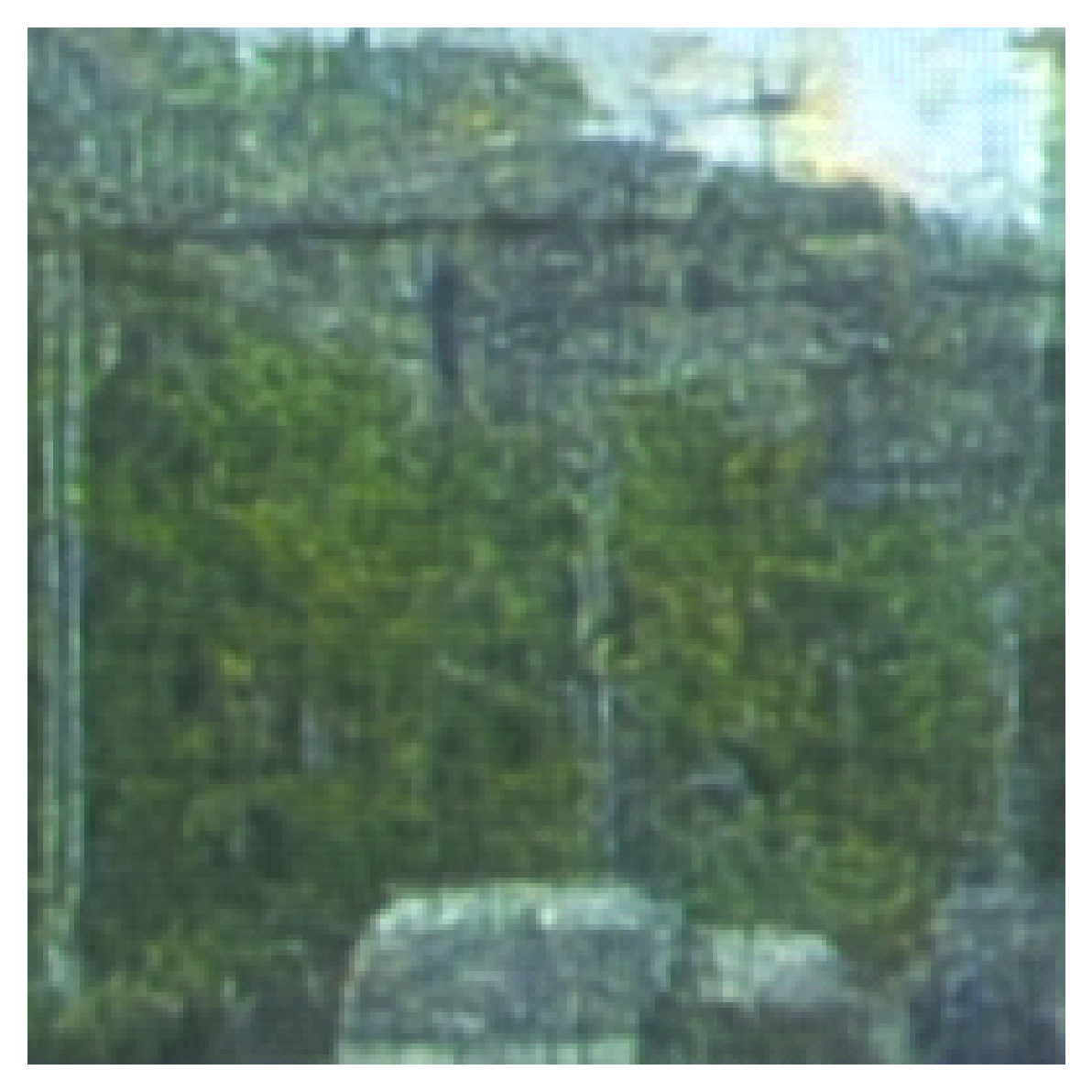}
     
 \end{subfigure}  
 \hfill
 \begin{subfigure}{0.09\textwidth}
     \includegraphics[width=\textwidth]{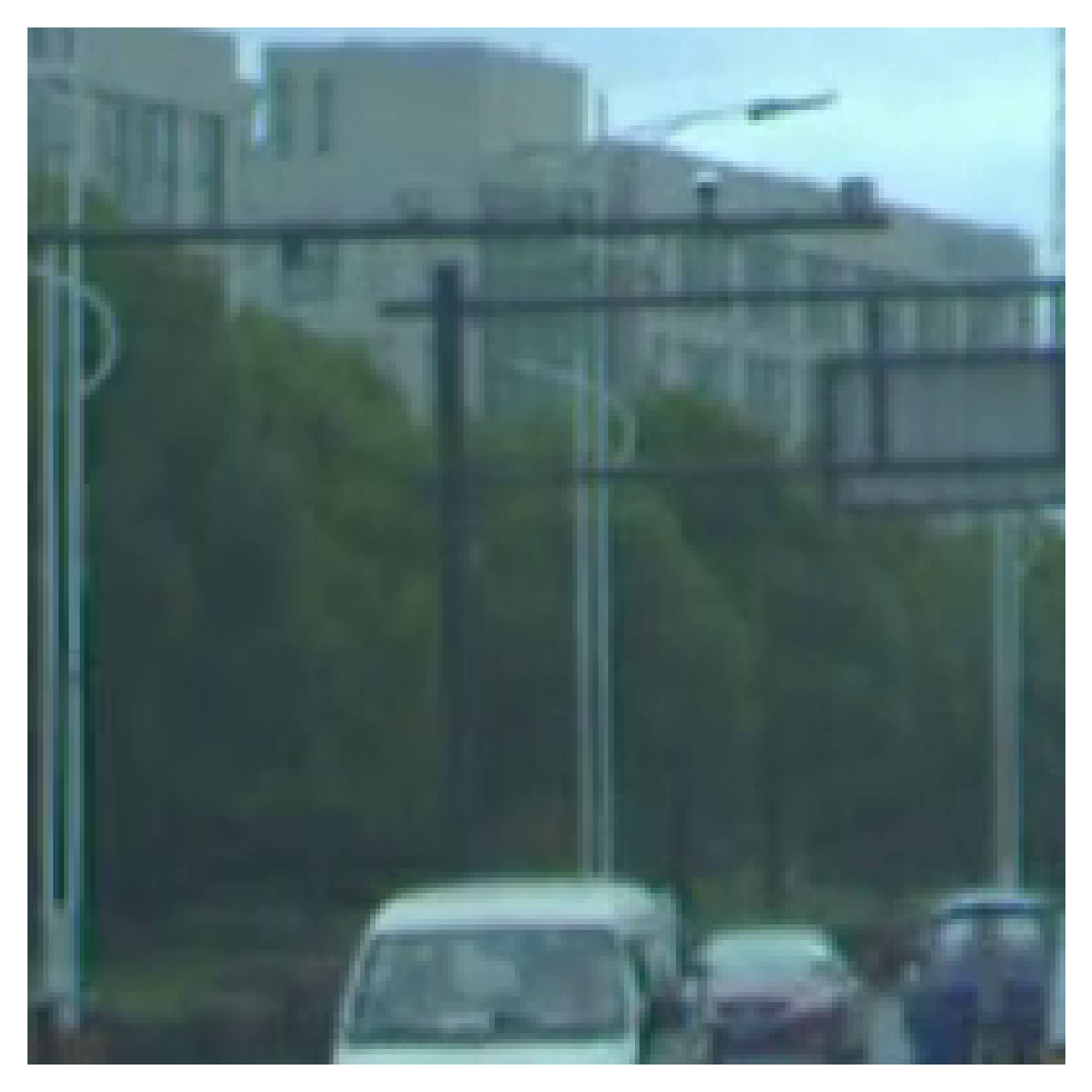}
     
 \end{subfigure}  
 \smallskip
 \begin{subfigure}{0.09\textwidth}
     \includegraphics[width=\textwidth]{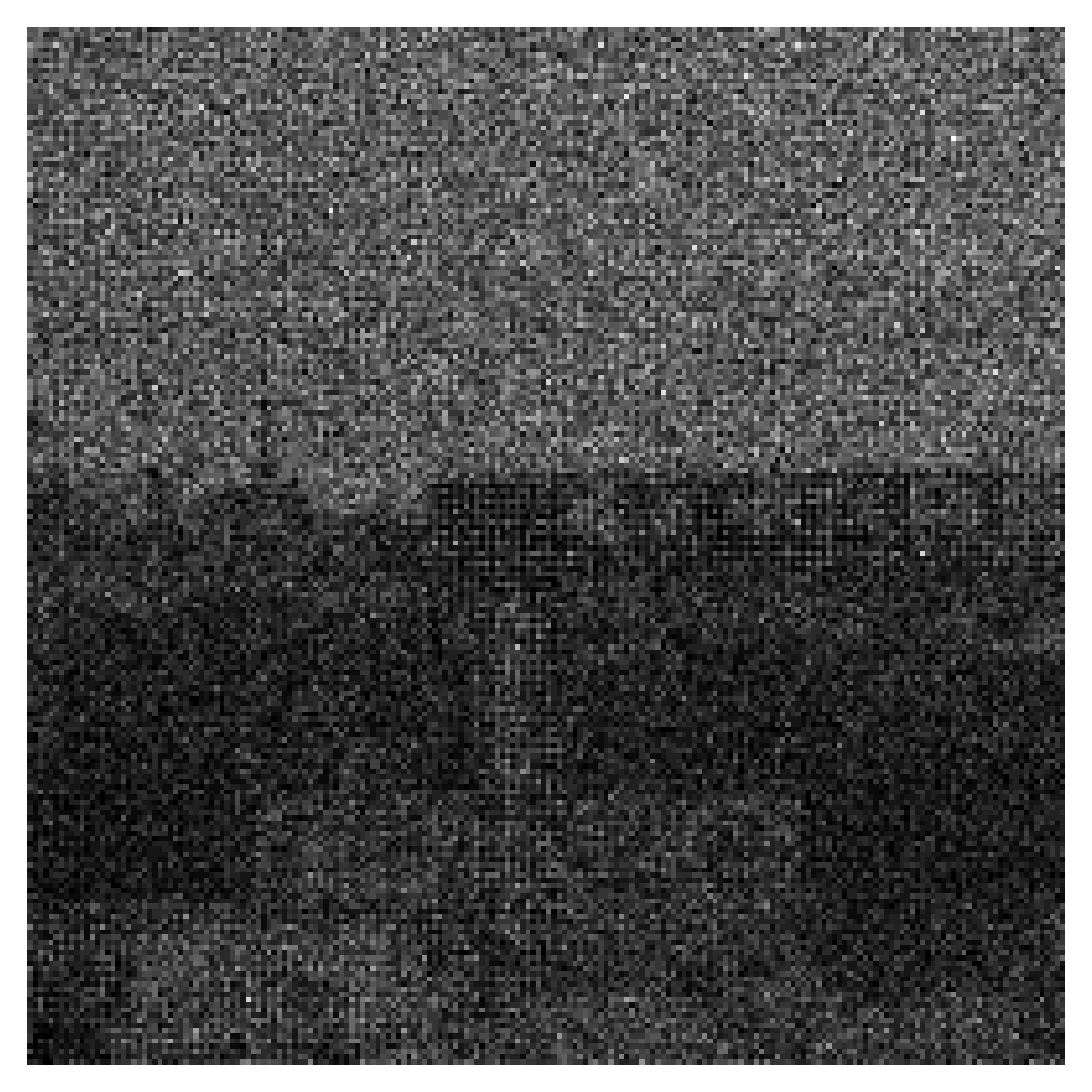}
     
 \end{subfigure}
 \hfill
 \begin{subfigure}{0.09\textwidth}
     \includegraphics[width=\textwidth]{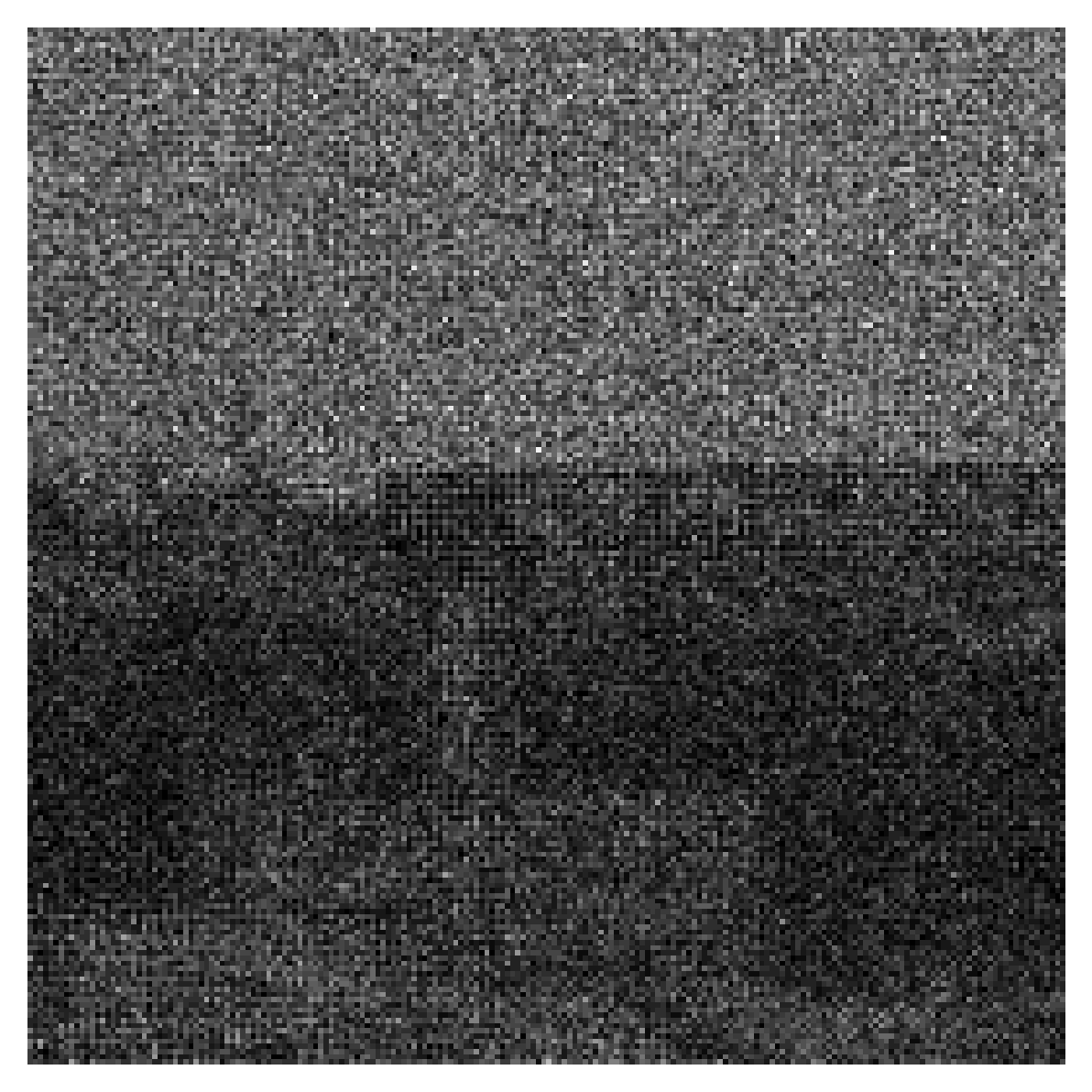}
     
 \end{subfigure}
 \hfill
 \begin{subfigure}{0.09\textwidth}
     \includegraphics[width=\textwidth]{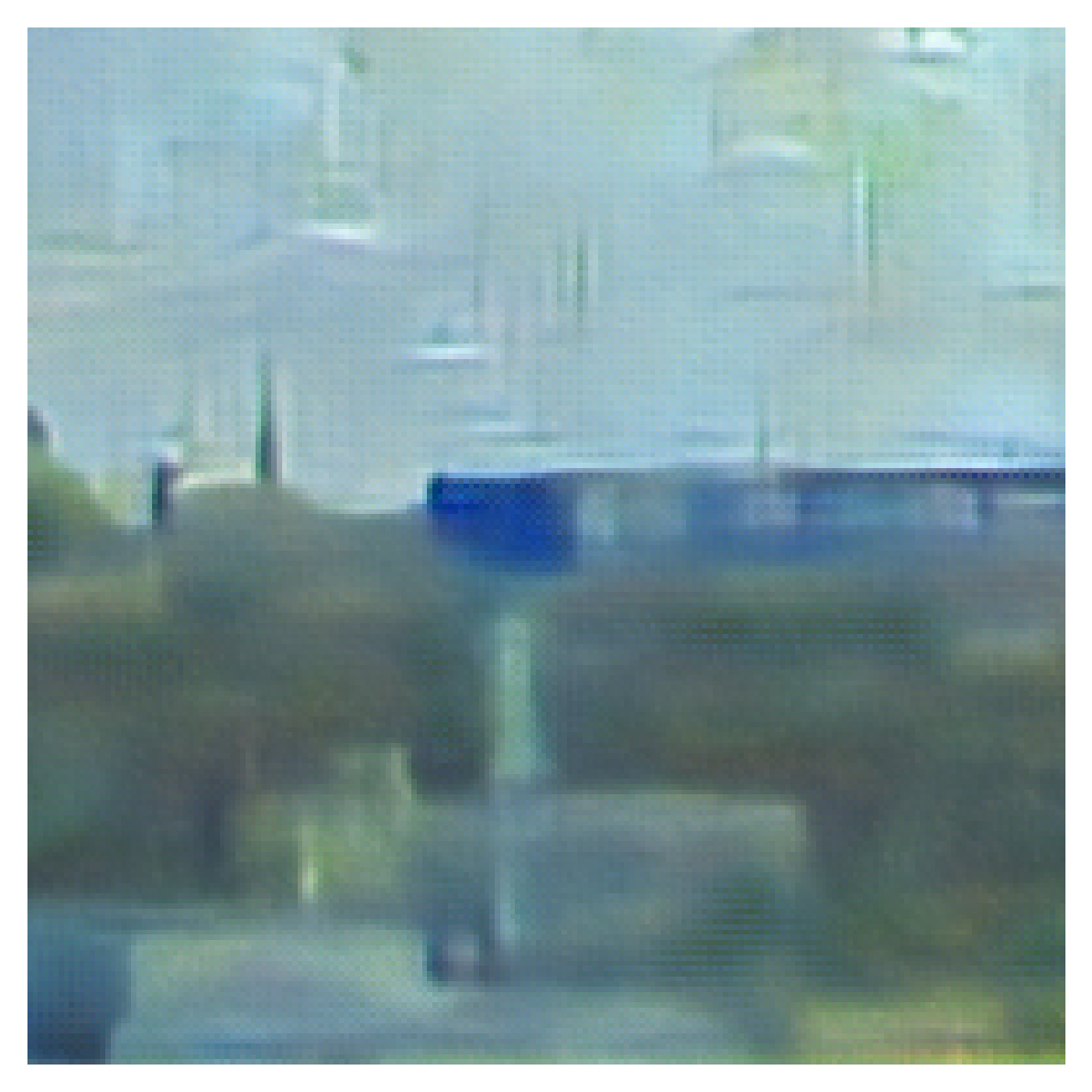}
     
 \end{subfigure}
 \hfill
 \begin{subfigure}{0.09\textwidth}
     \includegraphics[width=\textwidth]{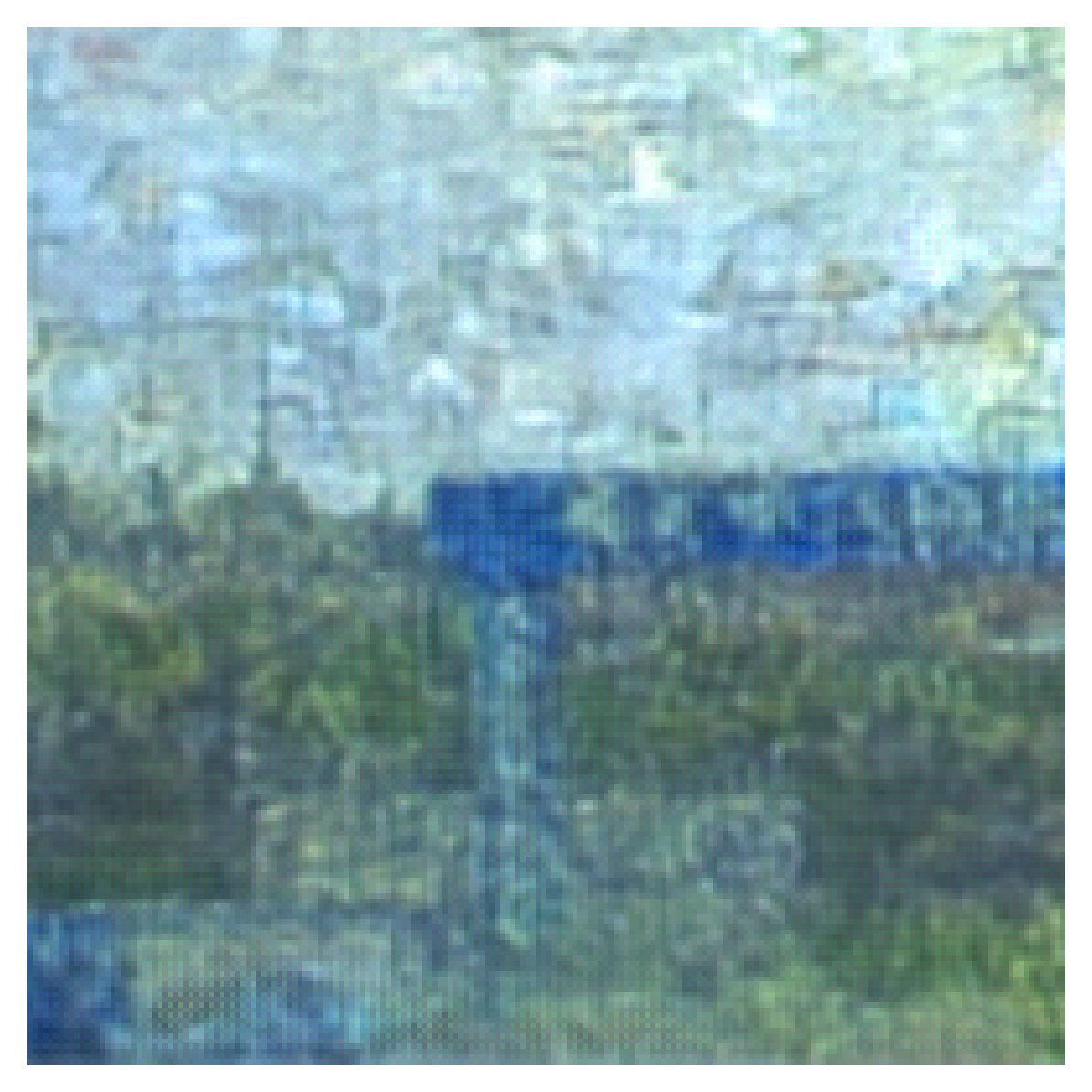}
     
 \end{subfigure}  
 \hfill
 \begin{subfigure}{0.09\textwidth}
     \includegraphics[width=\textwidth]{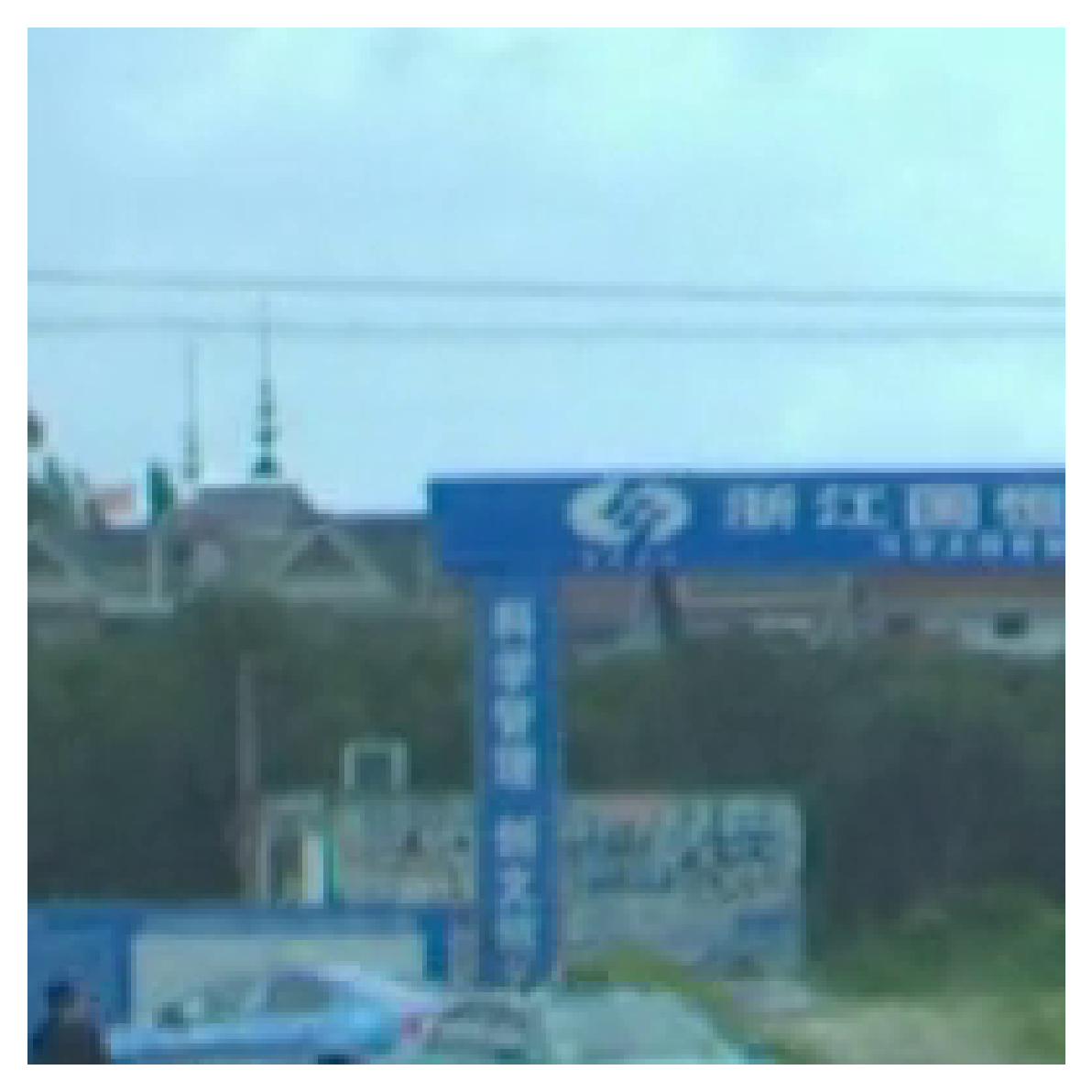}
     
 \end{subfigure} 
 \smallskip
 \begin{subfigure}{0.09\textwidth}
     \includegraphics[width=\textwidth]{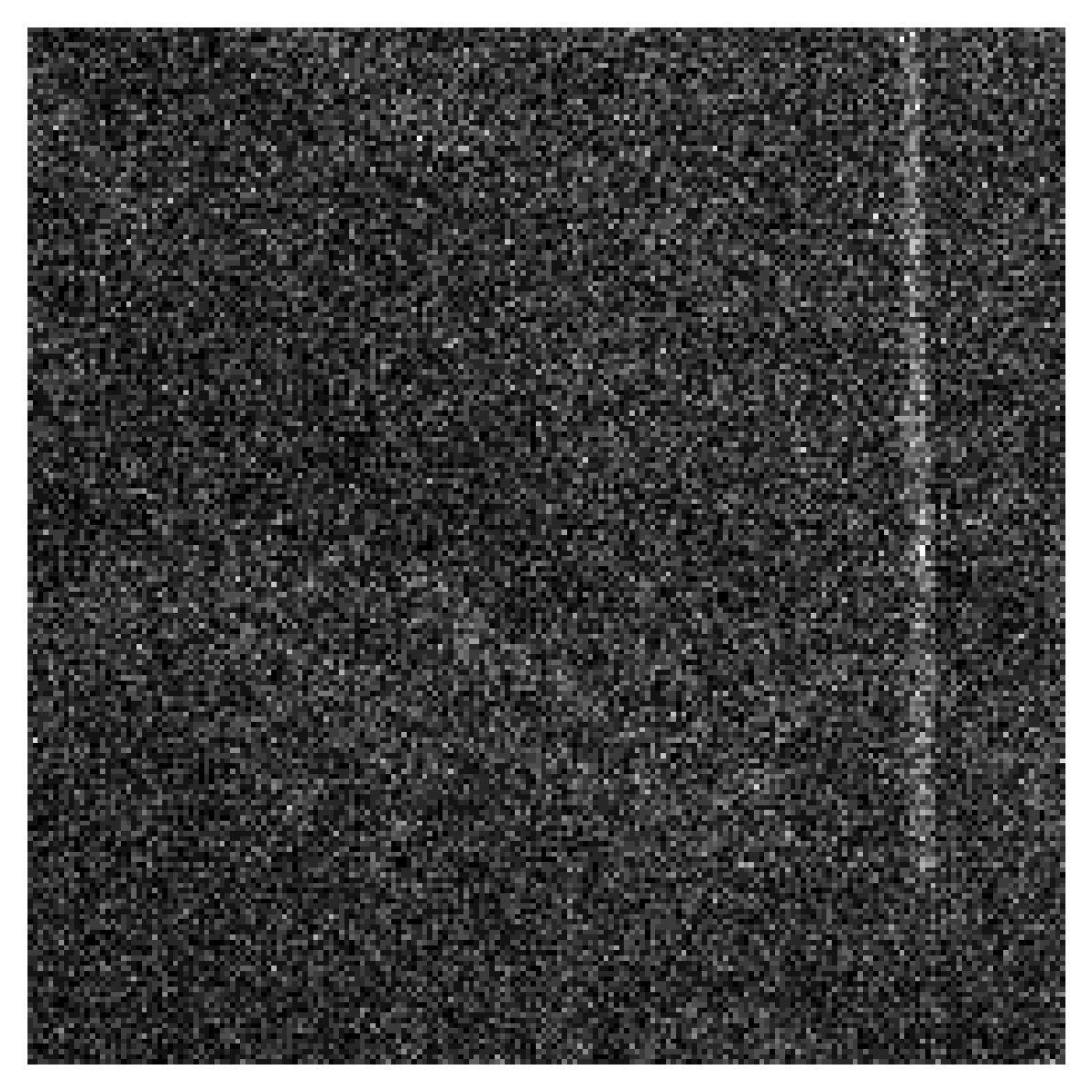}
     
 \end{subfigure}
 \hfill
 \begin{subfigure}{0.09\textwidth}
     \includegraphics[width=\textwidth]{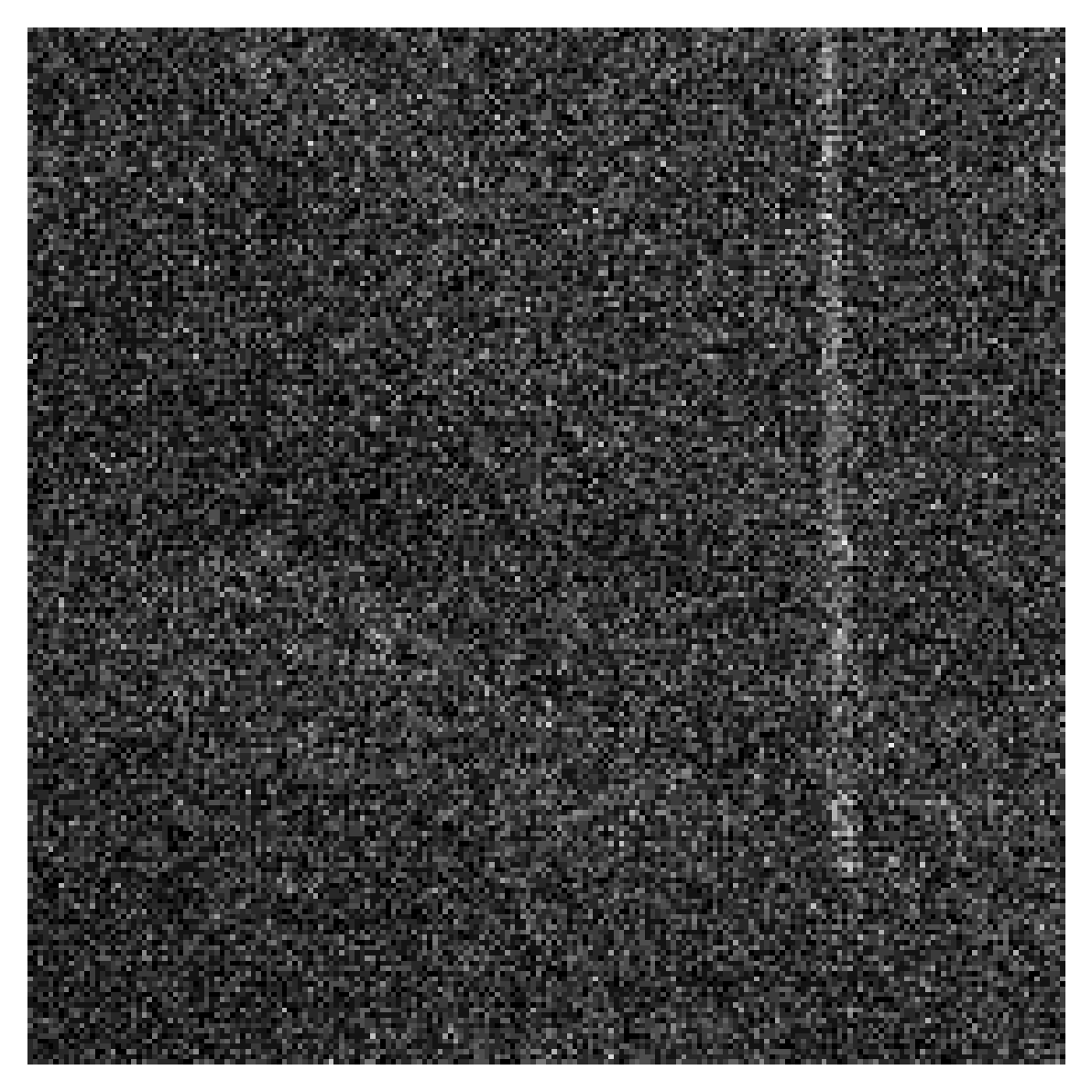}
     
 \end{subfigure}
 \hfill
 \begin{subfigure}{0.09\textwidth}
     \includegraphics[width=\textwidth]{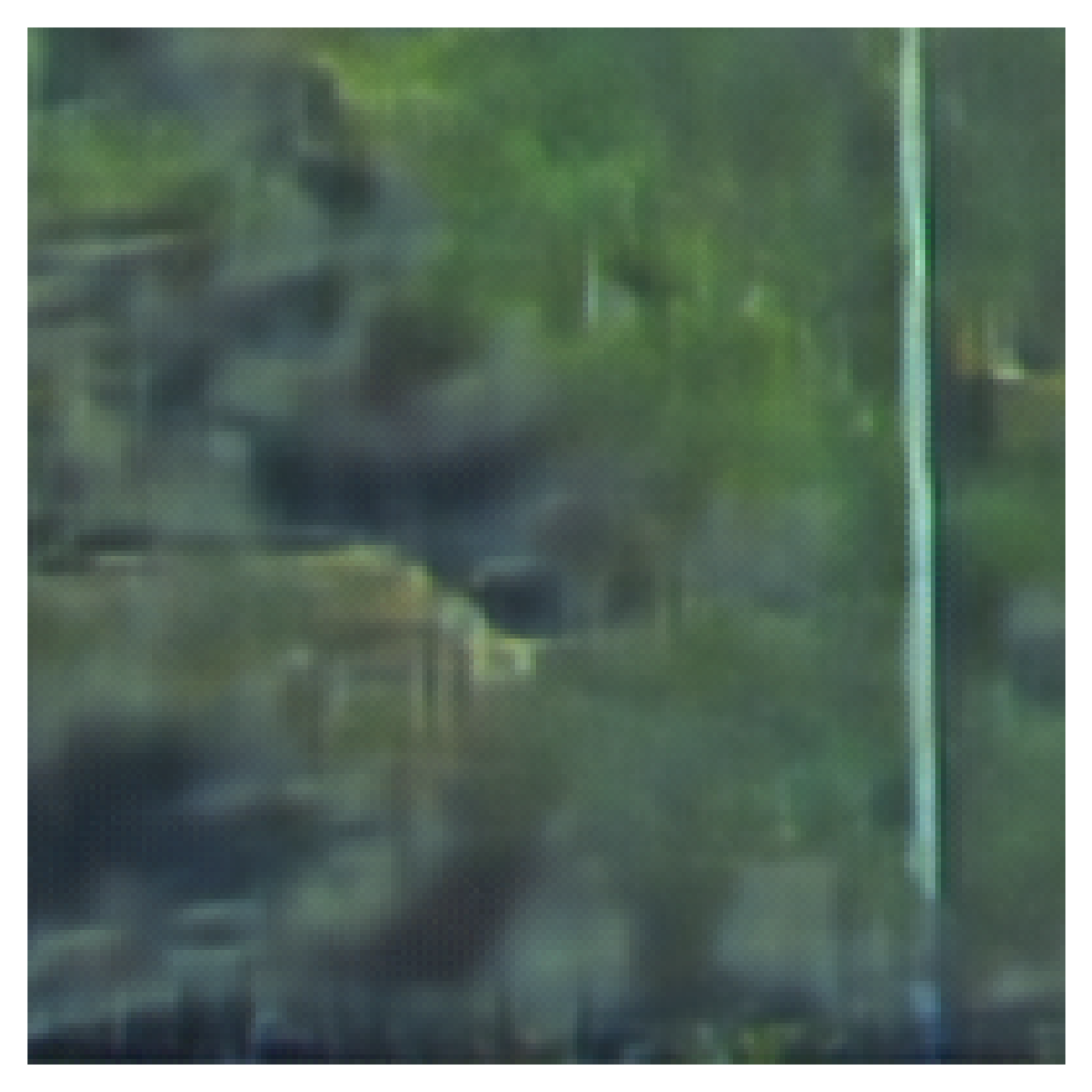}
     
 \end{subfigure}
 \hfill
 \begin{subfigure}{0.09\textwidth}
     \includegraphics[width=\textwidth]{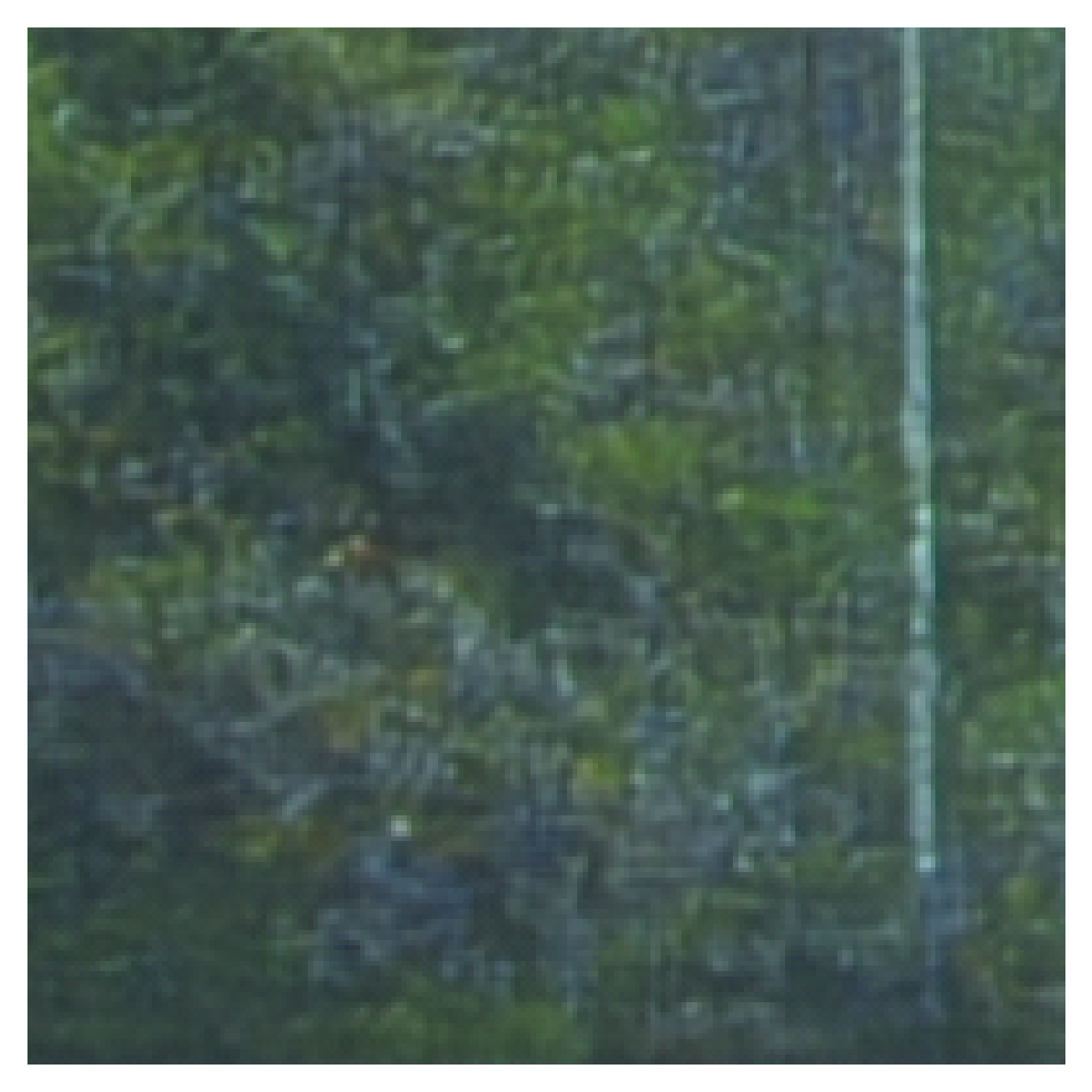}
     
 \end{subfigure}  
 \hfill
 \begin{subfigure}{0.09\textwidth}
     \includegraphics[width=\textwidth]{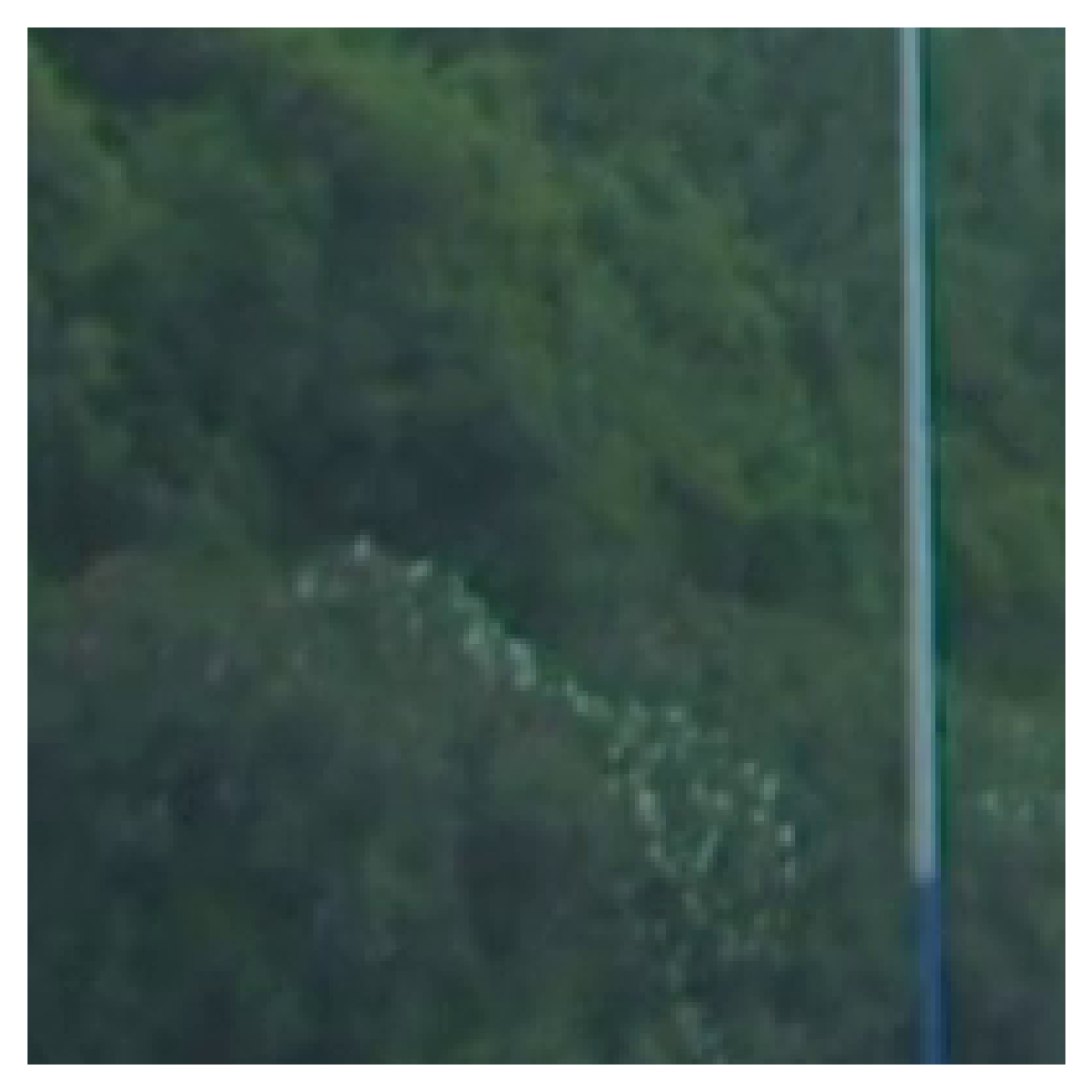}
     
 \end{subfigure}

\vspace{-10pt}
 \caption{Evaluation on drivingStereo \cite{yang2019drivingstereo} dataset with sparse disparity map. From left to right: left mosaicked noisy image $M$, right mosaicked noisy image $S$, DemosaicNet \cite{gharbi2016deep}, ours proposed in \cref{subsec: Stereo DemosaicNet}, ground truth $I$.}
 \label{fig: denoising results drivingstereo}
\vspace{-10pt}
\end{figure}

To assess our hypothesis, we make use of DemosaicNet \cite{gharbi2016deep} as a benchmark for our model since ours is an extended version of DemosaicNet. We make use of DemosaicNet architecture PyTorch implementation by the author \footnote{\url{https://github.com/mgharbi/demosaicnet}}. We extended on original DemosaicNet implementation by adding a noise layer in $F^0$. Since Authors in \cite{gharbi2016deep} trained their architecture on carefully selected dataset, we no longer be able to replicate their results. However, we consider two automobile datasets: KITTI 2015 \cite{menze2015object} and a large-scale drivingStereo \cite{yang2019drivingstereo}. Because of the nature of stereo automobile datasets as they include color bias (i.e., almost half of the image is a sky where its color is either blue or white or both), and given that we do not have as sufficient data in the stereo case as in other datasets, especially in automobile datasets, we expect DemosaicNet results on KITTI 2015 and drivingStereo could be lower than what authors stated in their work.

We trained the original model on KITTI 2015 for 2,000 epochs, with a decaying learning rate by a factor of 10 every 100 epochs. The model converged with a validation PSNR of 24.47 dB. We trained Stereo DemosaicNet in two stages as explained in \cref{subsec: procedure} with 2,246,146 trainable parameters. Our model converged with validation PSNR of 26.79 dB, with an improvement in the quality of the reconstructed image by 9.45\%. Results obtained by DemosaicNet and our model have been presented in \cref{fig: denoising results kitti}. We considered drivingStereo \cite{yang2019drivingstereo} with large-scale stereo images. As we trained the model for the large-scale dataset. The model has converged with a validation PSNR of 29.59 dB. We trained stereo DemosaciNet and we reached a validation PSNR of 32.02 dB, with an improvement in the quality of the reconstructed image by 8.2\%. Results for drivingStereo dataset have been shown in \cref{fig: denoising results drivingstereo}. Table \ref{tab: results} summarizes results and comparisons between DemosaicNet and our model inspired by the StereoISP framework.

\subsection{Ablation Study}
\subsubsection{Same input with different noise}
\label{subsec: Same input}

\begin{figure}
 \begin{subfigure}{0.09\textwidth}
     \includegraphics[width=\textwidth]{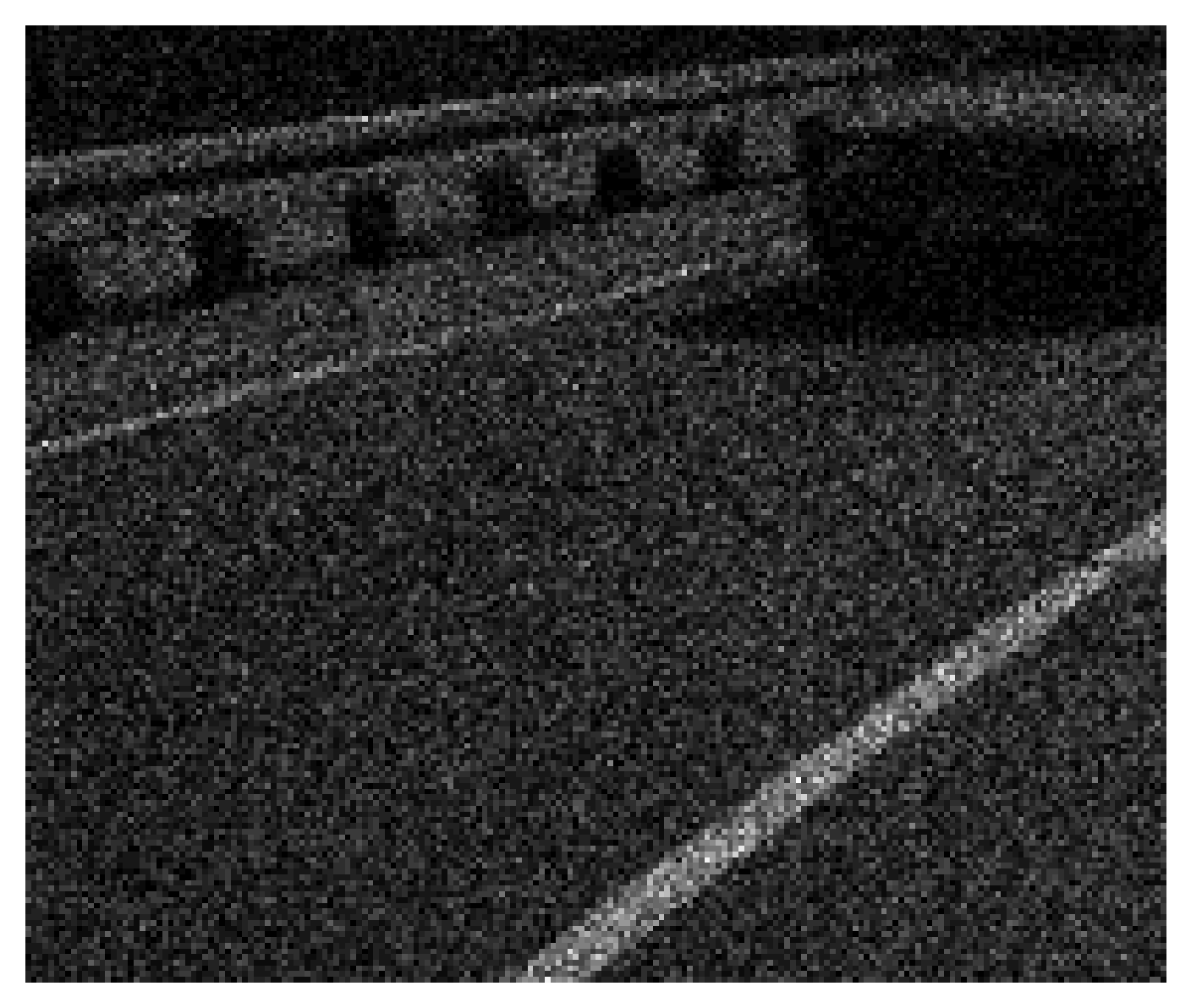}
     
 \end{subfigure}
 \hfill
 \begin{subfigure}{0.09\textwidth}
     \includegraphics[width=\textwidth]{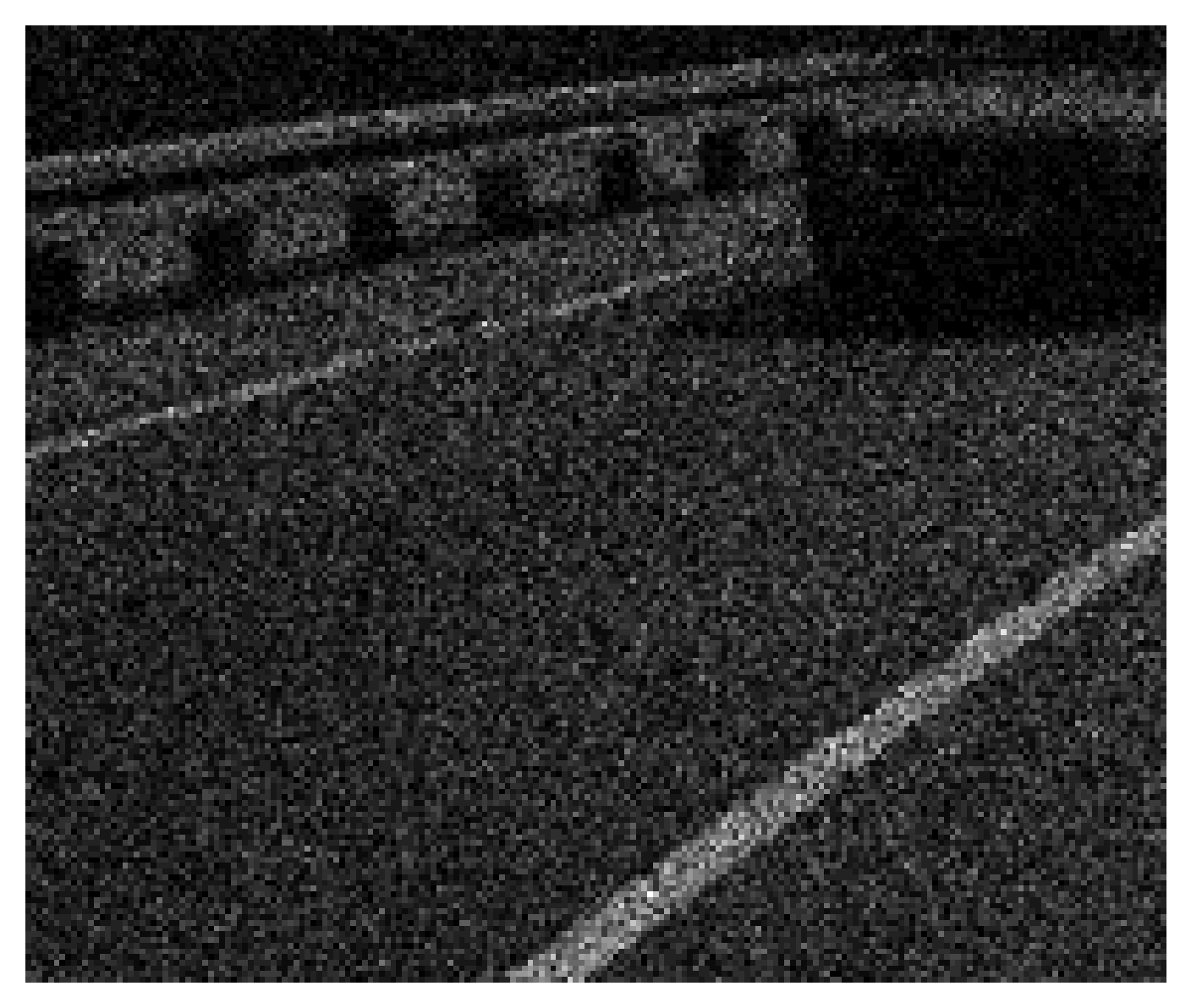}
     
 \end{subfigure}
 \hfill
 \begin{subfigure}{0.09\textwidth}
     \includegraphics[width=\textwidth]{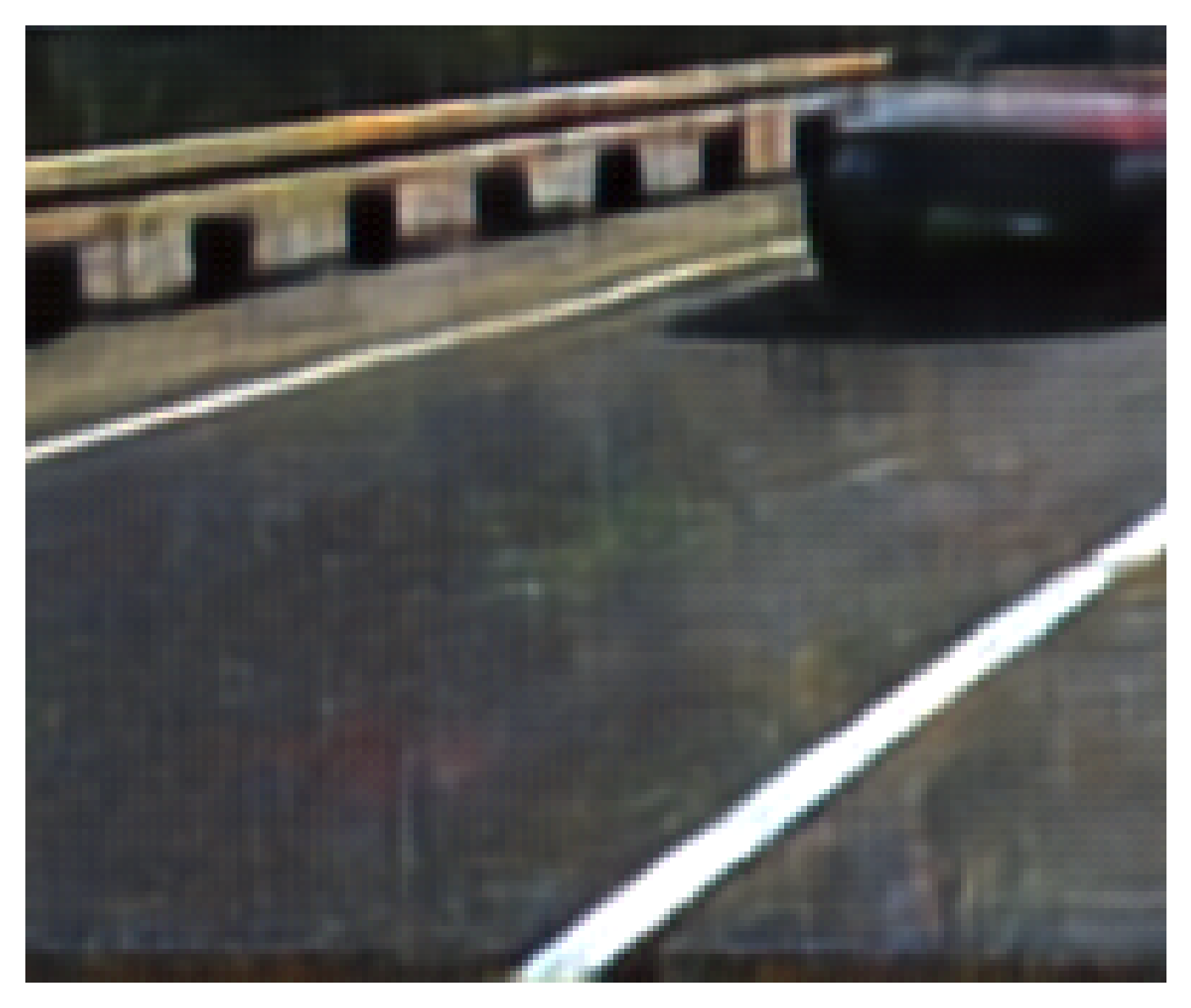}
     
 \end{subfigure}
 \hfill
 \begin{subfigure}{0.09\textwidth}
     \includegraphics[width=\textwidth]{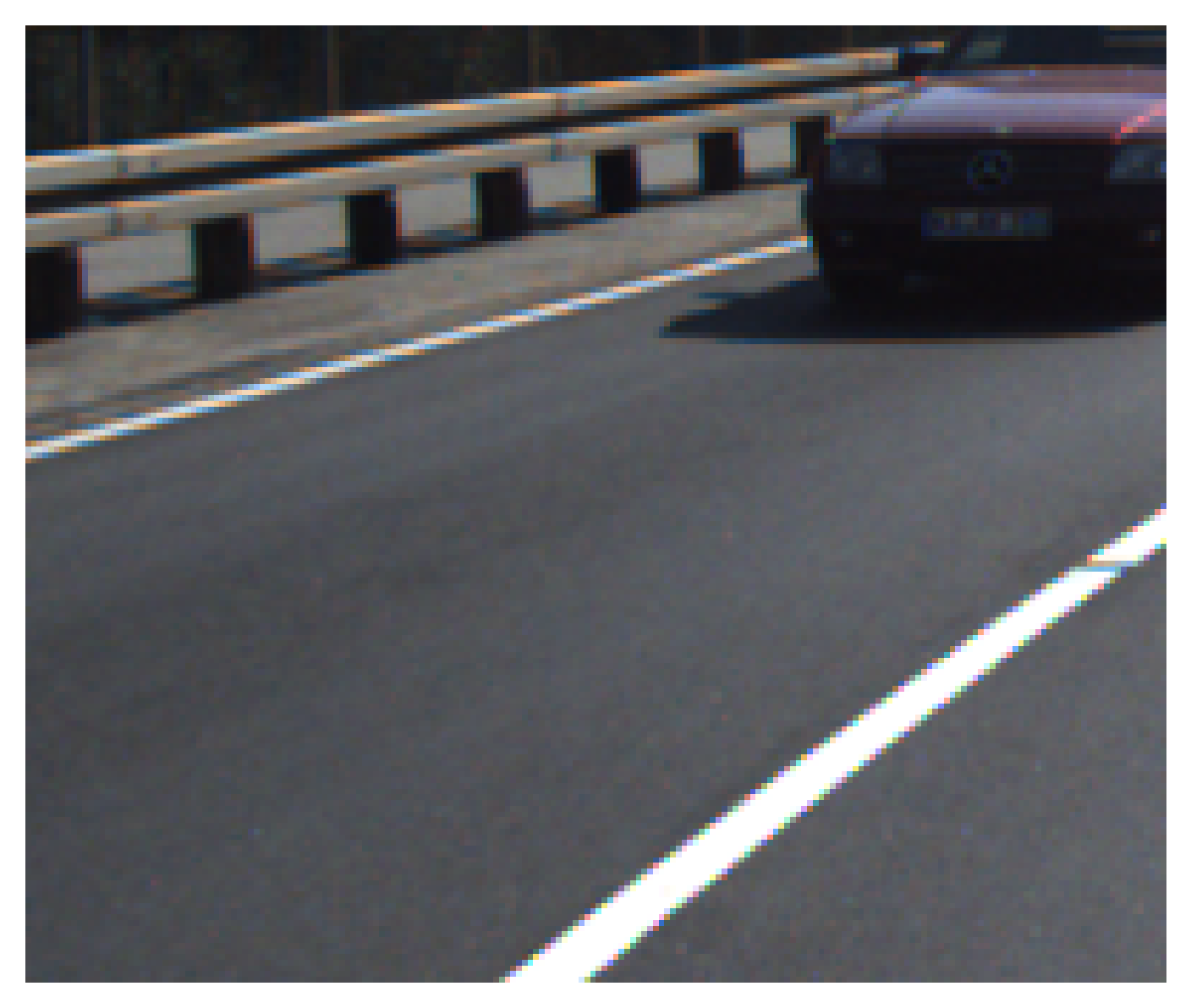}
     
 \end{subfigure}  
 \hfill
 \begin{subfigure}{0.09\textwidth}
     \includegraphics[width=\textwidth]{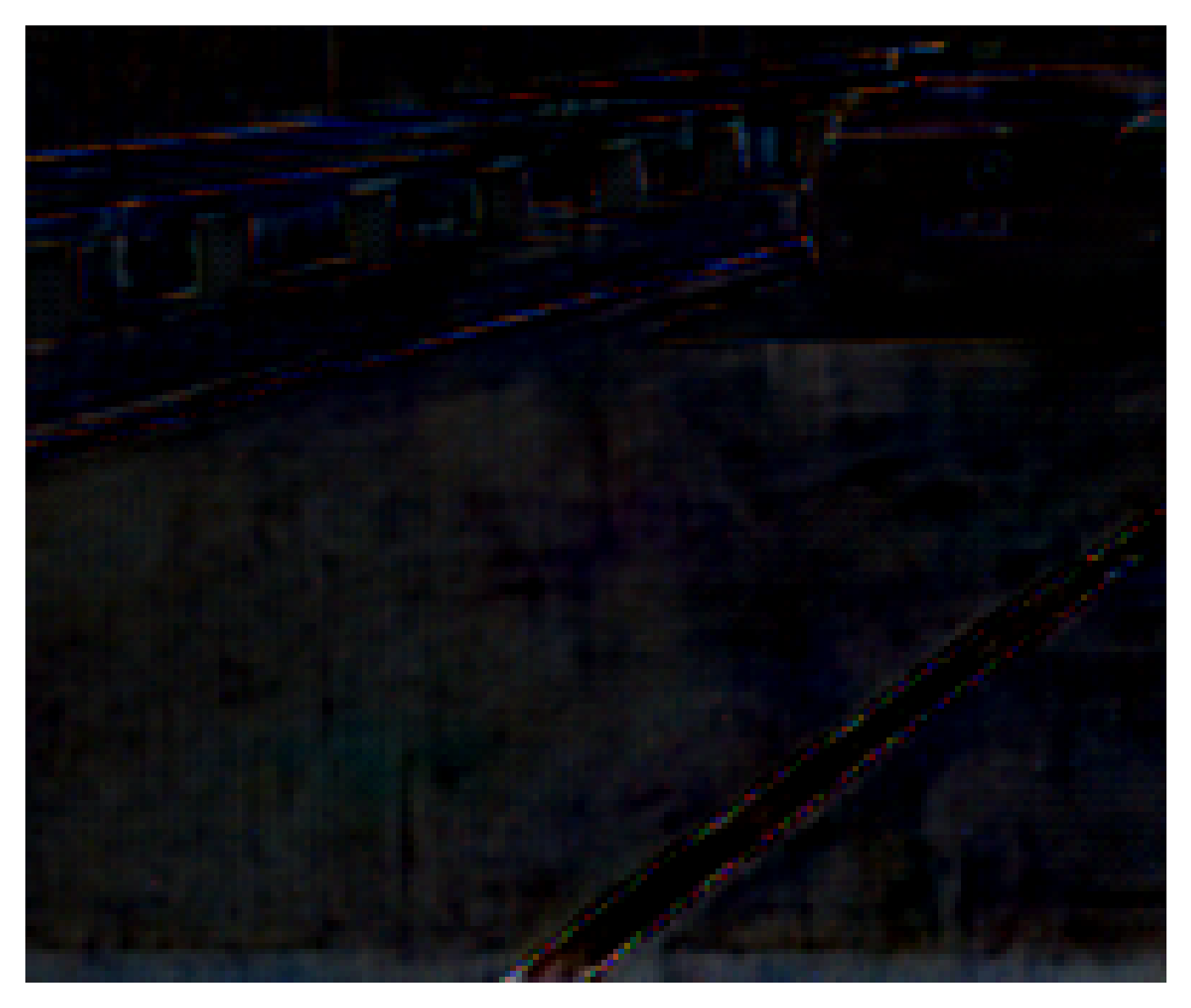}
     
 \end{subfigure}  
 \smallskip
 \begin{subfigure}{0.09\textwidth}
     \includegraphics[width=\textwidth]{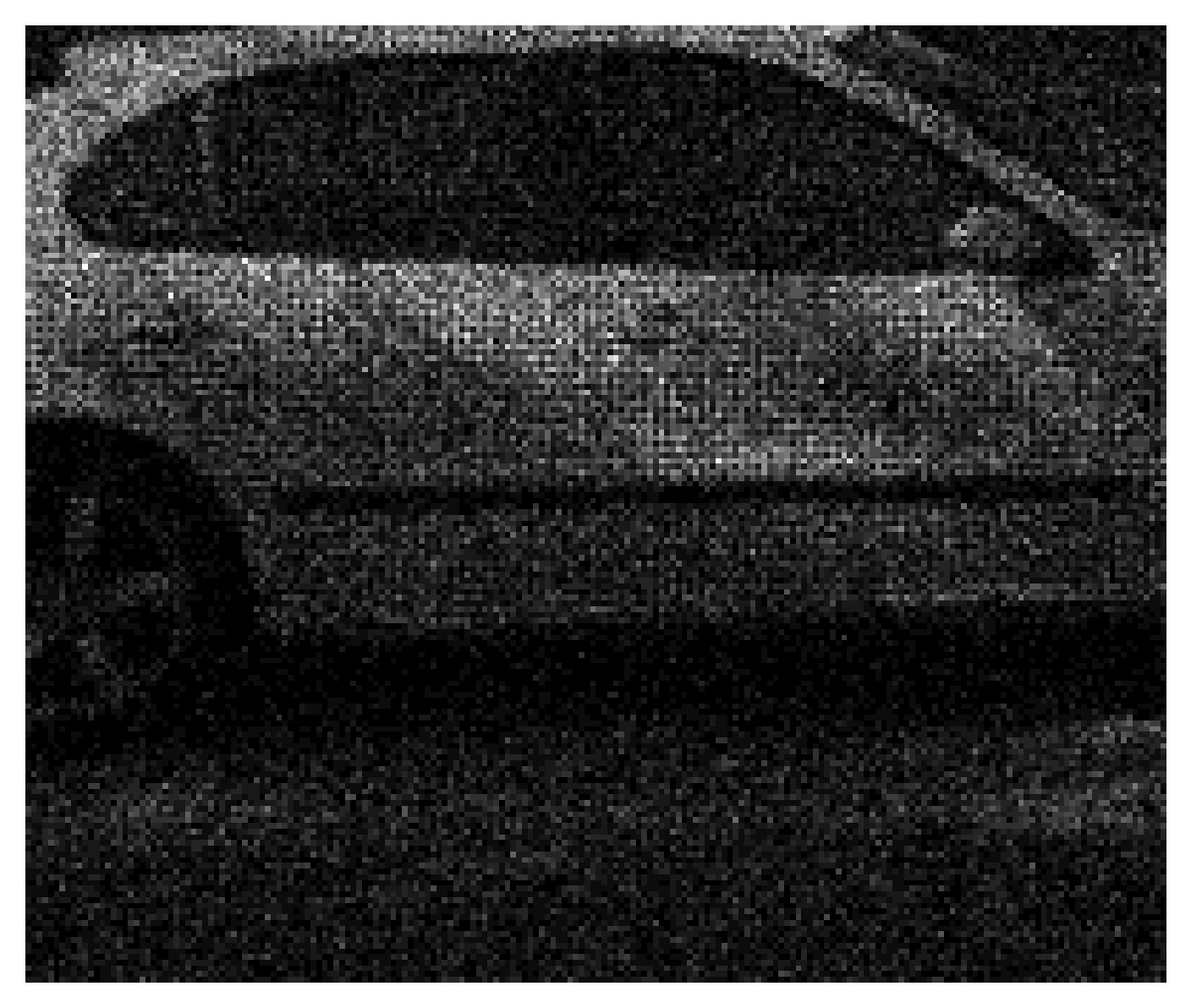}
     
 \end{subfigure}
 \hfill
 \begin{subfigure}{0.09\textwidth}
     \includegraphics[width=\textwidth]{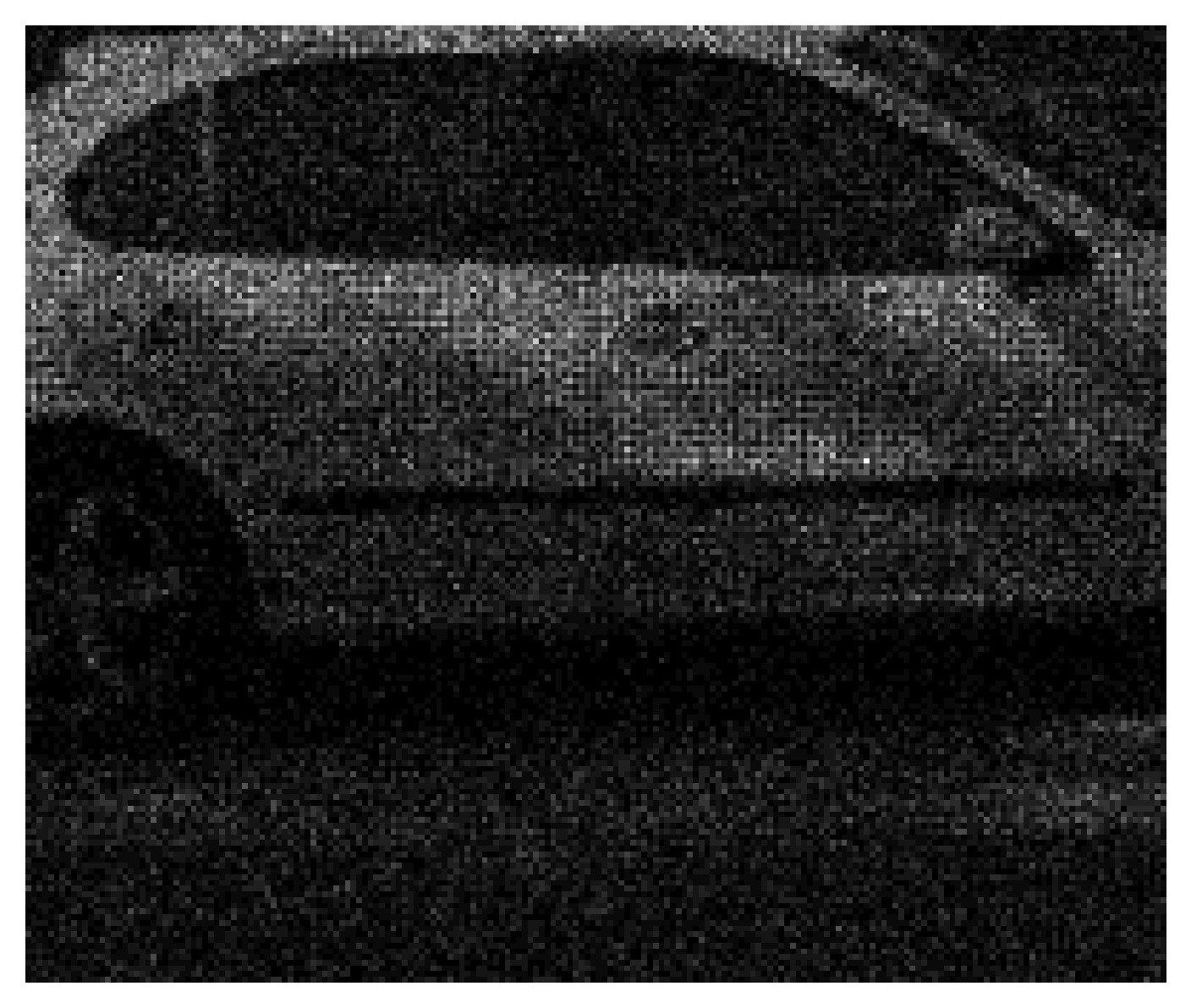}
     
 \end{subfigure}
 \hfill
 \begin{subfigure}{0.09\textwidth}
     \includegraphics[width=\textwidth]{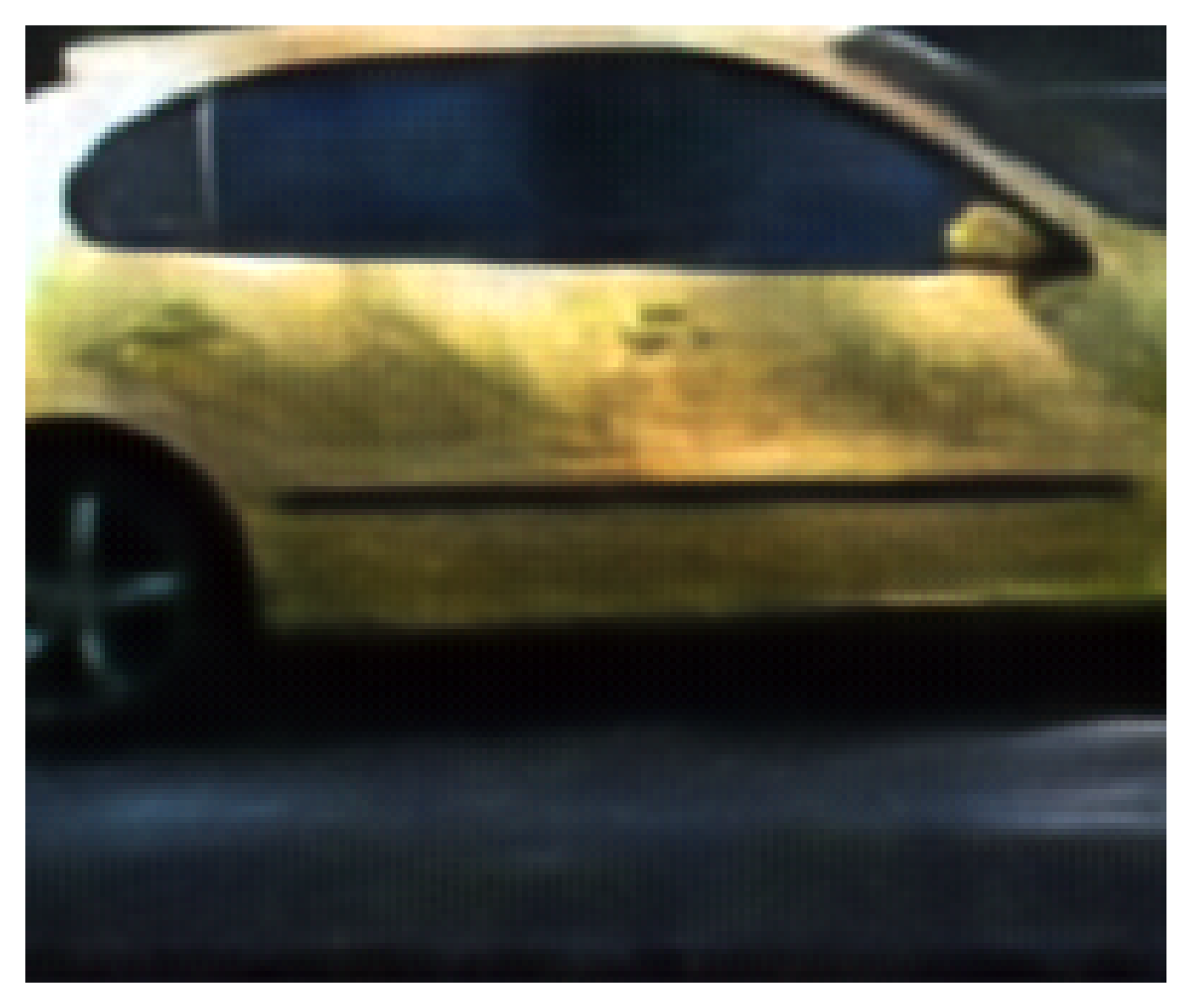}
     
 \end{subfigure}
 \hfill
 \begin{subfigure}{0.09\textwidth}
     \includegraphics[width=\textwidth]{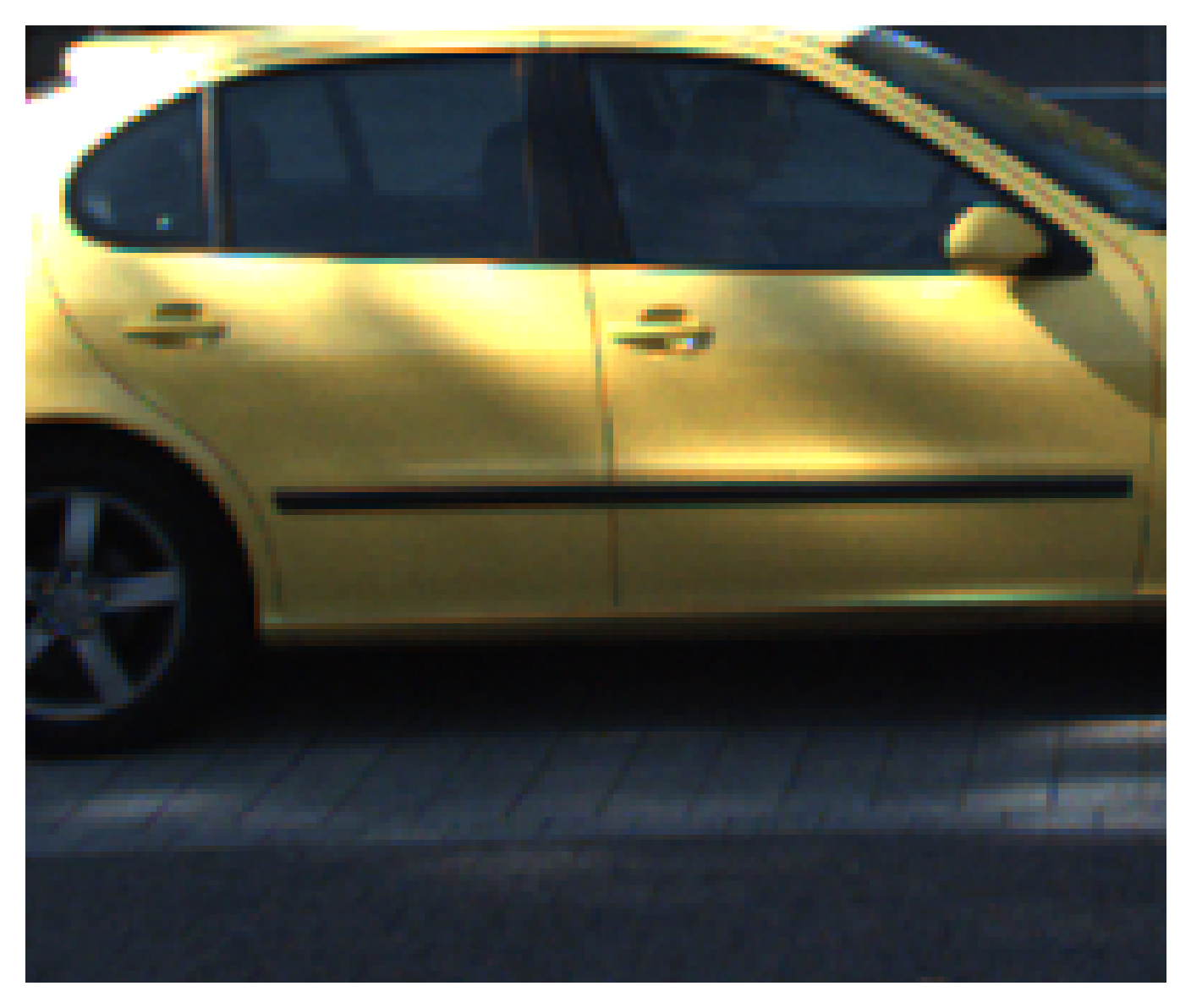}
     
 \end{subfigure}  
 \hfill
 \begin{subfigure}{0.09\textwidth}
     \includegraphics[width=\textwidth]{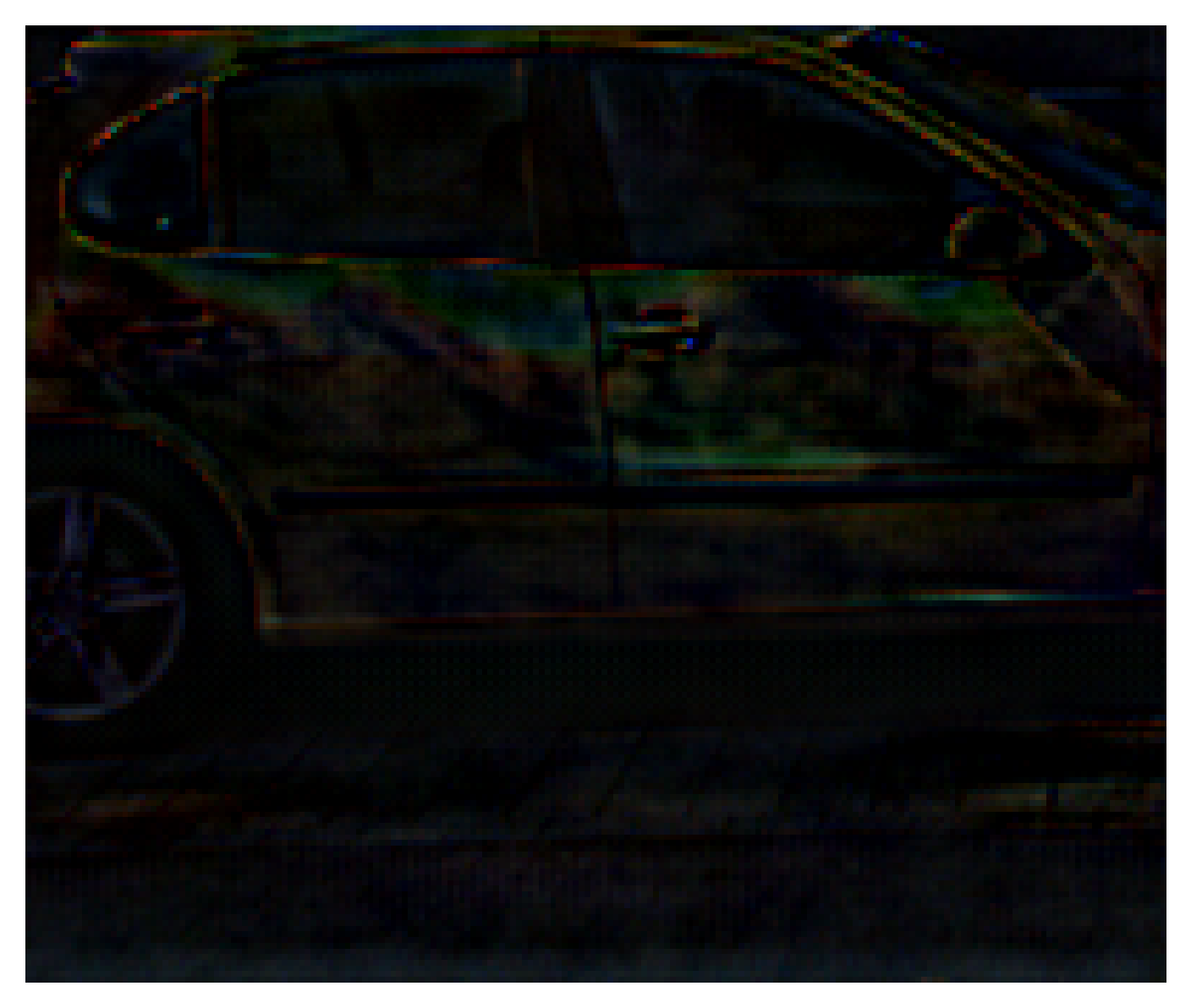}
     
 \end{subfigure} 
 \medskip
 \begin{subfigure}{0.09\textwidth}
     \includegraphics[width=\textwidth,height=.85\linewidth]{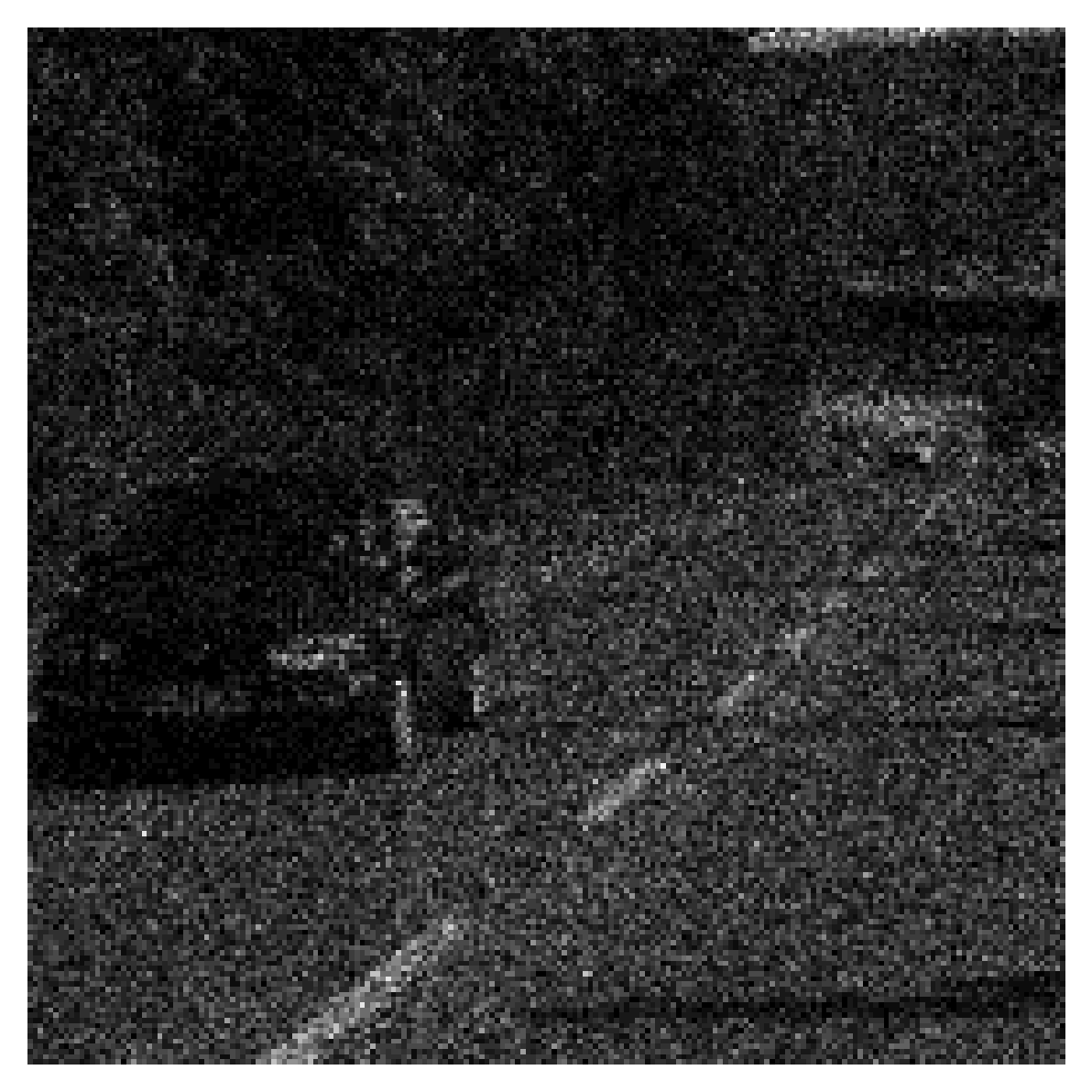}
     
 \end{subfigure}
 \hfill
 \begin{subfigure}{0.09\textwidth}
     \includegraphics[width=\textwidth,height=.85\linewidth]{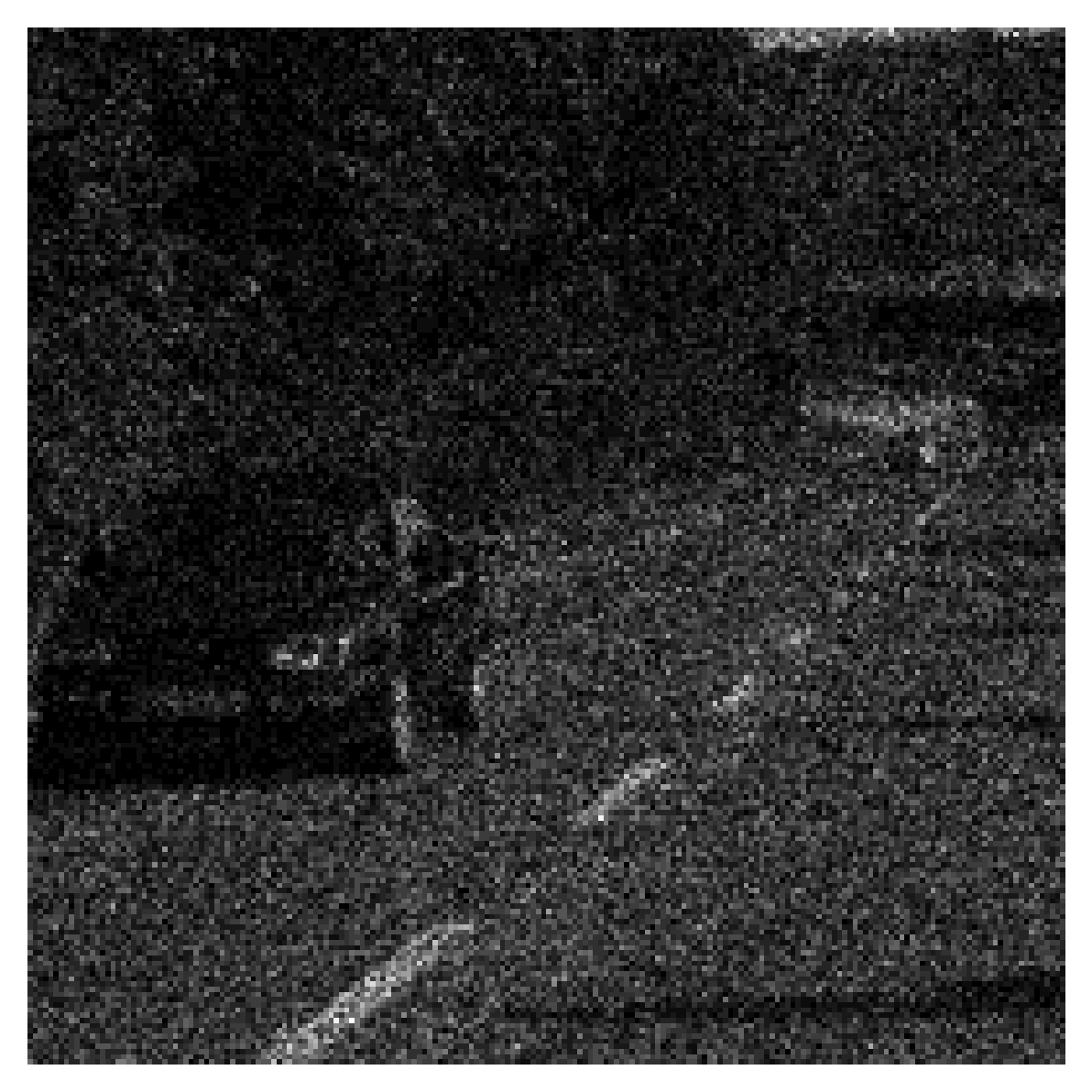}
     
 \end{subfigure}
 \hfill
 \begin{subfigure}{0.09\textwidth}
     \includegraphics[width=\textwidth,height=.85\linewidth]{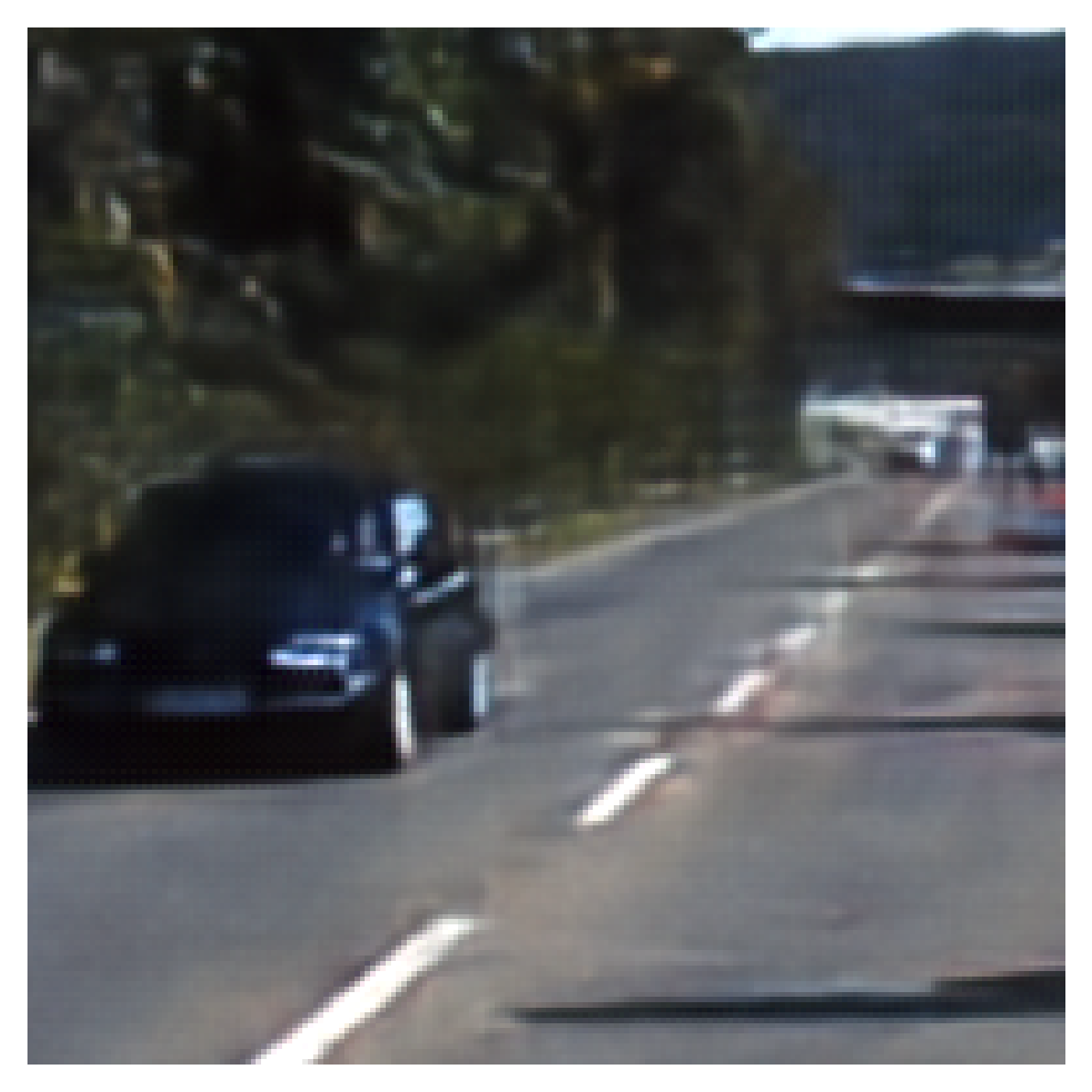}
     
 \end{subfigure}
 \hfill
 \begin{subfigure}{0.09\textwidth}
     \includegraphics[width=\textwidth,height=.85\linewidth]{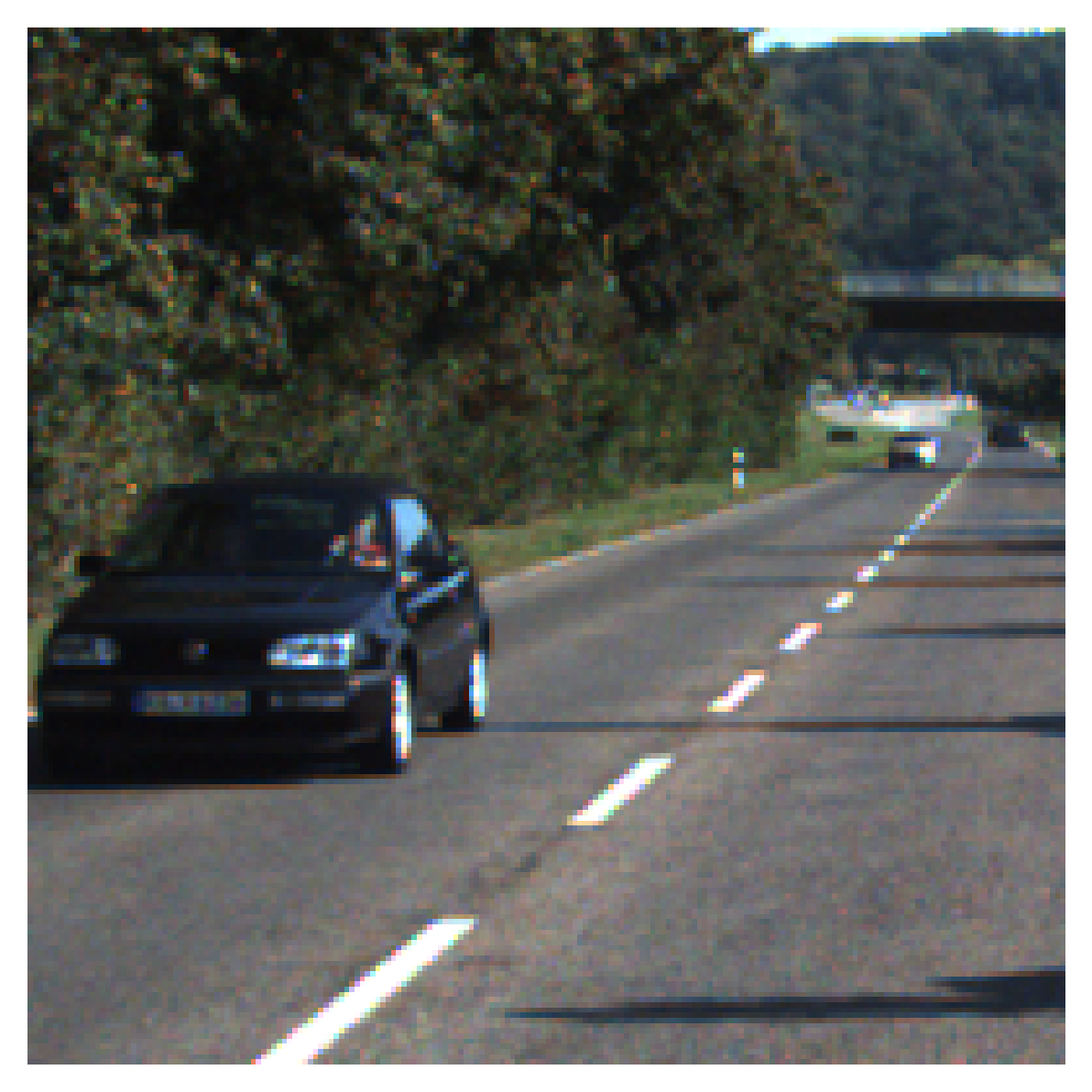}
     
 \end{subfigure}  
 \hfill
 \begin{subfigure}{0.09\textwidth}
     \includegraphics[width=\textwidth,height=.85\linewidth]{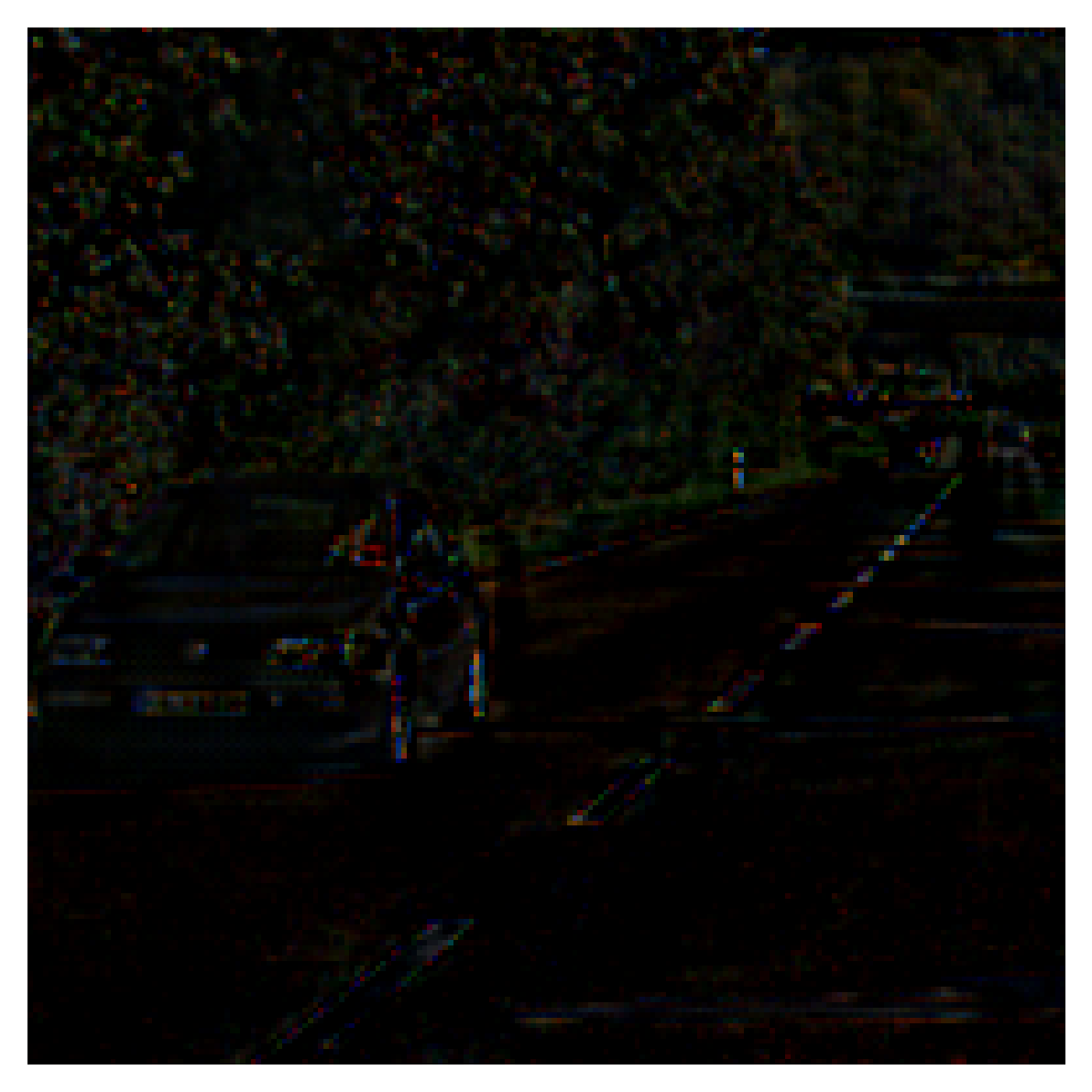}
     
 \end{subfigure}

\vspace{-10pt}
\caption{Study the upper bound of StereoISP frame work on KITTI 2015 \cite{menze2015object}. We input same images with different structured noise in a place of $M$ and $W$ in stereo DemosaicNet, i.e., the ideal case where disparity between two images is zero. From left to right: primary images $M$, warped image at disparity map = 0, denoised image, ground truth, error in reconstructions. }
 \label{fig: same image different noise results kitti}
\vspace{-10pt}
\end{figure}

\begin{figure}
 \begin{subfigure}{0.09\textwidth}
     \includegraphics[width=\textwidth,height=.85\linewidth]{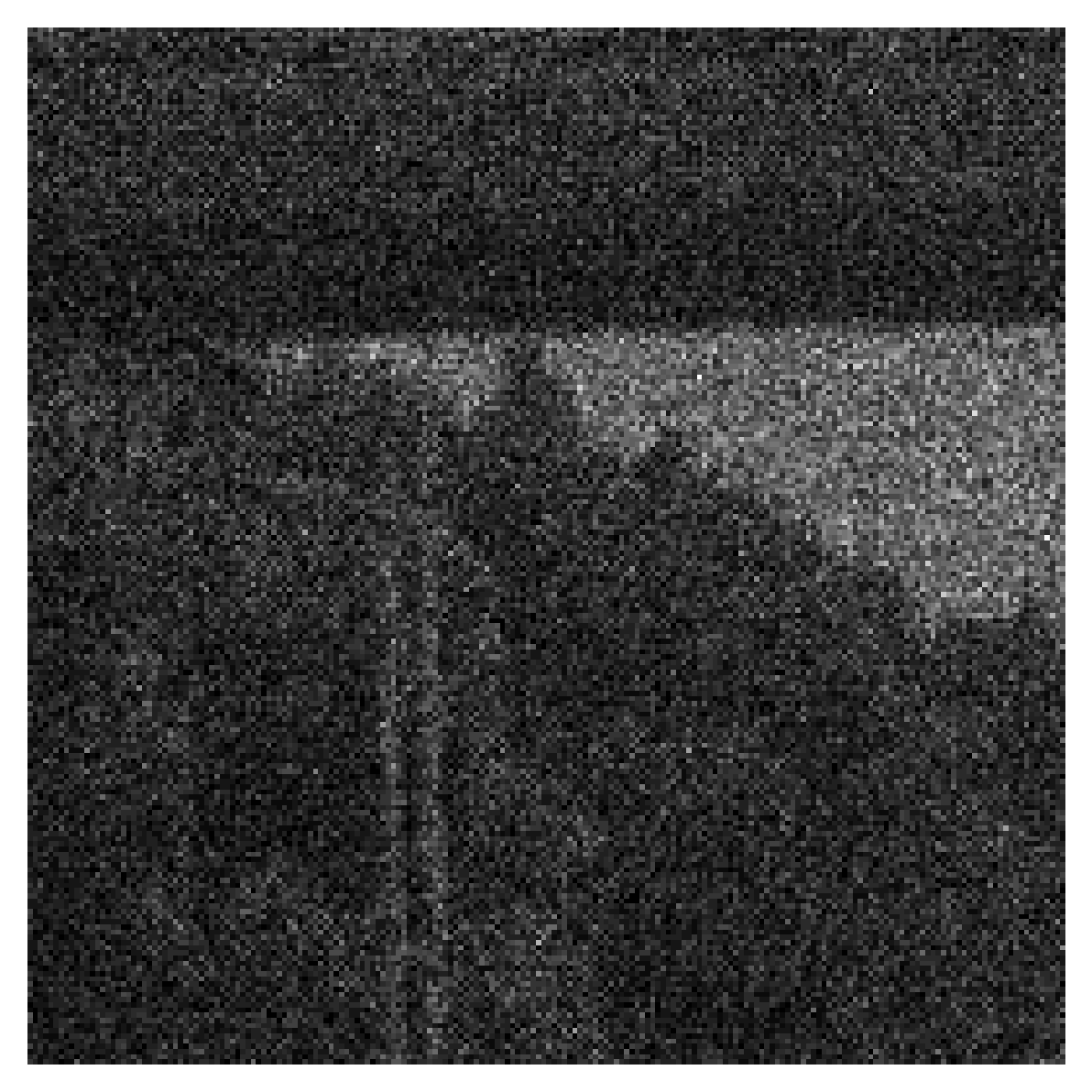}
     
 \end{subfigure}
 \hfill
 \begin{subfigure}{0.09\textwidth}
     \includegraphics[width=\textwidth,height=.85\linewidth]{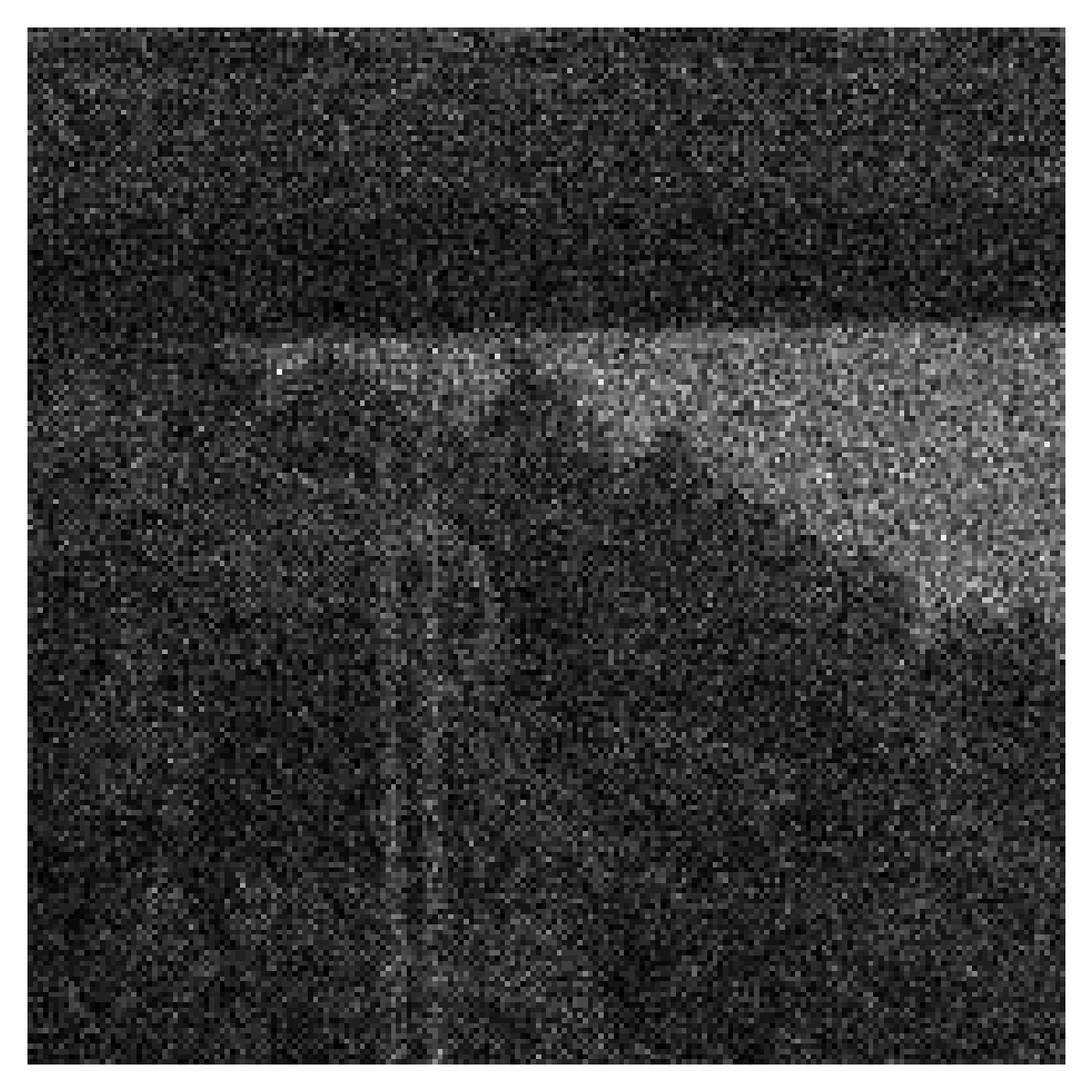}
     
 \end{subfigure}
 \hfill
 \begin{subfigure}{0.09\textwidth}
     \includegraphics[width=\textwidth,height=.85\linewidth]{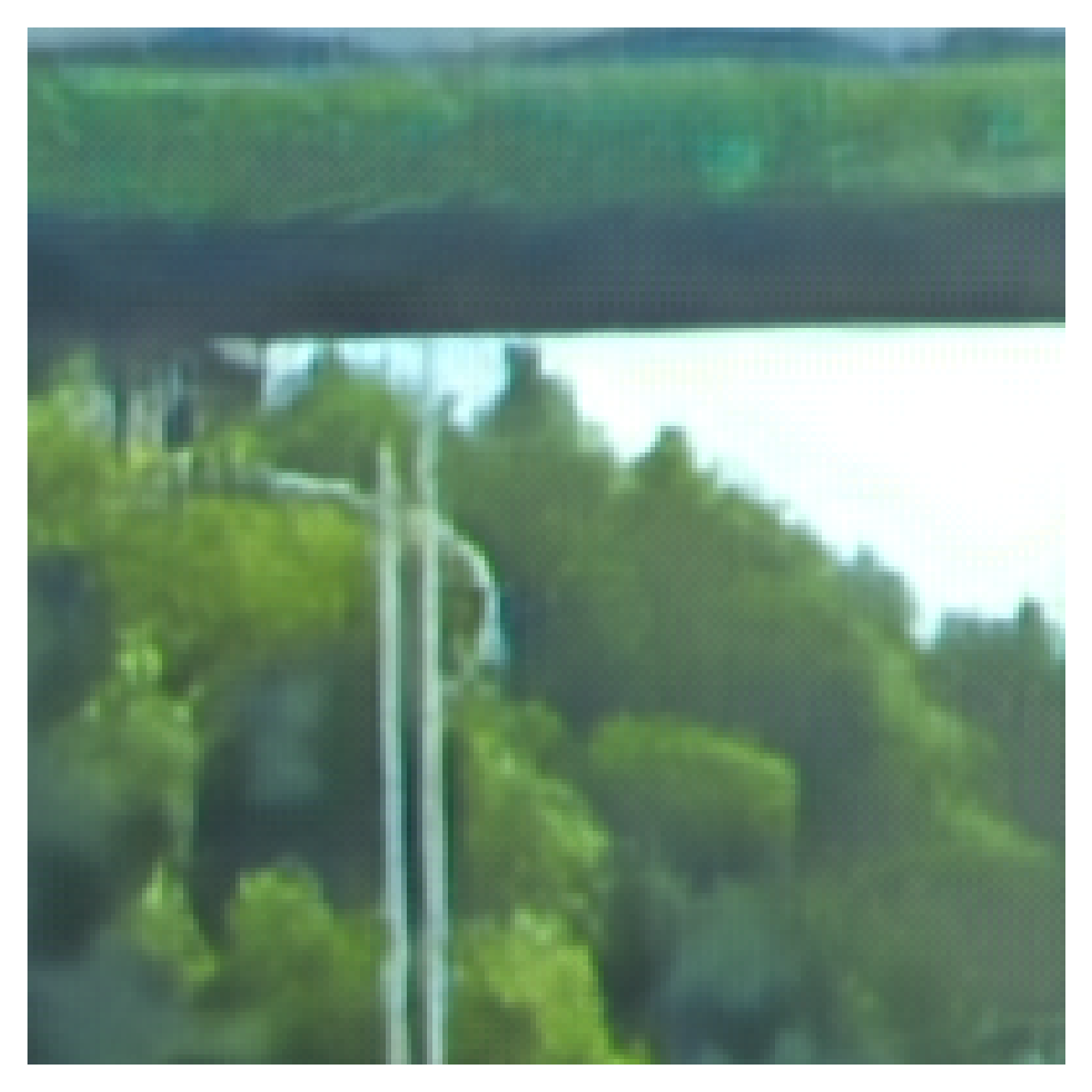}
     
 \end{subfigure}
 \hfill
 \begin{subfigure}{0.09\textwidth}
     \includegraphics[width=\textwidth,height=.85\linewidth]{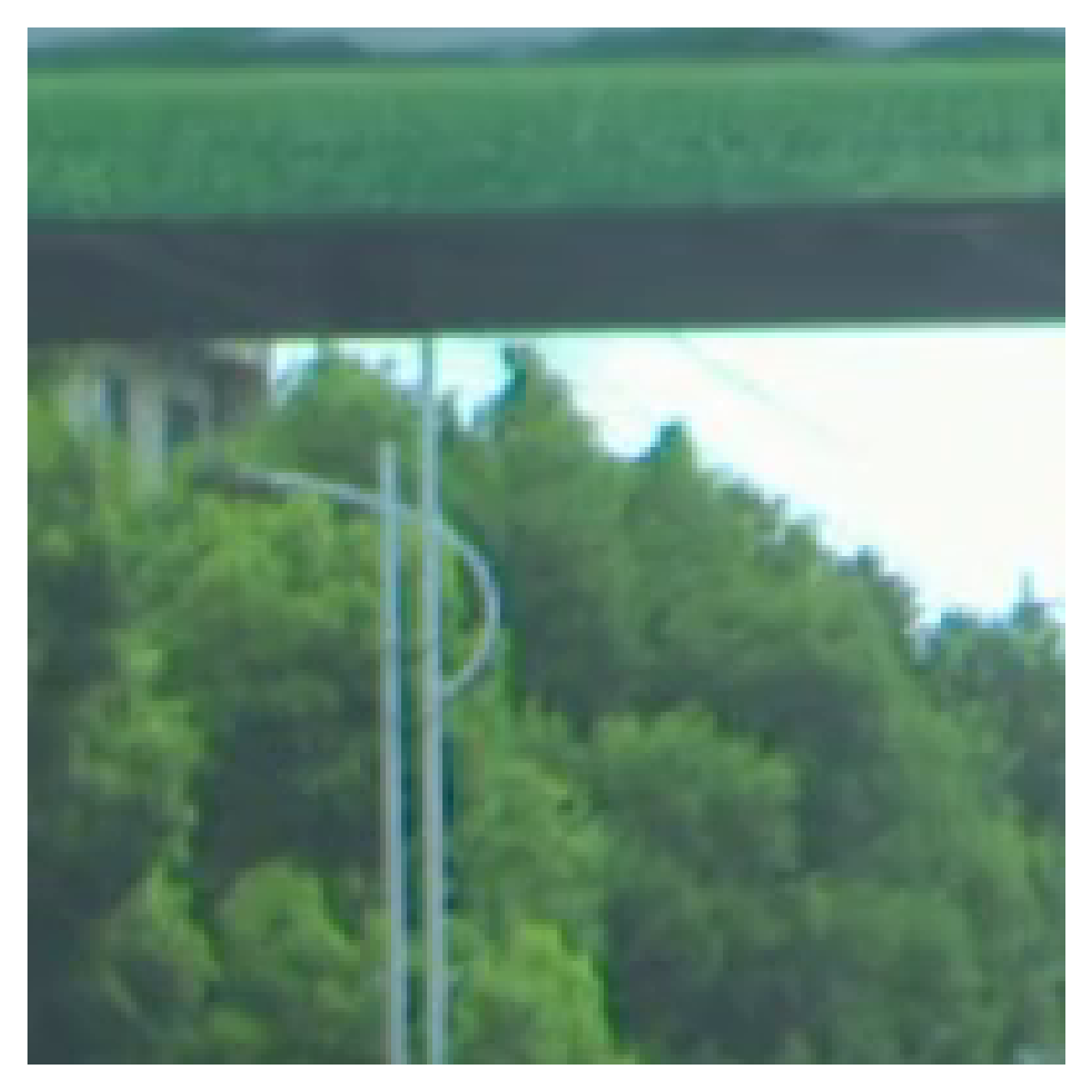}
     
 \end{subfigure}  
 \hfill
 \begin{subfigure}{0.09\textwidth}
     \includegraphics[width=\textwidth,height=.85\linewidth]{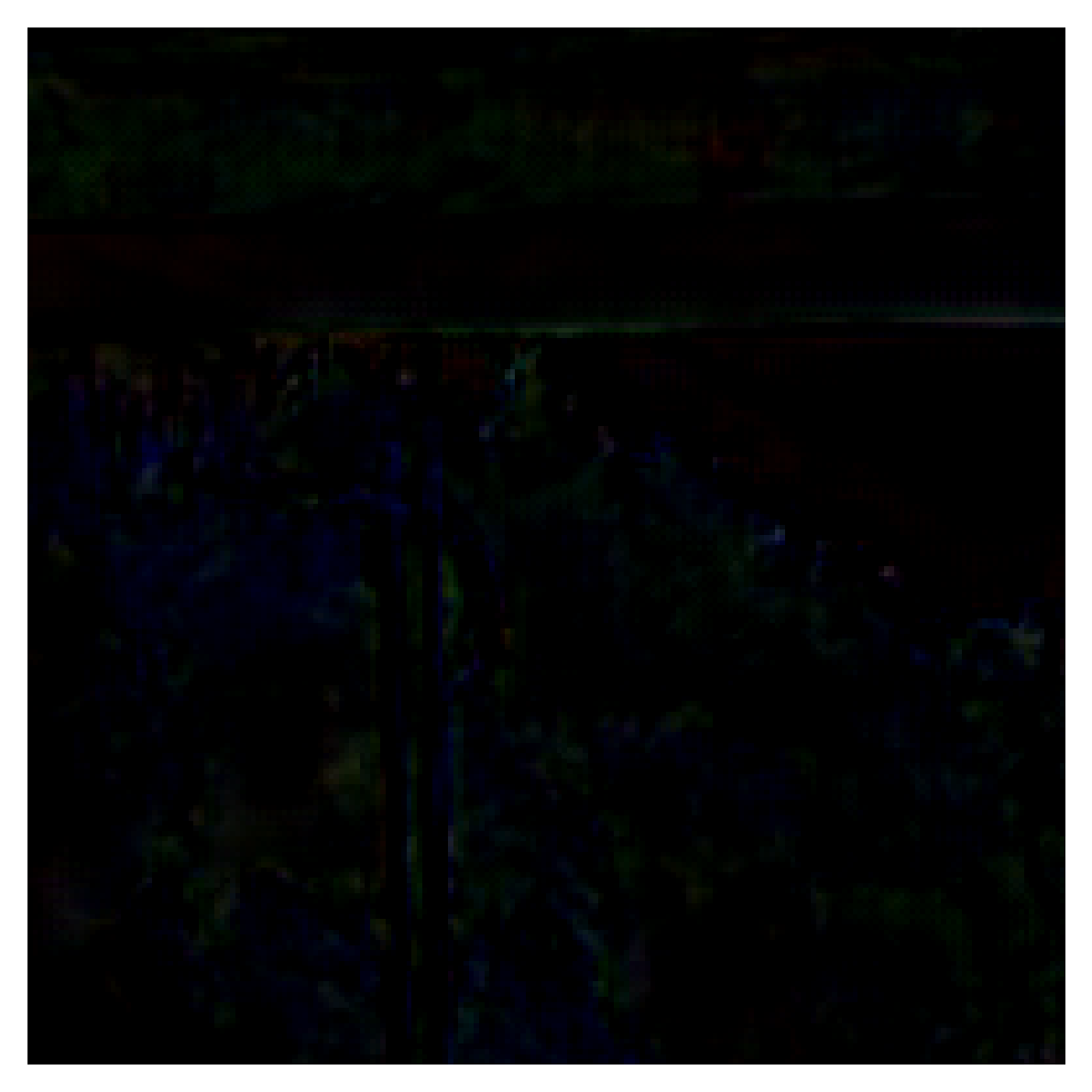}
     
 \end{subfigure}  
 \smallskip
 \begin{subfigure}{0.09\textwidth}
     \includegraphics[width=\textwidth,height=.85\linewidth]{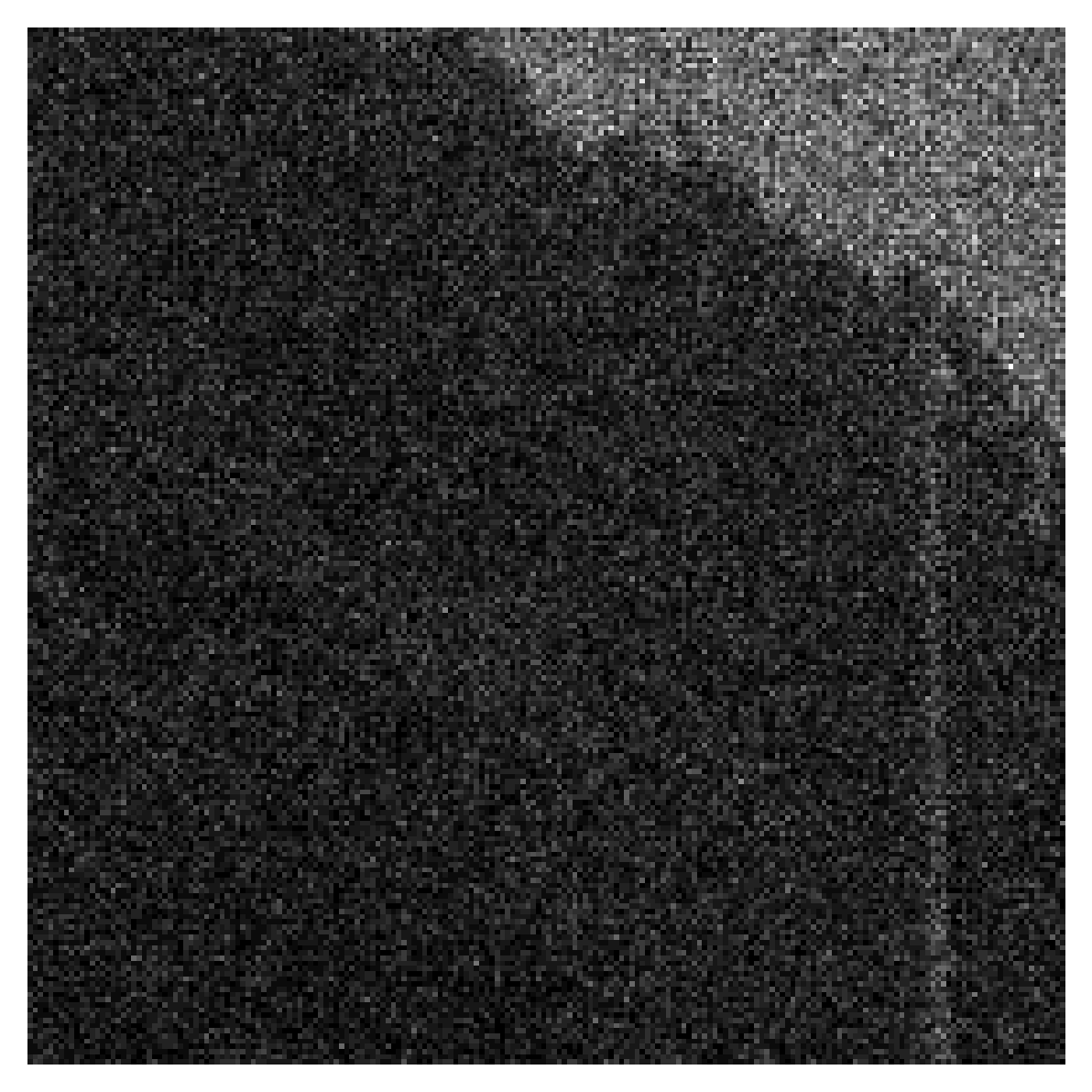}
     
 \end{subfigure}
 \hfill
 \begin{subfigure}{0.09\textwidth}
     \includegraphics[width=\textwidth,height=.85\linewidth]{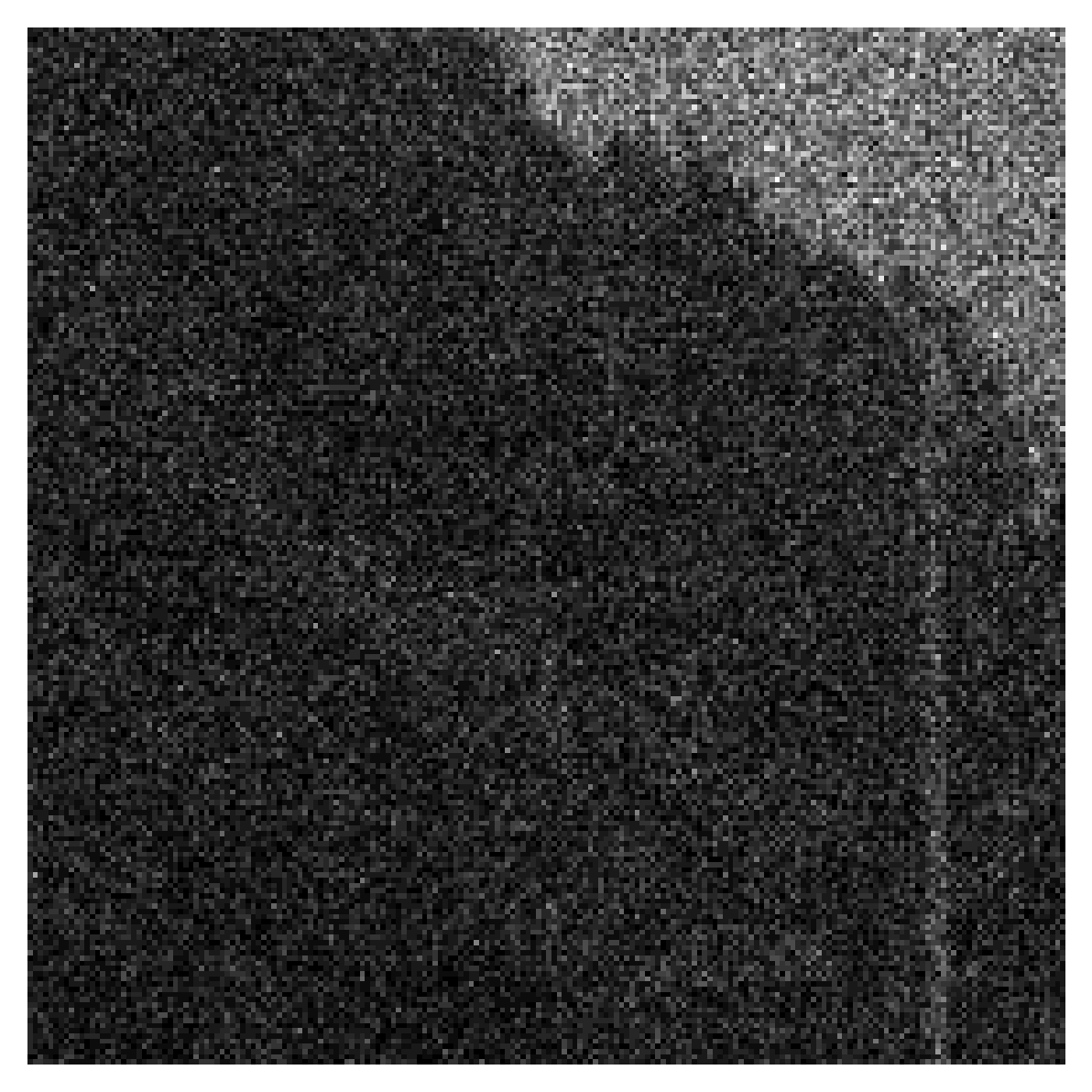}
     
 \end{subfigure}
 \hfill
 \begin{subfigure}{0.09\textwidth}
     \includegraphics[width=\textwidth,height=.85\linewidth]{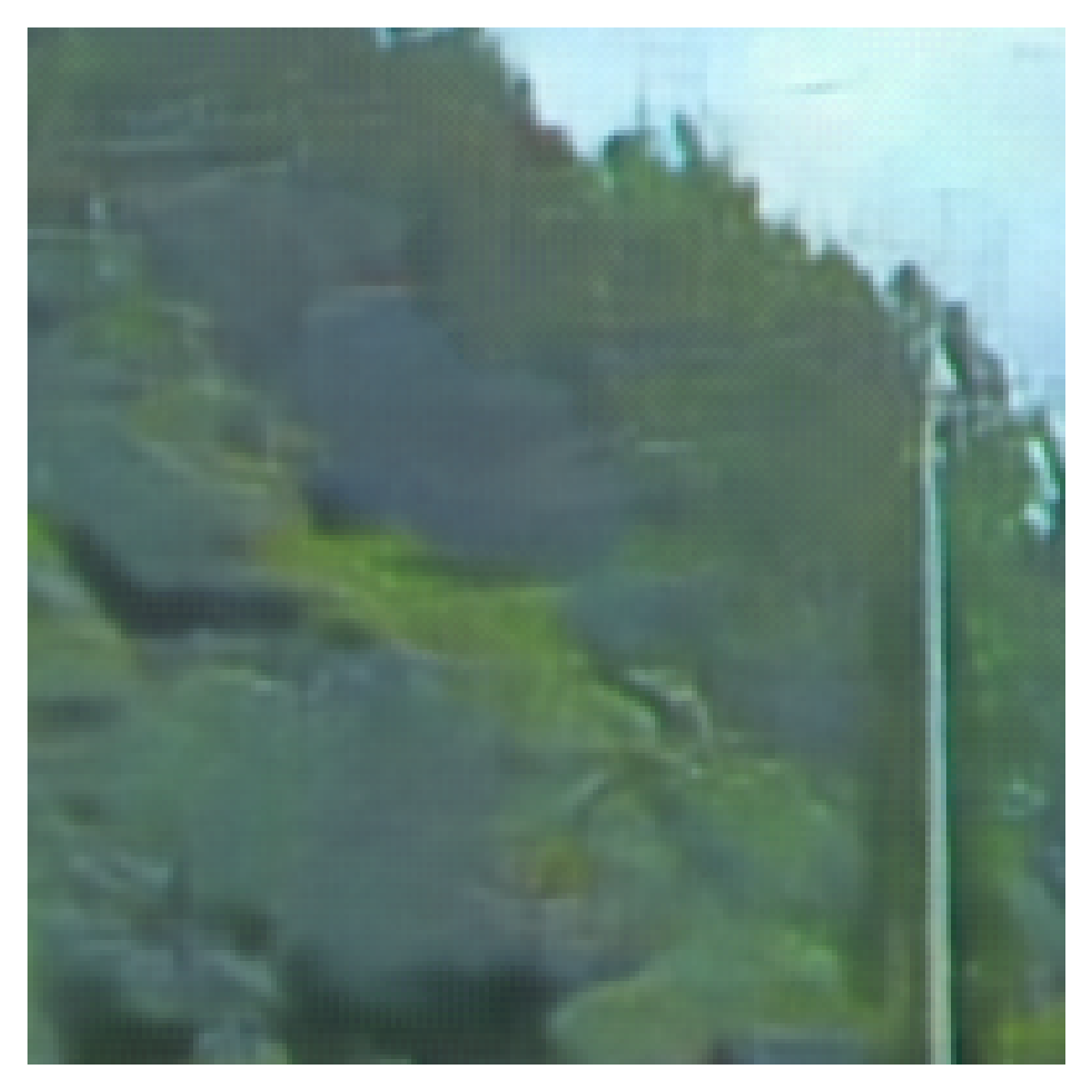}
     
 \end{subfigure}
 \hfill
 \begin{subfigure}{0.09\textwidth}
     \includegraphics[width=\textwidth,height=.85\linewidth]{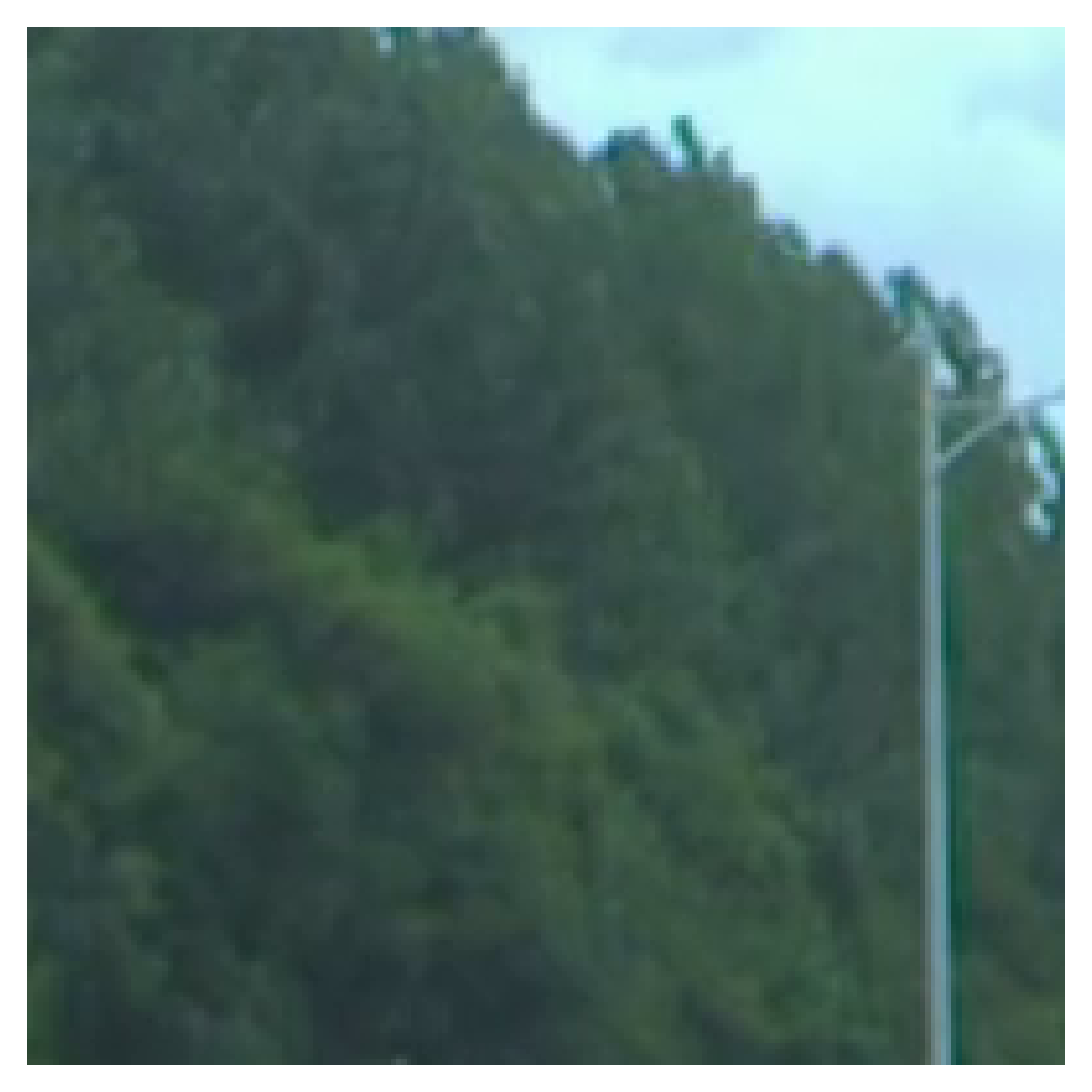}
     
 \end{subfigure}  
 \hfill
 \begin{subfigure}{0.09\textwidth}
     \includegraphics[width=\textwidth,height=.85\linewidth]{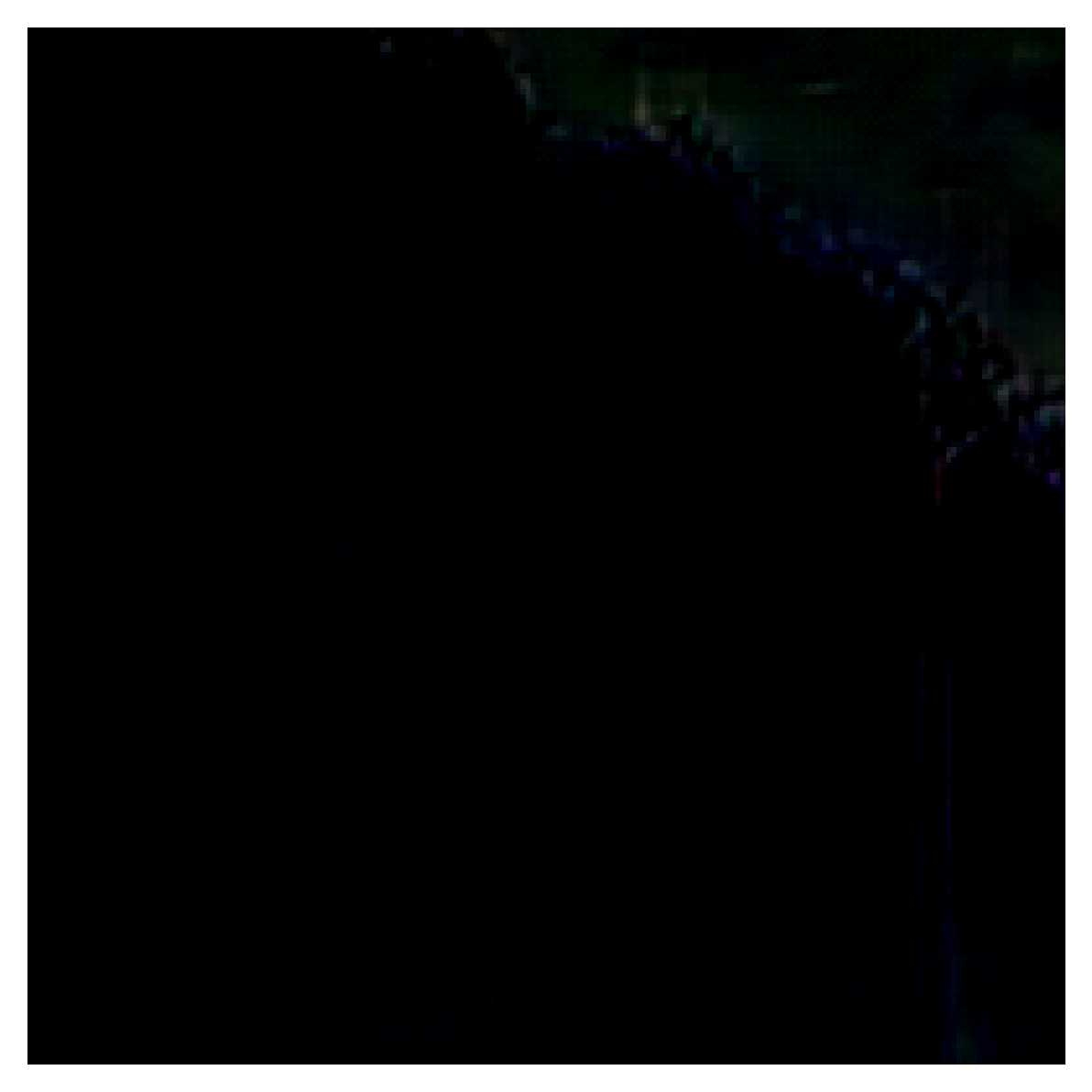}
     
 \end{subfigure} 
 \medskip
 \begin{subfigure}{0.09\textwidth}
     \includegraphics[width=\textwidth,height=.85\linewidth]{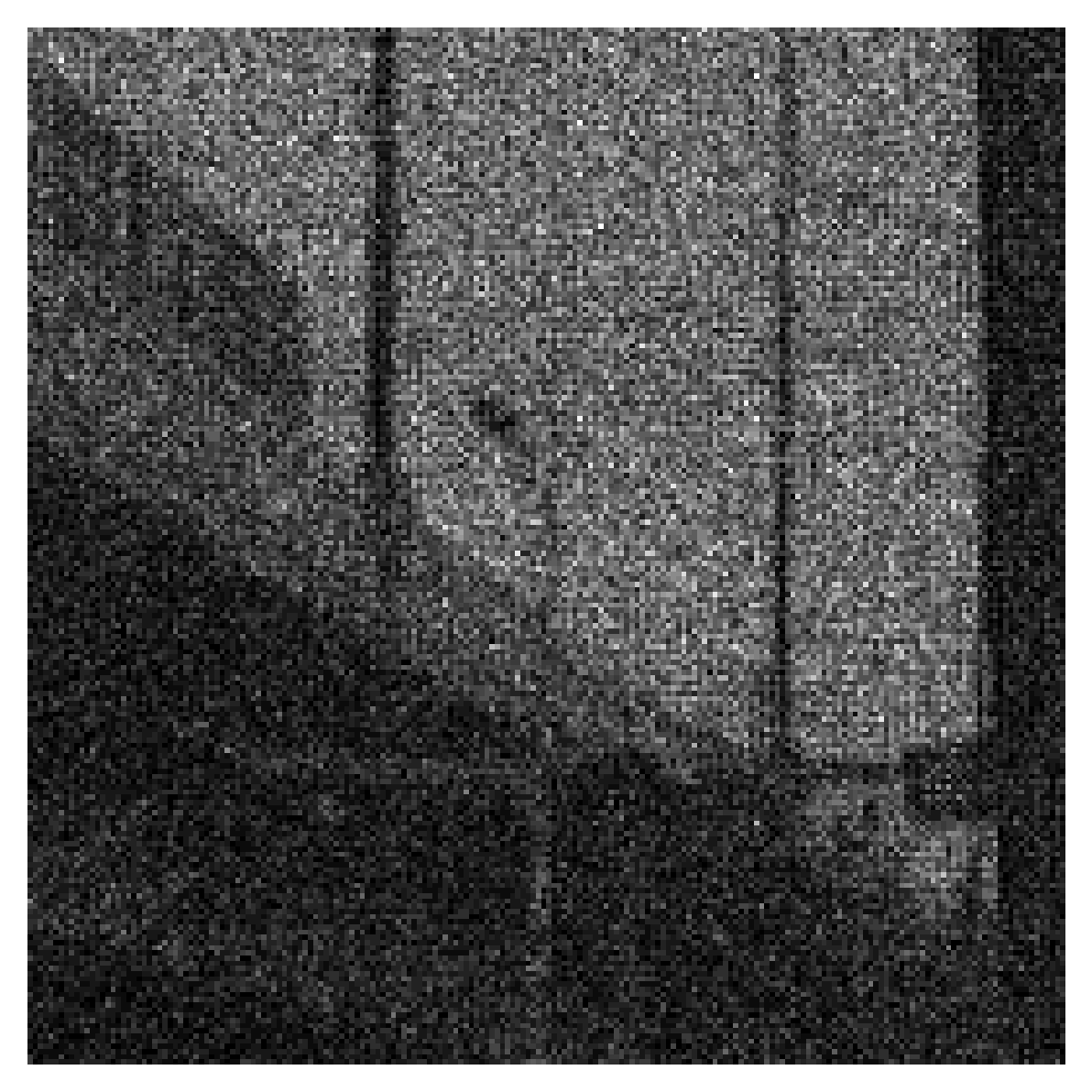}
     
 \end{subfigure}
 \hfill
 \begin{subfigure}{0.09\textwidth}
     \includegraphics[width=\textwidth,height=.85\linewidth]{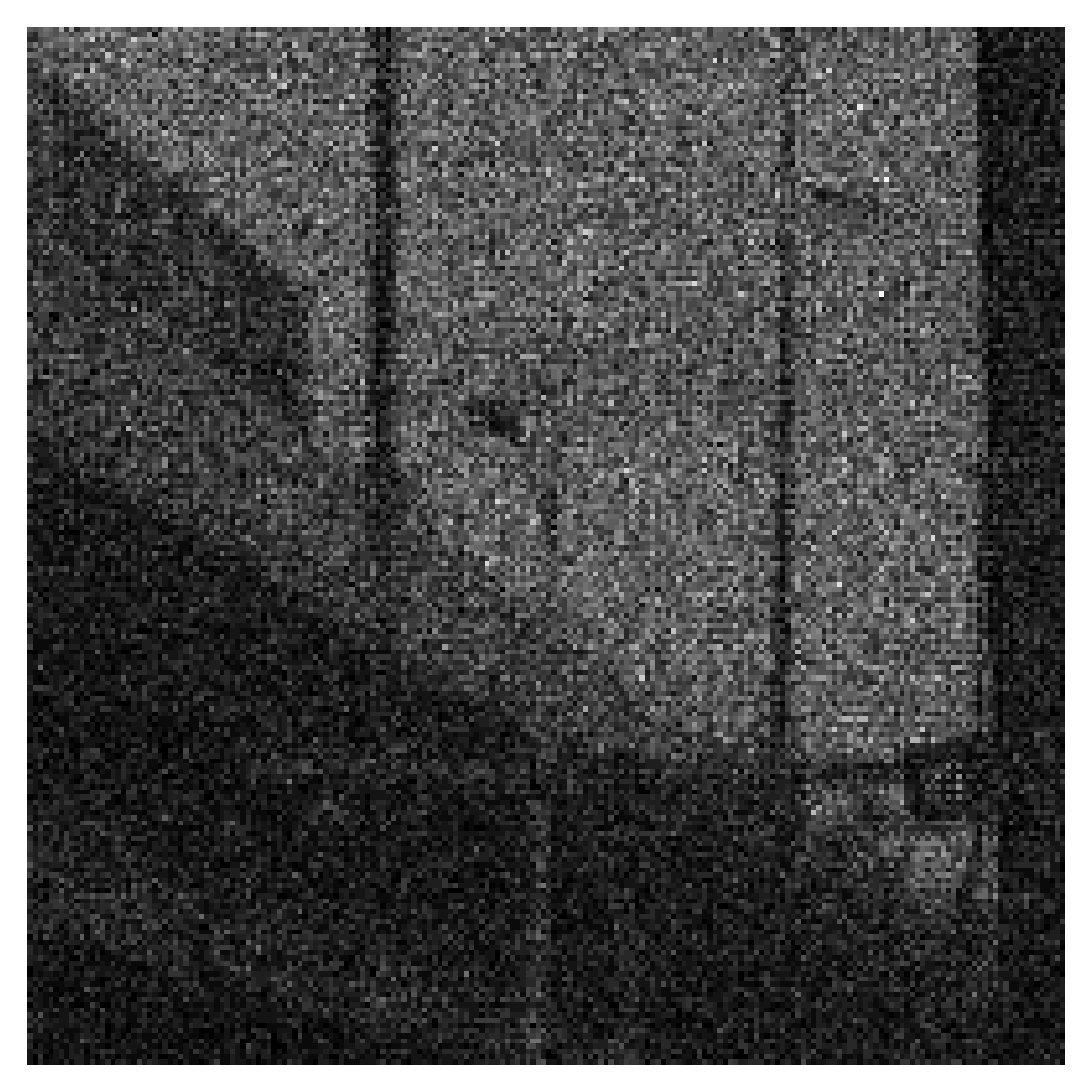}
     
 \end{subfigure}
 \hfill
 \begin{subfigure}{0.09\textwidth}
     \includegraphics[width=\textwidth,height=.85\linewidth]{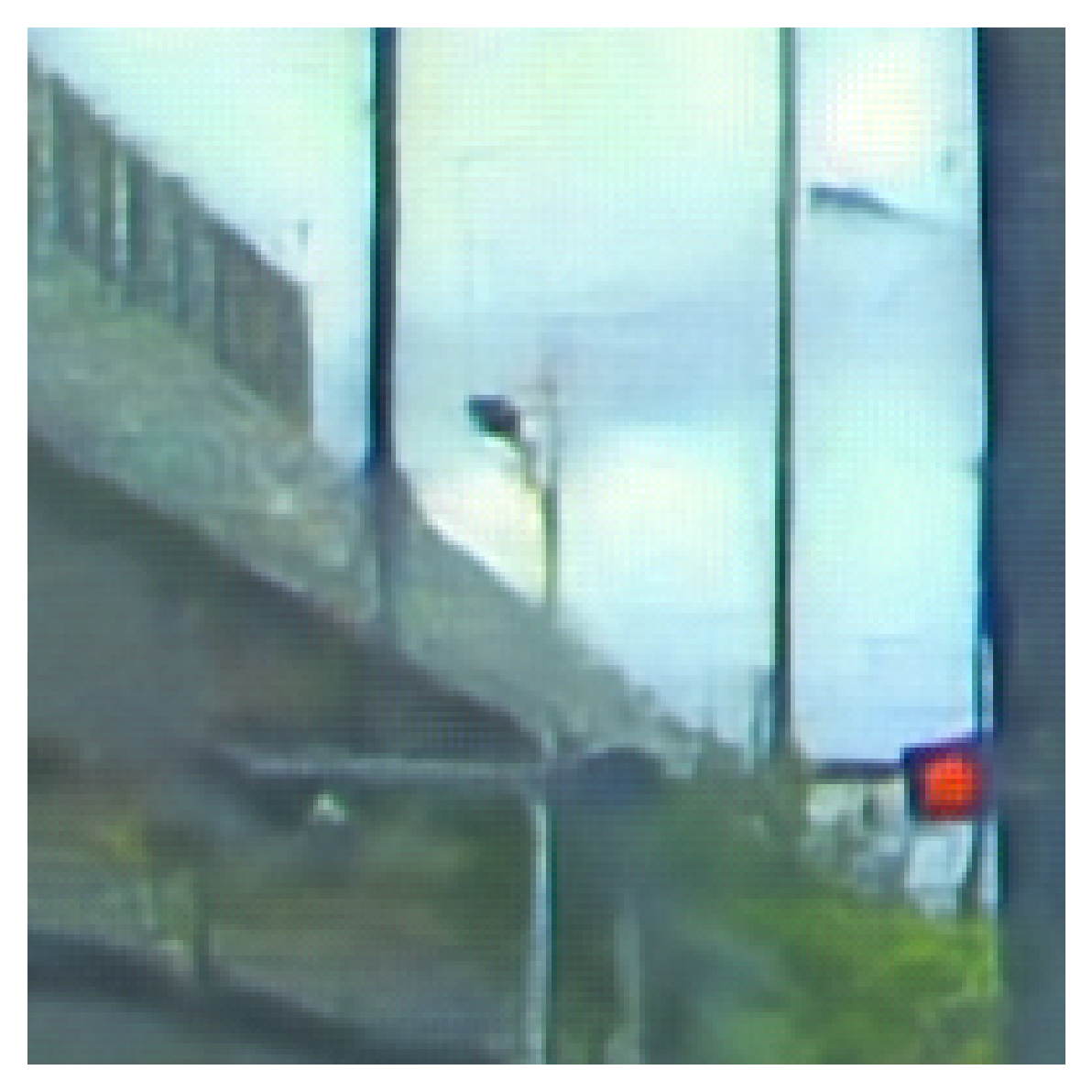}
     
 \end{subfigure}
 \hfill
 \begin{subfigure}{0.09\textwidth}
     \includegraphics[width=\textwidth,height=.85\linewidth]{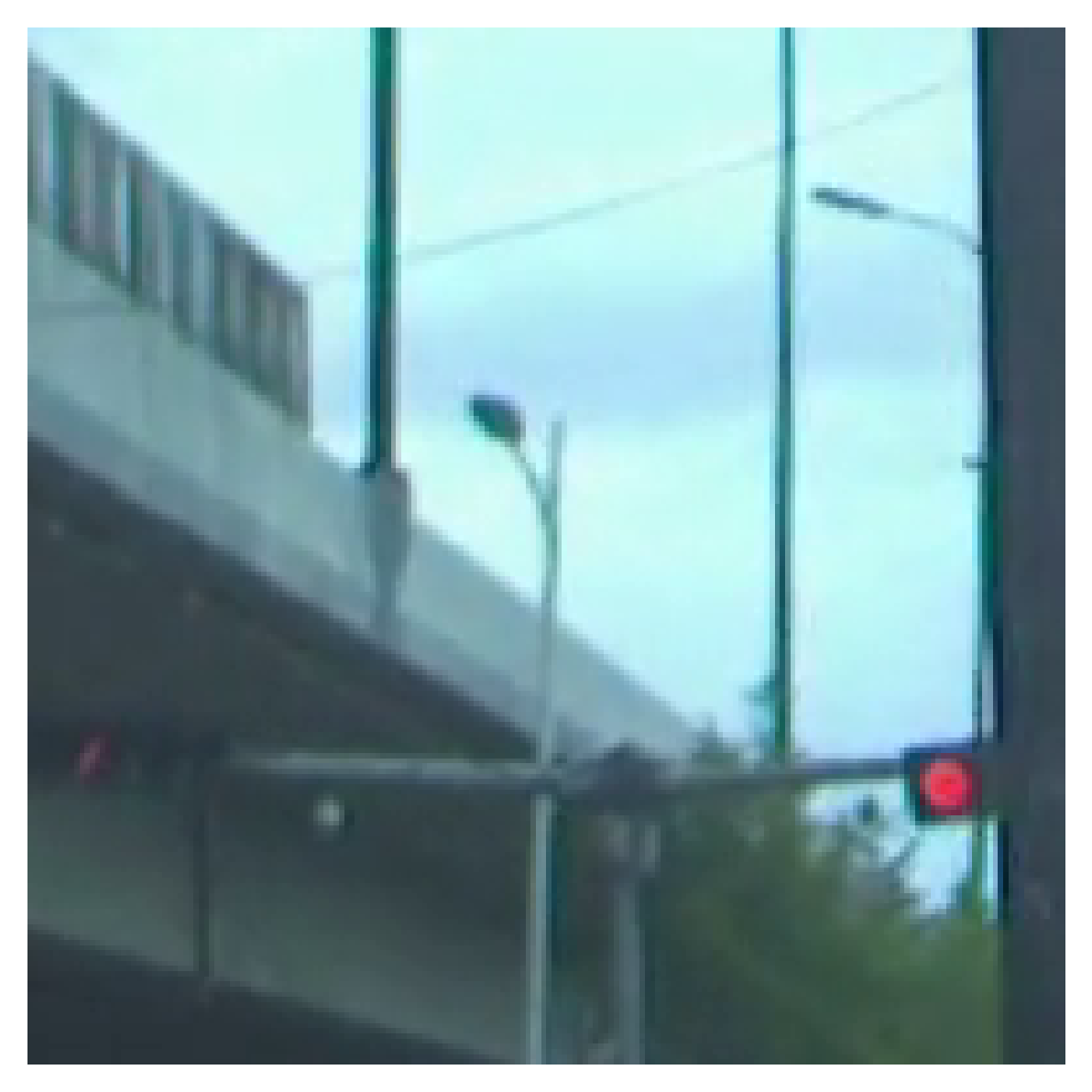}
     
 \end{subfigure}  
 \hfill
 \begin{subfigure}{0.09\textwidth}
     \includegraphics[width=\textwidth,height=.85\linewidth]{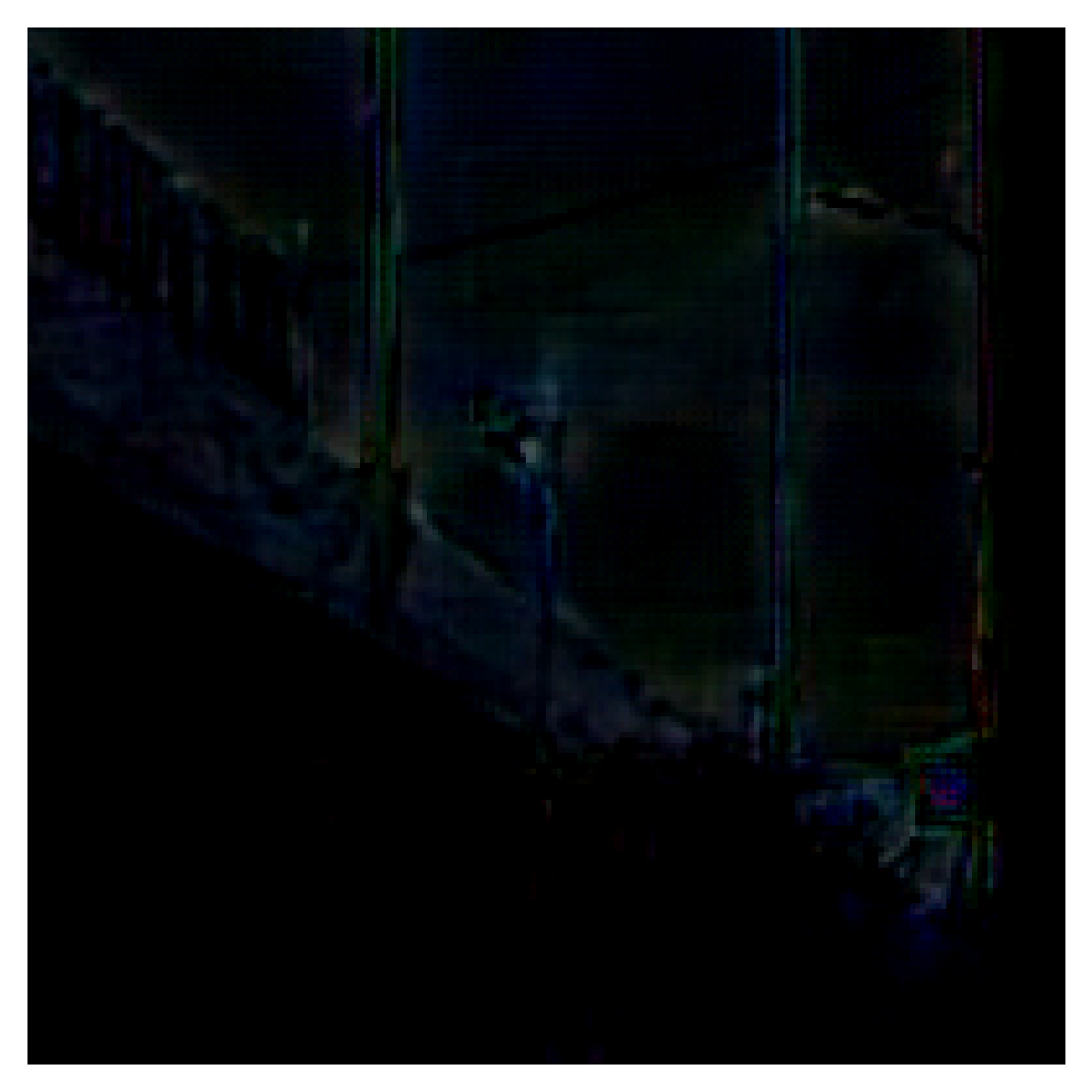}
     
 \end{subfigure}

\vspace{-10pt}
\caption{Study the upper bound of StereoISP frame work on drivingStereo \cite{yang2019drivingstereo}. We input same images with different structured noise in a place of $M$ and $W$ in stereo DemosaicNet, i.e., the ideal case where disparity between two images is zero. From left to right: primary images $M$, warped image at disparity map = 0, denoised image, ground truth, error in reconstructions. }
 \label{fig: same image different noise results drivingstereo}
\vspace{-10pt}
\end{figure}

We assume that we can get the maximum quality of the image if the disparity between the two images equals 0. Therefore, we trained Stereo DemosaicNet to find out the upper bound of the model. We corrupt the same image with different noise structures with the same noise level $\sigma$. The model successfully performed demosaicing and denoising as shown in figures \ref{fig: same image different noise results kitti} and \ref{fig: same image different noise results drivingstereo}. It shows that an error between two images has appeared only on the edges of objects. Although this case is not realistic as shot noise happened in cameras, it became useful to test the maximum image quality that the model can provide.
\vspace{-10pt}

\subsubsection{Image-pair without warping}

\label{subsec: without warping}

\begin{table}
  \centering
  
  \begin{tabular}{@{}lc@{}c@{}}
    \toprule
    \toprule
    & KITTI 2015 & drivingStereo\\
    \toprule
    Method & \multicolumn{2}{c}{PSNR (dB)} \\
    \midrule
    Baseline \cite{gharbi2016deep} & 24.47 & 29.59\\
    Unwarped (\cref{subsec: without warping}) & 24.37 (-0.41\%) & -\\
    \textbf{Ours} (\cref{subsec: Stereo DemosaicNet})  & \textbf{26.79 (+9.48\%)}& \textbf{32.02 (+8.2\%)} \\
    Same (\cref{subsec: Same input}) & 27.38 (+11.89\%) & 32.61 (+10.21\%)\\
    \bottomrule
    \bottomrule
  \end{tabular}
  \caption{Ablation study Stereo DemosaicNet model based on StereoISP framework.}
    
  \label{tab: ablation study}
\end{table}

We study the effect of image warping block $\mathcal{W(\cdot)}$ by removing it and passing a stereo-pair directly to the stereo demosaicing/denoising module $\mathcal{D_MD_N}$. In this case, the module is responsible to perform three major tasks: (1) estimate the disparity between the image-pair. (2) Warp the secondary image onto the primary image. (3) integrate the warped image with the primary image. As stereo DemosaicNet was designed to handle denoising tasks, it was hard to achieve other tasks. In table \ref{tab: ablation study}, we quantified the performance of the ablation study and our proposed model. It is not surprising that feeding the stereo-pair into the model not only does not change the image quality, but it degrades the performance below the baseline model \cite{gharbi2016deep}.

{\small
\bibliographystyle{ieee_fullname}
\bibliography{egbib}
}

\end{document}